\documentclass{article}

\usepackage{standalone}
\usepackage{geometry}
\usepackage{natbib}

\usepackage{amsmath,amssymb,amsfonts,amsthm}
\usepackage{graphicx}
\usepackage{booktabs}
\usepackage{multirow}
\usepackage{caption}
\usepackage{longtable}
\usepackage{float}
\usepackage{enumitem}
\usepackage{mathrsfs}
\usepackage{array,rotating}
\usepackage{hyperref}
\usepackage{xr}
\usepackage{bm}
\hypersetup{
	colorlinks=true,
	linkcolor=blue,
	citecolor=red,
	urlcolor=cyan
}

\newtheorem{theorem}{Theorem}
\newtheorem{proposition}[theorem]{Proposition}

\newtheorem{corollary}[theorem]{Corollary}
\theoremstyle{definition}

\newtheorem{remark}{Remark}

\newtheorem{innercustomlm}{Lemma}

\newcommand{\cvp}{\overset{p}{\to}}
\newcommand{\cvd}{\overset{d}{\to}}

\newenvironment{customlm}[1]
{\renewcommand\theinnercustomlm{#1}\innercustomlm}
{\endinnercustomlm}

\newtheorem{innercustomth}{Theorem}
\newenvironment{customth}[1]
{\renewcommand\theinnercustomth{#1}\innercustomth}
{\endinnercustomth}

\newtheorem{innercustomass}{Assumption}
\newenvironment{customass}[1]
{\renewcommand\theinnercustomass{#1}\innercustomass}
{\endinnercustomass}

\newtheorem{innercustomcor}{Corollary}
\newenvironment{customcor}[1]
{\renewcommand\theinnercustomcor{#1}\innercustomcor}
{\endinnercustomass}

\newtheorem{innercustompro}{Proposition}
\newenvironment{custompro}[1]
{\renewcommand\theinnercustompro{#1}\innercustompro}
{\endinnercustompro}

\renewcommand{\thetable}{\arabic{table}}
\renewcommand{\thefigure}{\arabic{figure}}
\renewcommand{\thesection}{\arabic{section}}
\renewcommand{\theequation}{\arabic{equation}}

\usepackage{appendix}

\begin{document}
	
	


\renewcommand{\baselinestretch}{2}

\markright{ \hbox{\footnotesize\rm Statistica Sinica
}\hfill\\[-13pt]
\hbox{\footnotesize\rm
}\hfill }

\markboth{\hfill{\footnotesize\rm Kun Yi AND Lucy Xia} \hfill}
{\hfill {\footnotesize\rm Robust Inference for Seamless II/III Trial} \hfill}

\renewcommand{\thefootnote}{}


\fontsize{12}{14pt plus.8pt minus .6pt}\selectfont 
\vspace{0.8pc}

\begin{center}
	\large\bf Model-robust Inference for Seamless II/III Trials
	
	\vspace{2pt}
	
	\large\bf with Covariate Adaptive Randomization
	
	\vspace{.4cm}
	
	Kun Yi and Lucy Xia$^{\ast}$\footnotemark
	
	\vspace{.4cm}
	
	\textit{Department of ISOM, HKUST}
\end{center}

\footnotetext{$^{\ast}$\,Corresponding author: lucyxia@ust.hk}

\vspace{.55cm}
\fontsize{9}{11.5pt plus.8pt minus.6pt}\selectfont


\begin{quotation}
\noindent {\it Abstract:} Seamless phase II/III trials have become a cornerstone of modern drug development, offering a means to accelerate evaluation while maintaining statistical rigor. However, most existing inference procedures are model-based, designed primarily for continuous outcomes, and often neglect the stratification used in covariate-adaptive randomization (CAR), limiting their practical relevance. In this paper, we propose a unified, model-robust framework for seamless phase II/III trials grounded in generalized linear models (GLMs), enabling valid inference across diverse outcome types, estimands, and CAR schemes. Using $Z$-estimation, we derive the asymptotic properties of treatment effect estimators and explicitly characterize how their variance depends on the underlying randomization procedure. Based on these results, we develop adjusted Wald tests that, together with Dunnett’s multiple-comparison procedure and the inverse–$\chi^2$ combination method, ensure valid overall Type~I error. Extensive simulation studies and a trial example demonstrate that the proposed model-robust tests achieve superior power and reliable inference compared to conventional approaches.

\vspace{9pt}
\par
\end{quotation}\par

\def\thefigure{\arabic{figure}}
\def\thetable{\arabic{table}}

\renewcommand{\theequation}{\thesection.\arabic{equation}}

\fontsize{12}{14pt plus.8pt minus .6pt}\selectfont

\section{Introduction}


The integration of seamless phase II/III clinical trials into modern drug development represents a transformative shift toward accelerating therapeutic evaluation while maintaining statistical rigor \citep{jennison2007adaptive,hampson2015optimizing}. Driven by regulatory mandates to streamline clinical research \citep{FDA2006}, seamless designs  are introduced to combine dose selection (phase II) and confirmatory analysis (phase III) under a single protocol. By eliminating the traditional 6-12 month gap between phases, these  designs reduce operational delays, lower costs, and enable continuous patient monitoring, thereby enhancing both pharmaceutical profitability and patient access to novel therapies \citep{prowell2016seamless}. However, their adoption introduces formidable statistical challenges, including inflated Type~I error rates due to treatment selection and multiplicity \citep{bauer2010selection}, and the need to reconcile adaptive designs with covariate-adaptive randomization (CAR), an essential mechanism for maintaining covariate balance in modern trials \citep{shao2010theory}.

Existing methodologies for seamless trials focus predominantly on continuous outcomes, relying on linear models and conventional Wald-tests  for inference \citep{liu2002unified, schmidli2006confirmatory}. However, clinical endpoints are often discrete (e.g., binary tumor response rates, ordinal disease severity scores) or mixed (e.g., survival data with censoring). Even when continuous measures are collected, practitioners frequently dichotomize them for interpretability, a practice that invalidates assumptions underpinning traditional analyses \citep{sverdlov2015modern}. This mismatch between methodological assumptions and real-world data structures creates a critical gap. While generalized linear models (GLMs) theoretically accommodate non-Gaussian outcomes \citep{mccullagh2019generalized}, their application in seamless trials under CAR remains underexplored. Model misspecification, such as omitting confounders or mis-specifying link functions, can introduce bias, reduce power, and compromise Type I error control. Furthermore, diverse estimands (e.g., average treatment effect, log relative risk, log odds ratio) demand flexible frameworks to ensure valid inference across metrics, a need unmet by current literature. This paper aims to develop a model-robust, universal framework capable of adapting to mixed outcome types and diverse estimands in seamless phase II/III trials. To complete this framework, we first discuss the trial designs that should be considered.

In the existing literature, discussions of seamless phase~II/III designs have largely focused on complete randomization. For example, following the approach of \cite{bauer1999combining}, people had employed the closure principle \citep{bretz2006confirmatory,marcus1976closed}, combination tests \citep{bauer1994evaluation}, and multiple testing procedures \citep{simes1986improved, dunnett1955multiple} to control the family-wise Type I error rate. Additionally, \cite{liu2002unified} provided a robust theoretical foundation for general two-stage adaptive designs, while \cite{koenig2008adaptive} proposed the adaptive Dunnett test based on the conditional error rate \citep{muller2001adaptive}. 

However, considering only the complete randomization is overly simplified. In practice, CAR procedures are widely implemented to mitigate covariate imbalances and improve statistical power, especially in trials with important prognostic factors \citep{baldi2011covariate, hu2012asymptotic}. Specifically, CAR is often implemented through stratified permuted block (STRPB)  designs \citep{zelen1974randomization} or minimization-based methods \citep{pocock1975sequential, taves1974minimization}, which enhance trial efficiency by dynamically balancing baseline covariates. Despite its widespread use in practice, CAR has only recently been formally considered in the analysis of seamless trials \citep{ma2022seamless}. \cite{ma2022seamless} pioneered this integration by developing a CAR-adjusted linear model and demonstrated that adjusted test statistics, such as replacing pooled variance with model-robust estimators, can restore Type I error control and improve power under CAR. Nevertheless, their reliance on linear models limits applicability to continuous outcomes, overlooking the complexity of discrete or mixed endpoints.

This paper bridges these gaps by proposing a unified framework for seamless phase II/III trials that harmonizes CAR with GLMs. Our contributions are threefold. First, we develop a model-robust Z-estimation approach under GLMs, ensuring consistent and asymptotically normal estimators for treatment effects across outcome types. This generality accommodates diverse estimands, from relative risk to log odds ratios, without requiring full parametric correctness. Second, we integrate CAR into seamless designs, extending the theoretical foundations established by \cite{ma2022seamless}. By deriving the asymptotic properties of test statistics under CAR, including covariance structures for correlated outcomes, we enable valid hypothesis testing and error control, addressing FDA concerns over inflated variances and conservative inference. Third, we demonstrate practical impact through extensive simulations and a hypothetical trial example, illustrating how our framework outperforms conventional methods in power and Type I error control.

The remainder of this paper is organized as follows. Section \ref{sec:framework} presents the general framework for seamless phase~II/III trials with CAR. Section \ref{sec:theory} develops a model-robust Z-estimation framework under the GLM setting and establishes the consistency and asymptotic normality of the resulting estimators, supported by a precise decomposition of the standard error. Section \ref{sec:application} specializes these general results to three commonly used estimands. 
Section \ref{sec:numerical} evaluates the finite-sample performance of the proposed methods through extensive simulations, comparing Type~I error control and power across diverse scenarios. Section \ref{sec:realdata} illustrates the practical utility of our framework in a hypothetical trial for treating alopecia areata. Finally, Section \ref{sec:discussion} concludes with a discussion of the broader implications and potential extensions. By unifying CAR with GLMs in seamless designs, this work advances statistical theory and offers a regulatory-compliant toolkit for the next generation of clinical research.

\subsection{Notation}

Let \([K] := \{1,2,\dots,K\}\) denote an index set. For a matrix \(\boldsymbol{M} \in \mathbb{R}^{K \times K}\), let \([\boldsymbol{M}]_{(k,k)}\) denote its \((k,k)\)-th entry, and for a vector \(\boldsymbol{x} \in \mathbb{R}^q\), let \(\|\boldsymbol{x}\| = \sqrt{\boldsymbol{x}^\top \boldsymbol{x}}\) denote the Euclidean norm; $\bm{1}_K$ and $\bm{0}_K$ denote the $K$-dimensional vectors of ones and zeros, respectively. For a square matrix \(\boldsymbol{M} \in \mathbb{R}^{q \times q}\), the spectral norm is \(\|\boldsymbol{M}\| = \sup \{ \|\boldsymbol{M}\boldsymbol{x}\| : \|\boldsymbol{x}\|=1, \boldsymbol{x} \in \mathbb{R}^q\}\). A sequence of random vectors \(\boldsymbol{X}_n\) satisfies \(\boldsymbol{X}_n = o_p(\bm{1})\) if \(\boldsymbol{X}_n \cvp \boldsymbol{0}\) and \(\boldsymbol{X}_n = O_p(\bm{1})\) if \(\|\boldsymbol{X}_n\|\) is bounded in probability. For a vector-valued function \(\boldsymbol{f}(\boldsymbol{\theta}) = (f_1(\boldsymbol{\theta}),\dots,f_m(\boldsymbol{\theta}))^\top: \mathbb{R}^p \to \mathbb{R}^m\), the Jacobian is \(\dot{\boldsymbol{f}}(\boldsymbol{\theta}) = \partial \boldsymbol{f}(\boldsymbol{\theta})/\partial \boldsymbol{\theta} \in \mathbb{R}^{m \times p}\), where the $j$-th row equals the gradient \(\dot{\boldsymbol{f}}_j(\boldsymbol{\theta}) = \partial f_j(\boldsymbol{\theta})/\partial \boldsymbol{\theta}\), and the Hessian of component \(f_j\) is \(\ddot{\boldsymbol{f}}_j(\boldsymbol{\theta}) = \partial^2 f_j(\boldsymbol{\theta})/\partial \boldsymbol{\theta}\partial \boldsymbol{\theta}^\top \in \mathbb{R}^{p \times p}\).  Finally, for a vector \(\boldsymbol{a} = (a_1,\dots,a_K)^\top\), we write \(\operatorname{diag}\{a_1,\dots,a_K\}\) for the diagonal matrix with entries $a_k$ on the diagonal and zeros elsewhere. $\boldsymbol{e}_k$ denotes the $k$-th standard basis vector in the Euclidean space. $\cvp$ and $\cvd$ denote convergence in probability and distribution, respectively.

\section{Seamless Phase II/III Trial with CAR}\label{sec:framework}
\subsection{An overview of the procedure}
 We work under a general framework for seamless phase II/III trials that accommodates various treatment effect metrics beyond the traditional average treatment effect. Let $\boldsymbol{Y} = (Y(0),\dots, Y(K))^\top$, $\bm{\mu} = (\mu_0, \mu_1, \ldots, \mu_K) := \mathbb{E}[\boldsymbol{Y}] \in\mathbb{R}^{K+1}$ with \(\mu_0\) and \(\mu_k\) represent the average potential outcome in the control arm and the \(k\)-th treatment arm for \(k \in [K]\), respectively. Instead of focusing solely on the difference \(\mu_k - \mu_0\), our framework allows treatment effects to be expressed as functions of these parameters. Specifically, for a chosen function \( g \), we define the treatment effect metric as \(\bm{\delta} = (\delta_1, \ldots, \delta_K)\), where each component is given by \(\delta_k = g(\mu_k) - g(\mu_0)\). This flexible formulation includes common metrics as special cases: \textit{the log relative risk} ($\log$RR) when \(g\) is the logarithm, \textit{the log odds ratio} (LOR) when \(g\) is the logit function, and \textit{the average treatment effect} (ATE) when \(g\) is the identity. Such flexibility enables efficient design and analysis, accommodating different targets. Building on this framework, the seamless phase II/III trial proceeds as follows:

\begin{itemize}
\item \textsf{Phase II (e.g. Stage 1):} Identification of promising treatments by testing
\[
H_0^1: \bm{\delta} = \bm{0}_K \quad \mathrm{versus} \quad H_1^1: \delta_k > 0 \quad \mathrm{ for \ some }\ k \in [K],
\]
based on estimator $\widehat{\bm{\delta}}=(\widehat{\delta}_1, \ldots, \widehat{\delta}_K)$ constructed using data from \(N_1\) patients. Dunnett’s test \citep{dunnett1964new} is used to control family-wise Type I error while accounting for multiplicity.

\item \textsf{Phase III (e.g. Stage 2):} Selection of the final treatment 
\[
k^* = \arg\max_{1 \leq k \leq K} \widehat{\delta}_k,
\]
and testing based on an additional \(N_2\) patients:
\[
H_0^2: \delta_{k^*} = 0 \quad \mathrm{versus} \quad H_1^2: \delta_{k^*} > 0.
\]  
\end{itemize}
Conclusions are drawn by combining evidence from both stages using the inverse chi-square method \citep{bauer1994evaluation}.

\subsection{Covariate-adaptive randomization}

Recall that the numbers of patients enrolled in Stage 1 and Stage 2 are denoted by \(N_1\) and \(N_2\), respectively.
For subject \(i\in[N_j]\) and $j \in\lbrace 1,2\rbrace$, let \(\bm X_i=(X_{i,1},\ldots,X_{i,q})^\top\in\mathbb R^q\) collect baseline covariates, and define the treatment-assignment vector
\[
\bm T_i=(T_i^0,T_i^1,\ldots,T_i^K)^\top\in\{0,1\}^{K+1},\qquad 
\sum_{k=0}^K T_i^k=1,
\]
where \(T_i^k=1\) indicates assignment to arm \(k\in\{0,\ldots,K\}\).
Following \citet{hu2006theory}, let \(\mathcal T_i\), \(\mathcal X_{i,\mathrm{ex}}\), \(\mathcal X_i\), and \(\mathcal Y_i\) denote the \(\sigma\)-algebras generated by the first \(i\) treatment assignments, the discretized covariates used in randomization, the baseline covariates, and the observed outcomes, respectively. Define
\[
\mathcal F_i=\mathcal X_{i+1,\mathrm{ex}}\otimes \mathcal T_i\otimes \mathcal X_i\otimes \mathcal Y_i,
\qquad
\boldsymbol{\phi}_i=\mathbb E[\bm T_i \mid \mathcal F_{i-1}],
\]
so \(\boldsymbol{\phi}_i\) is the (vector) assignment probability for subject \(i\) given past assignments and observed information, allowing the randomization covariates \(\bm X_{i,\mathrm{ex}}\) to be a subset of the outcome-relevant baseline covariates \(\bm X_i\).
Forming discrete strata \(S_i=S(\bm X_{i,\mathrm{ex}})=s\in\{1,\ldots,s_{\mathrm{max}}\}\) from \(\bm X_{i,\mathrm{ex}}\), CAR procedures sequentially assign patients to minimize a weighted imbalance score that combines overall imbalance, marginal imbalances of the $j$-th discrete (or discretized) covariate (for \(j\in[q]\)), and within-stratum imbalances, thereby improving covariate balance beyond what is achievable with complete randomization \citep{hu2023multi}.

Our theory and numerical studies cover three lines of CAR schemes. Classical stratified permuted-block randomization (e.g. STRPB) is effective when each stratum has ample sample size \citep{zelen1974randomization,hallstrom1988imbalance}, but its performance can degrade as the number of covariates (or levels) increases and strata fragment. To balance a relatively large number of covariates covariates, minimization-based methods such as the Pocock and Simon's procedure (\textsf{PS}) \citep{taves1974minimization,pocock1975sequential} targeting marginal imbalance directly, though its within-stratum imbalance may remain insufficiently controlled \citep{toorawa2009use}.
Building on these ideas, \citet{hu2012asymptotic} proposed extensions (\textsf{HH}) that simultaneously temper overall, marginal, and within-stratum imbalance.
Comprehensive reviews of CAR methods and their asymptotic properties can be found in \citet{rosenberger2008handling}, \citet{hu2012asymptotic}, \citet{hu2014adaptive}, and \citet{rosenberger2015randomization}.

\subsection{Z-estimation, Wald statistic and Dunnett's test}\label{subsec:dunnett}
Having established the design framework, we describe in detail the estimation and inference procedures that utilize the collected data. For each treatment arm $k$ with $0 \leq k \leq K$, let $Y_i(k)$ denote the potential outcome for the $i$-th patient if assigned to the $k$-th treatment. After treatment assignment $\bm{T}_i$, the observed outcome is  
\[
Y_i = \sum_{k=0}^K T_i^k Y_i(k).
\]  
To estimate the treatment effect vector $\bm{\delta} = (\delta_1, \ldots, \delta_K)^\top$, we follow the $Z$-estimation framework by \citet{van2000asymptotic}. Let $\bm{\theta}\in\mathbb{R}^{K+1+q} $ denote the full parameter vector, comprising a $(K+1)$-dimensional parameter of interest, on which $\bm{\delta}$ depends, and a $q$-dimensional nuisance parameter vector. Then the $Z$-estimator $\widehat{\bm{\theta}}$ is defined as the solution to the following $(K+1+q)$-dimensional estimating equation:
\begin{equation}  \sum_{k=0}^K\sum_{i=1}^n T_i^k \bm{\psi}^k(Y_i(k), \bm{X}_i; \bm{\theta}) = \bm{0}_{K+1+q}, \label{eq.estimation_equation} \end{equation}
where for $k\in0\cup[K]$
{\footnotesize
\begin{equation*} 
\bm{\psi}^0(Y_i(0), \bm{X}_i; \bm{\theta}):= \begin{pmatrix} 
h^0(\bm{X}_i;\bm{\theta}) - Y_i(0) \\ 
\bm{0}_{K} \\ 
\bm{\xi}^0(Y_i(0), \bm{X}_i; \bm{\theta}) 
\end{pmatrix},\quad \bm{\psi}^k(Y_i(k), \bm{X}_i; \bm{\theta}):= \begin{pmatrix} 
\bm{0}_{k} \\ 
h^k(\bm{X}_i;\bm{\theta}) - Y_i(k) \\ 
\bm{0}_{K-k} \\ 
\bm{\xi}^k(Y_i(k), \bm{X}_i; \bm{\theta}) 
\end{pmatrix}.
\end{equation*}
}
Here, $h^k(\bm{X}_i;\bm{\theta})$ denotes the working model for the estimation of the parameters and $\bm{\xi}^k(Y_i(k), \bm{X}_i; \bm{\theta})$ corresponds to additional estimation equations for the nuisance parameters. We define the targeted $\bm{\theta}^*$ as the solution to 
\[ \frac{1}{K+1}\sum_{k=0}^K  \mathbb{E}[\bm{\psi}^k(Y(k),\bm{X};\bm{\theta}^*)] = \bm{0}_{K+1+q}.\]
We further obtain $\widehat{\mu}_k$ and $\widehat{\bm{\delta}}$ by 
{\small
 \begin{equation}
 \label{eq:mu_hat}
\widehat{\mu}_k = \frac{1}{n} \sum_{i=1}^n h^k(\bm{X}_i; \widehat{\bm{\theta}}), \ k\in 0\cup[K];\quad \mathrm{and} \quad \widehat{\delta}_k = g(\widehat{\mu}_k) - g(\widehat{\mu}_0), \ \ k\in[K].
\end{equation} 
}
This formulation offers a flexible framework for estimation and subsequent analysis in various settings. Let us take the generalized linear model (GLM) \citep{mccullagh2019generalized} with canonical link $\gamma(\mu)$ as an example. Define $\bm{\theta} = (\iota_0,\cdots,\iota_K, \bm{\beta}_X^\top)^\top$ and we obtain $\widehat{\bm{\theta}}$ by taking $h^k(\bm{X}_i; \bm{\theta}) = \gamma^{-1}(\iota_k + \bm{X}_i^\top \bm{\beta}_X)$ and $\bm{\xi}^k(Y_i(k), \bm{X}_i; \bm{\theta}) = \bigl\lbrace Y_i(k) - h^k(\bm{X}_i; \bm{\theta})\bigr\rbrace \bm{X}_i$ in the estimating equation, and obtain $\widehat{\mu}_k$ and $\widehat{\delta}_k$ subsequently by (\ref{eq:mu_hat}). In a special case with $q=0$ where we do not include any information from $\bm{X}$, the $\bm{\xi}^k(\cdot)$ no longer exists, $\bm{\theta} = (\iota_0,\cdots,\iota_K)$ and $h^k(\bm{\theta})=\gamma^{-1}(\iota_k)$. Solving the estimating equations, we obtain the unadjusted estimators $\widehat{\bm{\delta}}_{\sf{unadj}} = (\widehat{\delta}_{ \sf{1,unadj}}, \ldots, \widehat{\delta}_{ \sf{K,unadj}})^\top$, where  
$\widehat{\delta}_{ \sf{k,unadj}} =  g(\bar{Y}_k) - g(\bar{Y}_0)$ with $\bar{Y}_k =n_k^{-1}\sum_{i=1}^{n} T_i^kY_i$  denoting the average of outcome in the $k$-th treatment arm, for $k\in 0\cup [K]$. To carry out the test for individual $H^1_{0,k}: \delta_k = 0$ in Stage 1, we define 
{\small
\[
g'(x) = \frac{d}{dx} g(x), \quad \bm{G} = \bigl(-g'(\mu_0) \bm{1}_K, \ \  \mathrm{diag} \{g'(\mu_1), \ldots, g'(\mu_K)\} \bigr) \in \mathbb{R}^{K \times (K+1)},
\]}
and let $\bm{\Gamma}$ be the sandwich variance covariance based on \eqref{eq.estimation_equation}. Then the model-robust Wald statistic is defined as 
\[
W^{\sf{robust}}_k = \frac{\widehat{\delta}_k}{\mathrm{se}_{\sf robust}(\widehat{\delta}_k)},\ \mathrm{where}\ \ \mathrm{se}^2_{\sf robust}(\widehat{\delta}_k) = \frac{1}{n} \left[ \widehat{\bm{G}} \widehat{\bm{\Gamma}}\widehat{\bm{G}} ^\top \right]_{(k,k)},
\]  
with $\widehat{\bm{\Gamma}}$ being the empirical plug-in estimate of $\bm{\Gamma}$. To control the Type~I error under the global null hypothesis \( H^1_0: \boldsymbol{\delta} = \boldsymbol{0}_K \), we compute  Stage~1  \(p\)-value evaluated    using Dunnett’s test as
{\small \[
P_1 = 1 - \mathbb{P}\bigl(Z_1 \le \max_{k} W_k, \dots, Z_K \le \max_{k} W_k \bigr), \ \text{with}\  (Z_1, \dots, Z_K)^\top \sim N\left(\boldsymbol{0}, \widehat{\boldsymbol{R}}\right),
\]}
where \( \widehat{\boldsymbol{R}} \) denotes the estimated correlation matrix derived from Theorem~\ref{thm:multiple_testing} and Remark~\ref{rmk:Rhat}, based on \( \widehat{\bm{\Gamma}} \) and \( \widehat{\bm{G}} \). Let \( P_2 \) denote the Stage~2 \(p\)-value corresponding to $H_0^2: \delta_{k^*} = 0$. The overall test combines evidence across the two stages using the inverse--\(\chi^2\) method, which rejects \( H_0 \) if $- \log(P_1 P_2) > \tfrac{1}{2} \chi^2_4(1 - \alpha)$,
where \( \chi^2_4(1 - \alpha) \) is the \((1 - \alpha)\)-quantile of the chi-squared distribution with four degrees of freedom.

Consider the logistic model for  illustration, with $\mathrm{logit}^{-1}(x)= [1+\exp(-x)]^{-1}$,
\[
\mathbb{E}[Y_i \mid \bm{X}_i, \bm{T}_i] 
= \mathrm{logit}^{-1}\!\left(\bm{\iota}^\top \bm{T}_i + \bm{X}_i^\top \bm{\beta}_X\right), 
\quad \bm{\iota} = (\iota_0, \ldots, \iota_K).
\]
Here we take $\gamma(x) = \mathrm{logit}(x)$, which yields 
\begin{align*}
    h^k(\bm{X}_i; \bm{\theta}) 
    &= \mathrm{logit}^{-1}\!\left(\iota_k + \bm{X}_i^\top \bm{\beta}_X\right), \\
    \bm{\xi}^k\!\left(Y_i(k), \bm{X}_i; \bm{\theta}\right) 
    &= \bigl[Y_i(k) - \mathrm{logit}^{-1}\!\left(\iota_k + \bm{X}_i^\top \bm{\beta}_X\right)\bigr]\bm{X}_i.
\end{align*}
For hypothesis testing, we set $g(x) = \log(x)$ when assessing log relative risks, 
and $g(x) = \log\!\left(x/(1-x)\right)$ when evaluating log odds ratios.

\section{Theoretical property}\label{sec:theory}
In this section, we first state the key assumptions and discuss their implications. 
We then present the main theoretical results, including the consistency of 
$\hat{\bm{\theta}}$ and $\hat{\bm{\mu}}$, the asymptotic normality of 
$\hat{\bm{\mu}}$ and $g(\hat{\bm{\mu}})$, and, finally, the asymptotic joint 
distribution of the Stage~1 test statistics, which establishes the validity of our 
model-robust tests.

\begin{customass}{1}[Data-generating process]\quad
\begin{enumerate}[leftmargin=*]
\item  \(\{ (\bm{X}_i,Y_i(0),\ldots,Y_i(K) \}_{i = 1}^n\) are independently identically distributed as \\ $(\bm{X},Y(0),\ldots, Y(K))$, with $\mathbb{E}\left[\|\bm{X}\|^2\right]<\infty$, $\max_{k\in 0\cup [K]}\mathrm{Var}[Y(k)]<\infty$.
   \item Let $\pi_{s}$ denote the proportion of observations in stratum $s$. For any \(s \in [s_{\mathrm{max}}]\), we have \(0 < \pi_{s} < 1\), and \(\sum_{s \in [s_{\mathrm{max}}]} \pi_{s} = 1\).

    \item Recall that \(\mathcal F_{i-1}=\mathcal X_{i,\mathrm{ex}}\otimes \mathcal T_{i-1}\otimes \mathcal X_{i-1}\otimes \mathcal Y_{i-1}\) and the assignment rule satisfies $\bm{\phi}_{i} = \mathbb{E}\left[\bm{T}_i|\mathcal F_{i-1}\right]$ for $i\in [n]$. Let us define the imbalance within stratum $s$ and treatment arm $k$ as $D_{n}^k(s)$ with the following decomposition 
    \begin{align*}
    D_{n}^k(s)&:=\sum_{i = 1}^{n}\left({T}_{i}^k - \frac{1}{K+1} \right)\mathbb{I}\{S_{i}=s\}\cr 
    &=\sum_{i = 1}^{n}M_{i}^k(s)+d_{n}^k(s), \quad s\in[s_{\mathrm{max}}],\quad k\in 0\cup[K].
    \end{align*}
   We assume that the CAR procedure satisfies
    \begin{enumerate}[leftmargin=*]
        \item \(\mathbb E[\lbrace d_{n}^k(s) \rbrace
    ^{2}]=o(n)\); and
        \item \(\{M_{i}^k(s)\}_{i = 1}^{n}\) is a sequence of bounded zero-mean martingale differences with respect to \(\mathcal{F}_{i - 1}\). Further with vectorized 
        \[
\bm M_i
:= \operatorname{vec}\!\Big(\big(M_i^k(s)\big)_{s\in[s_{\mathrm{max}}],\;k\in[K]}\Big)
\in \mathbb R^{s_{\mathrm{max}}(K+1)},
\]
the averaged conditional second moment converges in probability to a block covariance matrix:
\[
\frac{1}{n}\sum_{i=1}^{n}
\mathbb E\!\left[\bm M_i\,\bm M_i^\top \,\big|\, \mathcal F_{i-1}\right]
\ \cvp\ 
\boldsymbol{\Sigma}^{\sf{CAR}}
\ \in\ \mathbb R^{s_{\mathrm{max}}(K+1)\times s_{\mathrm{max}}(K+1)}.
\]

If we index $\boldsymbol{\Sigma}^{\sf{CAR}}$ by $(s,s')\in [s_{\mathrm{max}}]\times [s_{\mathrm{max}}]$ as a block matrix:
\[
\vspace{-10pt}
\boldsymbol{\Sigma}^{\sf{CAR}}
=\big[\,\boldsymbol{\Sigma}^{\sf{CAR}}(s,s')\,\big]_{s,s'\in [s_{\mathrm{max}}]},
\qquad
\boldsymbol{\Sigma}^{\sf{CAR}}(s,s') \in \mathbb R^{(K+1)\times (K+1)},
\]
and defined the $(k,k')$-th element in $\boldsymbol{\Sigma}^{\sf{CAR}}(s,s')$ as $\bm{\Sigma}^{\sf{CAR}}_{kk'}(s,s')$, then 

        \begin{equation*}
        \vspace{-10pt}
        \frac{1}{n}\sum_{i = 1}^{n}\mathbb E[M_{i}^k(s)M_{i}^{k'}(s') \mid \mathcal{F}_{i - 1}] \cvp \bm{\Sigma}^{\sf{CAR}}_{kk'}(s,s').
        \end{equation*}
    \end{enumerate}
\end{enumerate}
\label{Assumption 3.1}
\end{customass}

Assumption~\ref{Assump:Z-estimation} in the Supplementary Material outlines standard conditions for Z-estimation, analogous to those in \citet{wang2023model} and \citet{van2000asymptotic}, which ensure the consistency and asymptotic convergence of the estimator $\widehat{\bm{\theta}}$ to its population counterpart $\bm{\theta}^*$.

\begin{proposition}[Consistency of $\widehat{\bm{\theta}}$ and $\widehat{\bm{\mu}}$]
\label{prop:consistency}
Under Assumptions \ref{Assumption 3.1} and \ref{Assump:Z-estimation}, 
\[
\widehat{\bm{\theta}} - \bm{\theta}^* =O_p(n^{-1/2}),\qquad \widehat{\bm{\mu}}-\bm{\mu} =O_p(n^{-1/2}).
\]
\end{proposition}



Let $\boldsymbol{\Sigma}^{\sf{CR}}$ be the counterpart of $\boldsymbol{\Sigma}^{\sf{CAR}}$ under complete randomization. Beyond the consistency provided in Proposition \ref{prop:consistency}, in the following theorem, we derive the asymptotic normality of $\widehat{\bm\mu}$ and provide a clear decomposition of the limiting covariance structure to reflect the influence of covariate adjustment in the Z-estimation and the CAR procedure. 
\begin{theorem}[Asymptotic normality of $\widehat{\bm{\mu}}$]
\label{thm:mu_normality}
 Under Assumptions \ref{Assumption 3.1} and \ref{Assump:Z-estimation}, $\widehat{\bm{\mu}}$ in (\ref{eq:mu_hat}) satisfies:
\begin{equation}
\sqrt{n}(\widehat{\bm{\mu}}-\bm{\mu})\cvd N(\bm{0},\bm{\Gamma}),
\end{equation}
with $\boldsymbol{\Gamma} = \boldsymbol{\Gamma}_{\sf{base}}-\boldsymbol{\Gamma}_{\sf{adj}}-\boldsymbol{\Gamma}_{\sf{CAR,adj}}$. Here,
\begin{itemize}
\item $\boldsymbol{\Gamma}_{\sf{base}} = (K+1)\mathrm{diag}\left \{  \mathrm{Var}\left[Y(0)\right],\dots,\mathrm{Var}\left[Y(K)\right]\right \}$ represents the baseline covariance matrix that does not account for covariate adjustment or the CAR procedures. 
\item With $r^k(\boldsymbol{\theta}^*) = Y(k) - h^k(\boldsymbol{X},\boldsymbol{\theta}^*)$ and $\boldsymbol{r} = (r^0(\boldsymbol{\theta}^*), \ldots, r^K(\boldsymbol{\theta}^*))^\top$,
{\footnotesize
\begin{align*}
\boldsymbol{\Gamma}_{\sf{adj}} &=(K+1)\mathrm{diag}\left \{ \mathrm{Var}\left[Y(0)\right]-\mathrm{Var}\left[r^0(\boldsymbol{\theta}^*)\right],\dots,\mathrm{Var}\left[Y(K)\right]-\mathrm{Var}\left[r^K(\boldsymbol{\theta}^*)\right]\right \}\\
&\quad +\mathrm{Var}\left[\boldsymbol{r}\right]-\mathrm{Var}\left[\boldsymbol{Y}\right]
\end{align*}
}
captures the contribution of covariate adjustment in Z-estimation.
\item With $\boldsymbol{L}(s,k) = \mathbb E\left[r^{k}(\boldsymbol{\theta}^*)\mid S = s\right]\boldsymbol{e}_k$, recall the definition of $\boldsymbol{\Sigma}^{\sf{CAR}}_{kk'}(s,s')$ from Assumption \ref{Assumption 3.1},
{\footnotesize
\[
\boldsymbol{\Gamma}_{\sf{CAR,adj}} = (K+1)^2\sum_{k,k'\in 0 \cup [K]}\sum_{s,s'\in [s_{\mathrm{max}}]}\lbrace\boldsymbol{\Sigma}^{\sf{CR}}_{kk'}(s,s')-\boldsymbol{\Sigma}^{\sf{CAR}}_{kk'}(s,s')\rbrace \boldsymbol{L}(s,k)\boldsymbol{L}(s',k')^{\top}
\]} depicts the interplay between covariate adjustment in Z-estimation and the CAR procedures.
\end{itemize}
\end{theorem}

\begin{corollary}[Special cases]\quad
\begin{enumerate} \item Under CR, Assumption \ref{Assumption 3.1} holds with $\boldsymbol{\Sigma}_{kk'}^{\sf{CR}}(s,s') = \boldsymbol{\Sigma}_{kk'}^{\sf{CAR}}(s,s')$ for $s,s'\in [s_{\mathrm{max}}]$ and $k,k'\in 0 \cup [K]$, and $\boldsymbol{\Gamma}_{\sf{CAR,adj}} = 0$.
\item When $q = 0$ in the Z-estimation, $\boldsymbol{\Gamma}_{\sf{adj}} = 0$.
\item Under HH or STRPB,  $\boldsymbol{\Gamma}_{\sf{CAR,adj}}$ is positive definite as $\boldsymbol{\Sigma}^{\sf{CAR}}_{kk'}(s,s') = 0$ for all $k, k' \in 0 \cup[K]$ and $s,s' \in [s_\mathrm{max}]$.
\end{enumerate}
\end{corollary}In contrast, under PS , $\boldsymbol{\Sigma}^{\sf{CAR}}_{kk'}(s,s')$ does not have a closed-form and thus may not guarantee the positive definiteness of $\boldsymbol{\Gamma}_{\sf{CAR,adj}}$.
\begin{remark}\label{rmk:Gamma}  It is worth emphasizing that the simplest and widely used estimation procedure neither adjusts for covariates nor the impact of CAR. This corresponds to the unadjusted estimator $\widehat{\boldsymbol{\mu}}_{\sf{unadj}}$ and its asymptotic covariance matrix $\boldsymbol{\Gamma}_{\sf{unadj}} = \boldsymbol{\Gamma}_{\sf{base}}$. By contrast, for model-based methods, while they incorporate covariate adjustment in estimation, the conventional calculation of the asymptotic covariance still ignores the influence of CAR, effectively treating $\boldsymbol{\Gamma}_{\sf{CAR,adj}} = 0$. In this case, the estimator is $\widehat{\boldsymbol{\mu}}$, with $\boldsymbol{\Gamma}_{\sf{conv}} = \boldsymbol{\Gamma}_{\sf{base}} - \boldsymbol{\Gamma}_{\sf{adj}}.$ 
Our proposed model-robust method fully accounts for both covariate adjustment and CAR, yielding the same estimator $\widehat{\boldsymbol{\mu}}$ but with asymptotic covariance matrix $\boldsymbol{\Gamma} = \boldsymbol{\Gamma}_{\sf{base}} - \boldsymbol{\Gamma}_{\sf{adj}} - \boldsymbol{\Gamma}_{\sf{CAR,adj}}.$ More specifically, CAR procedures, such as HH or STRPB designs, improve covariate balance and guarantee that $\boldsymbol{\Sigma}^{\sf{CAR}}_{kk'}(s,s') = 0$ for all $k, k' \in \{0\} \cup [K]$ and $s, s' \in [s_{\mathrm{max}}]$, thereby further ensuring the positive definiteness of $\boldsymbol{\Gamma}_{\sf{CAR,adj}}$. A systematic comparison of the three approaches will be presented in Section \ref{sec:numerical} and \ref{sec:realdata}. Lastly, when the working model used for estimation and inference is correct, $\boldsymbol{\Gamma}_{\sf{CAR,adj}} = 0$ and thus the results from CR and CAR procedures are asymptotically the same.
\end{remark}

\begin{theorem}[Asymptotic normality of the general form $\widehat{\bm{\delta}}$]
\label{thm:contrast_normality}
Consider the parameter of interest \(\bm{\delta} = (g(\mu_1) - g(\mu_0), \ldots, g(\mu_K) - g(\mu_0))^\top\), and denote by \(\widehat{\bm{\delta}}\) its estimator obtained by replacing \(\bm{\mu}\) with \(\widehat{\bm{\mu}}\).
 Then under Assumptions \ref{Assumption 3.1} and \ref{Assump:Z-estimation},
\begin{equation}
\sqrt{n}(\widehat{\bm{\delta}} - \bm{\delta}) \cvd N(\bm{0}, \bm{G}\bm{\bm{\Gamma}}\bm{G}^\top),
\end{equation}
where $\bm{G} = \frac{\partial \bm{\delta}}{\partial \bm{\mu}}$ is the $K \times (K+1)$ Jacobian matrix of $\bm{\delta}$ evaluated at $\bm{\mu}$, and $\bm{\Gamma}$ is defined in Theorem~\ref{thm:mu_normality}.
\end{theorem}

To construct both the conventional and model-robust test statistics for inference on $\bm{\delta}$, we require an estimate of $\bm{G}\boldsymbol{\Gamma}\bm{G}^\top$. We first evaluate the Jacobian of $\bm{\delta}$ at $\widehat{\bm{\mu}}$ to obtain $\widehat{\bm{G}}$. Recall that
\[
\boldsymbol{\Gamma}
= \boldsymbol{\Gamma}_{\sf base}
  - \boldsymbol{\Gamma}_{\sf adj}
  - \boldsymbol{\Gamma}_{\sf CAR,adj}
= \boldsymbol{\Gamma}_{\sf conv}
  - \boldsymbol{\Gamma}_{\sf CAR,adj},
\]
where $\widehat{\boldsymbol{\Gamma}}_{\sf conv}
:= \widehat{\boldsymbol{\Gamma}}_{\sf base}
   - \widehat{\boldsymbol{\Gamma}}_{\sf adj}$ can be estimated directly using plug-in estimators. By contrast, estimating $\boldsymbol{\Gamma}_{\sf CAR,adj}$ is more involved, as $\boldsymbol{\Sigma}^{\sf CAR}_{kk'}(s,s')$ does not admit a closed form under PS. Following \citet{liu2023impacts}, we therefore employ a bootstrap procedure to estimate $\boldsymbol{\Sigma}^{\sf CAR}$ and subsequently construct $\widehat{\boldsymbol{\Gamma}}_{\sf CAR,adj}$. The resulting estimator is $\widehat{\boldsymbol{\Gamma}}
= \widehat{\boldsymbol{\Gamma}}_{\sf conv}
  - \widehat{\boldsymbol{\Gamma}}_{\sf CAR,adj}.$ Additional details on the construction and asymptotic properties are provided in Section~\ref{supp:asymp_delta} of the Supplementary Material. Finally, the standard error estimate of $\widehat{\delta}_k$ is obtained as the square root of the $k$th diagonal element of the estimated variance--covariance matrix of $\widehat{\bm{\delta}}$:
\[
\mathrm{se}^2_{\sf conv}(\widehat{\delta}_k)= n^{-1}[\widehat{\bm{G}}\,\widehat{\boldsymbol{\Gamma}}_{\sf conv}\,\widehat{\bm{G}}^\top]_{(k,k)},\quad \quad  \mathrm{se}^2_{\sf robust}(\widehat{\delta}_k)=n^{-1}[\widehat{\bm{G}}\,\widehat{\boldsymbol{\Gamma}}\,\widehat{\bm{G}}^\top]_{(k,k)}.
\]

\begin{proposition}[The consistency of $\widehat{\bm{\Gamma}}_{\sf{conv}}$ and $\widehat{\bm{\Gamma}}$]\label{prop:consistency}
Under Assumptions \ref{Assumption 3.1} and \ref{Assump:Z-estimation}, $\widehat{\bm{\Gamma}}_{\sf{conv}} \cvp \bm{\Gamma}_{\sf{conv}}$ and $\widehat{\bm{\Gamma}} \cvp \bm{\Gamma}$.
\end{proposition}

\begin{theorem}[Comparison between conventional and model-robust inference]
\label{thm:multiple_testing} Under Assumptions~\ref{Assumption 3.1} and~\ref{Assump:Z-estimation}, we consider the individual hypothesis test $H_{0}: \delta_k = 0$ versus $H_{1}: \delta_k > 0$. Define the individual conventional and model-robust Wald test statistics for $k \in [K]$ as
\begin{equation*}
W^{\sf{conv}}_k = \frac{\sqrt{n}\,\widehat{\delta}_k}{\sqrt{[\widehat{\bm{G}}\,\widehat{\bm{\Gamma}}_{\sf{conv}}\,\widehat{\bm{G}}^\top]_{(k,k)}}}, 
\quad \text{and} \quad 
W^{\sf{robust}}_k = \frac{\sqrt{n}\,\widehat{\delta}_k}{\sqrt{[\widehat{\bm{G}}\,\widehat{\bm{\Gamma}}\,\widehat{\bm{G}}^\top]_{(k,k)}}}.
\end{equation*}
$W^{\sf{robust}}_k$ always produces a valid type~I error by Theorem \ref{thm:contrast_normality} and Proposition \ref{prop:consistency}. Because $\boldsymbol{\Gamma}_{\sf conv} - \boldsymbol{\Gamma}$ is positive definite under HH and STRPB, the conventional statistic $W^{\sf conv}_k$ exhibits a reduced type~I error under these procedures.

\end{theorem}

\begin{theorem}[Asymptotic joint distribution of the $\{W^{\sf{robust}}_k\}_{k\in[K]}$]\label{thm:comparison}
    Under Assumptions \ref{Assumption 3.1} and \ref{Assump:Z-estimation},  we aim to conduct the following hypothesis test
\begin{align*}
\vspace{-10pt}
H^1_{0}&: \delta_1 = \delta_2=\cdots =\delta_K= 0,\\
H^1_{1}&: \exists k\in[K], \delta_k > 0.
\end{align*}
In particular, $(W^{\sf{robust}}_1,\ldots,W^{\sf{robust}}_K)^\top \cvd N(\bm{0}, \bm{R})$ with 
\begin{equation*}
\bm{R}_{(k,k')} = \frac{[\bm{G}\bm{\Gamma}\bm{G}^\top]_{(k,k')}}{\sqrt{[\bm{G}\bm{\Gamma}\bm{G}^\top]_{(k,k)}[\bm{G}\bm{\Gamma}\bm{G}^\top]_{(k',k')}}}, \quad k, k'\in[K].
\end{equation*}
\end{theorem}

\begin{remark}\label{rmk:Rhat} The plug-in estimator \( \widehat{\boldsymbol{R}} \) is constructed by replacing \( \boldsymbol{G} \) and \( \boldsymbol{\Gamma} \) with their empirical counterparts \( \widehat{\boldsymbol{G}} \) and \( \widehat{\boldsymbol{\Gamma}} \). 
The Stage~1 \(p\)-value is then computed using Dunnett’s test and combined with the Stage~2 \(p\)-value through the inverse--\(\chi^2\) method to obtain the overall test decision, as detailed in Section~\ref{subsec:dunnett}. By Theorem~\ref{thm:comparison}, each individual statistic \(W^{\sf robust}_k\) maintains valid Type~I error control; consequently, the combined procedures using Dunnett’s test and the inverse--\(\chi^2\) method also preserve overall Type~I error validity. However, since \(W^{\sf conv}_k\) tends to be conservative, the joint procedure may no longer guarantee exact Type~I error control.

\end{remark}
\vspace{-0.5cm}
\section{Application to important estimands}
\label{sec:application}

In clinical and epidemiological studies with binary outcomes and multiple treatment arms, it is often necessary to report treatment effects on scales that are both scientifically interpretable and compatible with standard modeling. Our framework is designed to be flexible: by specifying an appropriate $g(\cdot)$, we link $\bm{\mu}=(\mu_0,\mu_1,\ldots,\mu_K)^\top$ to $\bm{\delta}$ through the relation $\delta_k=g(\mu_k)-g(\mu_0)$. It encompasses three special cases that are particularly important in applications, log relative risk ($\log$RR), log odds ratio (LOR) and average treatment effect (ATE). Each arises from a distinct choice of $g$ yet is treated within a unified asymptotic framework.

We proceed by introducing each treatment effect measure $\bm{\delta}$ together with their associated $g$ and the Jacobian required by Theorem~\ref{thm:contrast_normality}. Throughout, $\widehat{\bm{\mu}}$ denotes plug-in estimators obtained from a working model (e.g., logistic regression), and $\bm{\Gamma}$ denotes the asymptotic covariance of $\sqrt{n}\widehat{\bm{\mu}}$.

\vspace{-0.2cm}
\subsection{Log relative risk ($\log$RR)}
\label{subsec:RR}

Log relative risk provides a multiplicative comparison standard in epidemiology and risk communication. For treatment arm $k$,
\[
\mathrm{RR}_k=\frac{\mu_k}{\mu_0},\qquad\text{and}
\quad \bm{\delta}=\bigl(\log \mu_1-\log \mu_0,\ldots,\log \mu_K-\log \mu_0\bigr)^\top.
\]
The log transformation stabilizes variability and ensures additivity in regression. A convenient working model is the logistic regression
\[
\mathbb{E}[Y_i \mid \bm{X}_i,\bm{T}_i]
=\mathrm{logit}^{-1}\!\left(\bm{\tau}^\top \bm{T}_i+\bm{\beta}_X^\top \bm{X}_i\right),
\]
and valid inference for 
$\log(\mathrm{RR}_k)$ follows from delta method and Theorem~\ref{thm:contrast_normality}.

\begin{proposition}
\label{prop:RR}
When $\log$RR is the target, $g(x)=\log x$ and
\[
\bm{G}_{\sf RR}=
\bigl(-\mu_0^{-1}\bm{1}_K, \ \  \mathrm{diag} \{  \mu_1^{-1}, \ldots,\mu_K^{-1}\} \bigr),
\qquad
\widehat{\bm{G}}_{\sf RR}=\bm{G}_{\sf RR}\big|_{\mu=\widehat{\mu}}.
\]
Thus Theorem~\ref{thm:contrast_normality} implies
\[
\sqrt{n} \left( \hat{\bm{\delta}}- \bm{\delta} \right)
\xrightarrow{d} N\!\left(0,\;\bm{G}_{\sf RR}\,\bm{\Gamma}\,\bm{G}_{\sf RR}^\top\right).
\]
\end{proposition}
\vspace{-0.2cm}
\subsection{Log Odds Ratio (LOR)}
\label{subsec:LOR}

The LOR is a natural scale for expressing effects in logistic regression, where model coefficients correspond to log-odds differences. For arm $k$,
\begin{align*}
\mathrm{LOR}_k&=\log\!\left(\frac{\mu_k}{1-\mu_k}\right)-\log\!\left(\frac{\mu_0}{1-\mu_0}\right),\\
\bm{\delta}&=\bigl(\log(\tfrac{\mu_1}{1-\mu_1})-\log(\tfrac{\mu_0}{1-\mu_0}),\ldots,\log(\tfrac{\mu_K}{1-\mu_K})-\log(\tfrac{\mu_0}{1-\mu_0})\bigr)^\top.
\end{align*}
Estimation utilizes the fitted logistic model, followed by the transformation of $\widehat{\mu}_k$ and calculation of variance using delta method.

\begin{proposition}
\label{prop:LOR}
For LOR, $g(x)=\log\!\bigl(x/(1-x)\bigr)$,
{\footnotesize
\begin{align*}
\bm{G}_{\sf LOR}=
\bigl([(\mu_0-1)^{-1}- \mu_0^{-1}] \bm{1}_K, \ \  \mathrm{diag} \{\mu_1^{-1}-(\mu_1-1)^{-1}, \ldots, \mu_K^{-1}-(\mu_K-1)^{-1} \}\bigr)
\end{align*}
}
and $\widehat{\bm{G}}_{\sf LOR}=\bm{G}_{\sf LOR}\big|_{\mu=\widehat{\mu}}$.
Consequently Theorem~\ref{thm:contrast_normality} implies
\[\sqrt{n} \left( \hat{\bm{\delta}}- \bm{\delta} \right)
\xrightarrow{d} N\!\left(0,\;\bm{G}_{\sf LOR}\,\bm{\Gamma}\,\bm{G}_{\sf LOR}^\top\right).
\]
\end{proposition}

Details on ATE are provided in Section~\ref{subsec:ATE} of the Supplementary Material. Taken together, the $\log$RR, LOR and ATE specifications illustrate how our unified inferential mechanism yields effect estimates and valid large-sample inference across the three scales most commonly reported in practice, while maintaining a common theoretical and computational pipeline.

\section{Numerical studies}\label{sec:numerical}
We conducted extensive simulation studies to examine the finite-sample performance of the proposed testing procedures. The simulations are structured as follows: we first outline the general setup common to all experiments, and then present two examples based on distinct data-generating models: M1 (logistic) and M2 (probit).  Within each example, we consider three working models (\(\mathcal{A}_0\), \(\mathcal{A}_1\) and \(\mathcal{A}_2\)), ranging from unadjusted to fully adjusted specifications, and compare the conventional Wald test with the new model-robust Wald test. We investigate four randomization procedures: complete randomization (CR), stratified permuted block design (STRPB), Pocock--Simon procedure (PS), and Hu and Hu's procedure (HH). Randomization is performed using two baseline covariates, $X_1$ and $X_2$, which will be specified later. For continuous covariates, values are discretized as 0 for negative values and 1 for positive values.

We consider a two-stage seamless trial design. In Stage~1, a total of \(420\) patients are sequentially enrolled and randomized among \(K=2\) experimental arms and a control arm. For each experimental arm \(k = 1, \ldots, K\), we compute a test statistic of the following form (both the conventional and model-robust versions, which differ in the specification of \(\mathrm{se}\), using $\boldsymbol{\Gamma}_{\sf{conv}}$ and $\boldsymbol{\Gamma}$ respectively):  
\[
W_k = \frac{\widehat{\delta}_k}{\mathrm{se}(\widehat{\delta}_k)}.
\]

The treatment arm with the largest statistic \(W_k\) (denoted \(k^\star\)) is selected as the most promising and advanced to Stage~2. In Stage~2, an additional \(500\) patients are enrolled and randomized between control and treatment \(k^\star\). At the end of the study, we test the global null hypothesis \(H_0:\boldsymbol{\delta}=\mathbf{0}_K\) by combining the stage-wise \(p\)-values from Stage~1 (obtained via Dunnett’s test) and Stage~2 using the inverse \(\chi^2\) combination method \citep{bauer1994evaluation}. We evaluate treatment effects using (\(\log\)RR), LOR, and ATE.  All simulation summaries are based on \(10{,}000\) replicates.

\subsection{Example 1: logistic model (M1)}\label{sec:simu1}

The data are generated from the logistic regression model, where $Y_i \sim \mathrm{Bernoulli}(p_i)$ with
$p_i = 1/(1+\exp(-\eta_i))=\text{logit}^{-1} (\eta_i)$, and
\[
\eta_i = \beta_0 + \iota_1 T_i^1 + \iota_2 T_i^2  
        + \beta_{1} X_{i,1} + \beta_{2} X_{i,2}.
\]
Covariates are generated as $X_{i,1} \sim \mathrm{Bernoulli}(1/2)$ and $X_{i,2} \sim \mathcal{N}(0,1)$. Parameters are set to $(\beta_0,\beta_{1},\beta_{2})=(-1,1,2)$. We evaluate three working models for estimation and inference with $p_i \sim \text{logit}^{-1} (\eta_i)$, i.e., $\gamma(\cdot)=\text{logit}(\cdot)$:
\begin{itemize}
    \item[\(\mathcal{A}_0\)] No covariates adjustment: $\eta_i = \beta_0 + \iota_1 T_i^1 + \iota_2 T_i^2 .$
    \item[\(\mathcal{A}_1\)] Adjust for $X_{i,1}$ only:
$\eta_i = \beta_0 + \iota_1 T_i^1 + \iota_2 T_i^2 +  + \beta_{1} X_{i,1}.$
    \item[\(\mathcal{A}_2\)] Adjust for both $X_{i,1}$ and $X_{i,2}$: $\eta_i = \beta_0 + \iota_1 T_i^1 + \iota_2 T_i^2 + \beta_{1} X_{i,1} + \beta_{2} X_{i,2}.$
\end{itemize}
Thus \(\mathcal{A}_0\) and \(\mathcal{A}_1\) are misspecified working models, while \(\mathcal{A}_2\) is correct. 

Table~\ref{tb:sim1_conv} in the Supplementary Material reports the conventional Wald test statistics computed without adjustment for CAR, across complete randomization and the three CAR procedures under different working models. As the impact of CAR is not accounted for, the standard errors are systematically inflated for all CAR schemes in the misspecified models \(\mathcal{A}_0\) and \(\mathcal{A}_1\); as a result, the nominal 5\% Wald tests become conservative (Type I error $<0.05$). By contrast, CR is unaffected and has valid Type I errors because it does not use covariates in assignment. Moreover, consistent with expectations, the correctly specified working model \(\mathcal{A}_2\) leads to valid Type I errors and yields smaller sampling variability for the estimators than the misspecified models \(\mathcal{A}_0\) and \(\mathcal{A}_1\). Finally, combining the asymptotically independent \(p\)-values from Stage 1 and 2 via the inverse–\(\chi^2\) method preserves overall type~I error control provided each phase maintains valid type~I error.

Using the $\log$RR as a representative estimand, Table~\ref{tb:sim1_logRR} contrasts the conventional Wald tests (ignoring the impact of CAR) with our model-robust Wald tests. The findings are as follows: 
(i) The model-robust standard error estimates properly account for the CAR schemes, yielding valid 5\% type~I error across all working models, while the conventional tests are conservative in misspecified models \(\mathcal{A}_0\), \(\mathcal{A}_1\). Accordingly, under the alternative hypothesis, the model-robust tests achieve substantially higher power, showing gains of at least 50\% under \(\mathcal{A}_0\) and 35\% under \(\mathcal{A}_1\) across all CAR procedures.  (ii) Under the correctly specified working model \(\mathcal{A}_2\), we have  $\boldsymbol{\Gamma}_{\sf CAR,adj}=0$ asymptotically (cf.~Remark~\ref{rmk:Gamma}).  Consequently, CR and the CAR procedures exhibit comparable performance. For the same reason, the conventional Wald test, which ignores $\boldsymbol{\Gamma}_{\sf CAR,adj}$ when constructing 
$\boldsymbol{\Gamma}$, performs similarly to the model-robust Wald test. (iii) Under misspecification (e.g., \(\mathcal{A}_0\) and \(\mathcal{A}_1\)), residuals retain covariate signal; better covariate balance improves covariance estimation by Theorem \ref{thm:mu_normality}, so CAR, by achieving a superior strata balance, outperforms CR in power. (iv) The model-robust standard errors are computed using the sandwich estimator, which does not rely on a particular generative model. This yields more accurate estimation and greater robustness, even when the working model is misspecified. Similar observations are found with LOR and ATE, with results summarized in Table \ref{tb:sim1_LOR} and \ref{tb:sim1_ATE} in the Supplementary Material.

\begin{figure}[H]  \captionsetup{font=footnotesize}
  \centering
  \includegraphics[width=\textwidth]{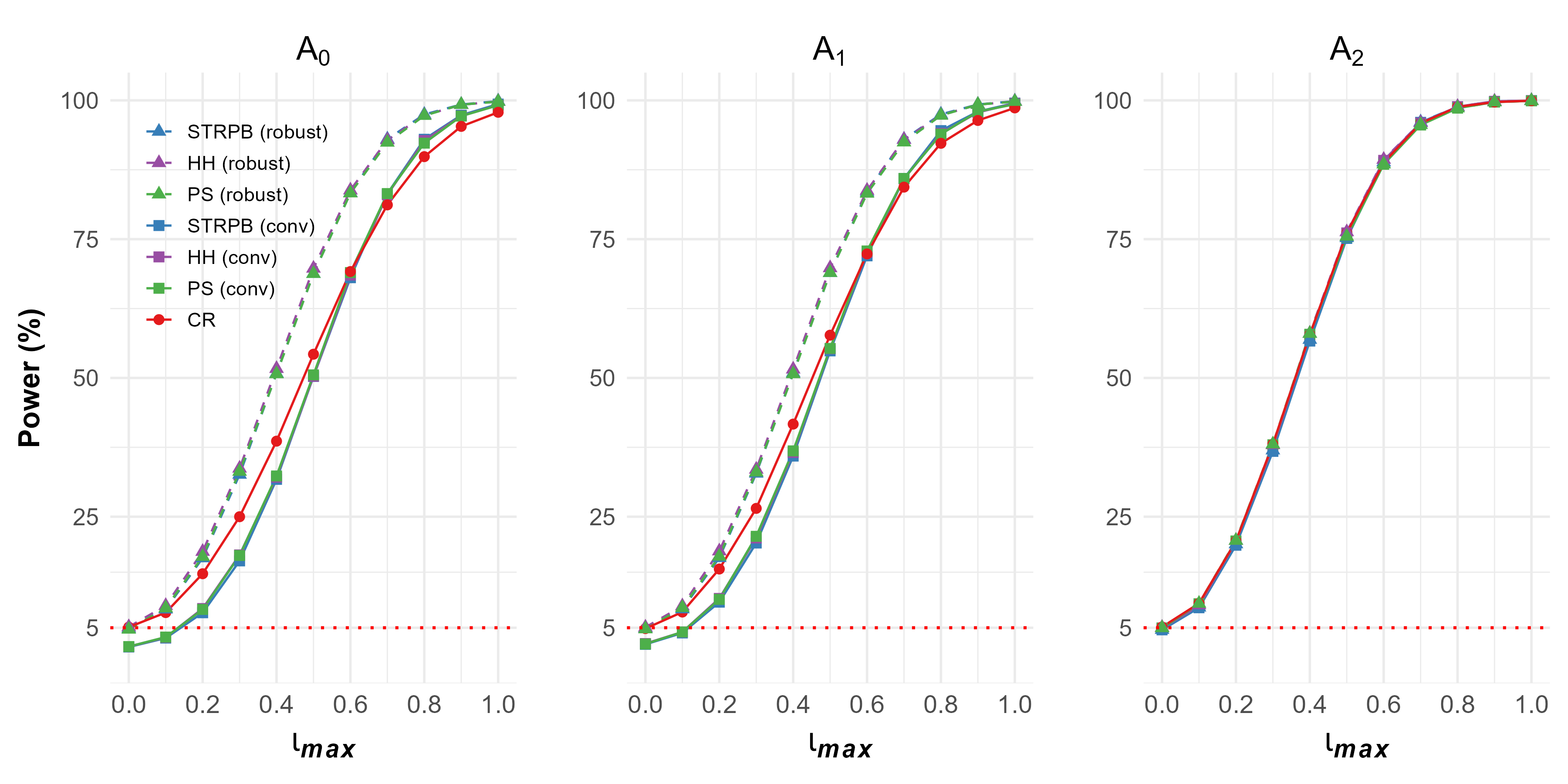}
 \caption{
   $\log$RR for Example~1, power comparison of conventional and model-robust tests 
under $(\mathcal{A}_0, \mathcal{A}_1, \mathcal{A}_2)$ across $\iota_{\max}$; 
the red dashed line denotes $\alpha=0.05$ under the null hypothesis.
  }
  \label{fig:sim1_logRR}
\end{figure}

To further illustrate our advantage, we vary $(\iota_1, \iota_2)$ systematically. Starting from $(0, 0)$, we sequentially increase them to $(0, 0.1)$, $(0.1, 0.2)$, and so on, up to $(0.9, 1)$. Let $\iota_{\max} = \max(\iota_1, \iota_2)$, and Figure~\ref{fig:sim1_logRR} displays the statistical power of the conventional and model-robust tests for $\log$RR under different randomization schemes as $\iota_{\max}$ increases. When the working models are misspecified (e.g., under $\mathcal{A}_0$ and $\mathcal{A}_1$), 
the model-robust tests achieve substantial power gains over their conventional counterparts. Although CR maintains valid Type~I error control when $\iota_{\max}=0$, its power remains lower than that of the model-robust tests. Results on LOR and ATE are shown in Figure \ref{fig:sim1_LOR} and \ref{fig:sim1_ATE}, which convey similar findings.

\begin{table}[H]
\captionsetup{font=scriptsize}
\centering
\scriptsize   
\renewcommand{\arraystretch}{0.35} 
\setlength{\tabcolsep}{3pt}        
\setlength{\aboverulesep}{0pt}     
\setlength{\belowrulesep}{0pt}     
\setlength{\abovetopsep}{0pt}      
\setlength{\belowbottomsep}{0pt}   
\caption{$\log$RR results for Example 1, the logistic model: Type I error rates and power (in $10^{-2}$) for Stage~1, ~2, and the combined analysis, across different working models and randomization schemes.}
\begin{tabular}{lll *{6}{c}}
\toprule
\textbf{} & \textbf{Procedure} & \textbf{Test} &
\multicolumn{3}{c}{\textbf{Type I}} & \multicolumn{3}{c}{\textbf{Power}} \\
\cmidrule(lr){4-6}\cmidrule(lr){7-9}
& & & \textbf{Stage 1} & \textbf{Stage 2} & \textbf{All} & \textbf{Stage 1} & \textbf{Stage 2} & \textbf{All} \\
\midrule
\multicolumn{3}{r}{ } & \multicolumn{3}{c}{$(\iota_1,\iota_2)=(0,0)$} & \multicolumn{3}{c}{$(\iota_1,\iota_2)=(0.3,0.4)$} \\
\midrule

\multirow{7}{*}{$\mathcal{A}_0$}
& CR &  & 5.15 & 5.02 & 5.11 & 21.72 & 29.62 & 38.61 \\

& STRPB & conv & 1.76 & 2.41 & 1.53 & 14.69 & 26.29 & 31.73 \\
& & robust & 4.82 & 4.88 & 4.89 & 28.29 & 39.10 & 50.82 \\

& HH & conv & 1.62 & 2.67 & 1.60 & 15.80 & 26.30 & 32.12 \\
& & robust & 4.99 & 5.42 & 5.15 & 29.53 & 39.34 & 51.67 \\

& PS & conv & 1.76 & 2.23 & 1.61 & 15.39 & 26.42 & 32.34 \\
& & robust & 4.76 & 4.86 & 4.71 & 28.29 & 39.16 & 50.72 \\
\midrule

\multirow{7}{*}{$\mathcal{A}_1$}
& CR &  & 5.13 & 5.05 & 4.85 & 23.68 & 31.71 & 41.68 \\

& STRPB & conv & 2.23 & 2.84 & 2.01 & 17.64 & 29.11 & 35.92 \\
& & robust & 4.82 & 4.93 & 4.91 & 28.23 & 39.16 & 50.91 \\

& HH & conv & 2.17 & 3.21 & 2.15 & 18.46 & 29.33 & 36.58 \\
& & robust & 5.05 & 5.38 & 5.08 & 29.57 & 39.41 & 51.56 \\

& PS & conv & 2.37 & 2.88 & 2.07 & 18.16 & 29.63 & 36.86 \\
& & robust & 4.86 & 4.84 & 4.77 & 28.37 & 39.15 & 50.72 \\
\midrule

\multirow{7}{*}{$\mathcal{A}_2$}
& CR &  & 5.49 & 5.22 & 5.14 & 33.06 & 43.64 & 57.79 \\

& STRPB & conv & 4.84 & 4.70 & 4.62 & 31.38 & 44.30 & 56.64 \\
& & robust & 4.96 & 4.71 & 4.66 & 31.72 & 44.41 & 56.91 \\

& HH & conv & 5.00 & 5.23 & 4.96 & 32.27 & 43.93 & 57.72 \\
& & robust & 5.13 & 5.24 & 5.05 & 32.56 & 43.95 & 58.03 \\

& PS & conv & 4.87 & 4.97 & 4.95 & 31.80 & 43.71 & 57.87 \\
& & robust & 4.92 & 4.99 & 4.97 & 31.96 & 43.75 & 57.99 \\
\bottomrule
\end{tabular}
\label{tb:sim1_logRR}
\end{table}
\vspace{-1cm}
\subsection{Example 2: probit model (M2)}\label{sec:simu2}

The data are generated from the probit regression model, where $Y_i \sim \mathrm{Bernoulli}(p_i)$ with $p_i = \Phi(\eta_i)$, $\Phi(\cdot)$ being the standard normal CDF and 
\[
\eta_i = \beta_0 + \iota_1 T_i^1 + \iota_2 T_i^2 
        + \beta_{1} X_{i,1}X_{i,2} 
        + \beta_{2} \exp(X_{i,1}+X_{i,2})+\beta_3X_{i,3}.
\vspace{-0.35cm}\]
 Covariates are generated as $X_{i,1},X_{i,2},X_{i,3} \sim \mathcal{N}(0,1)$ and parameters are set to $(\beta_0,\beta_{1},\beta_{2},\beta_3)=(-1,1,1,0.5)$. We consider three working models for analysis:
\begin{itemize}
    \item[\(\mathcal{A}_0\)] No covariates adjustment: $ \eta_i = \beta_0 + \iota_1 T_i^1 + \iota_2 T_i^2 $.
    \item[\(\mathcal{A}_1\)] Adjust for $X_{i,1}$ and $X_{i,2}$: $\eta_i = \beta_0 + \iota_1 T_i^1 + \iota_2 T_i^2 + \beta_{1} X_{i,1} + \beta_{2} X_{i,2}.$
    \item[\(\mathcal{A}_2\)] Adjust for $X_{i,1}$, $X_{i,2}$, and $X_{i,3}$: $\eta_i = \beta_0 + \iota_1 T_i^1 + \iota_2 T_i^2  
    + \beta_{1} X_{i,1} + \beta_{2} X_{i,2} + \beta_{3} X_{i,3}.$

\end{itemize}
 The key distinction between Examples~1 and~2 is that, in Example~2, all three working models \(\mathcal{A}_0\), \(\mathcal{A}_1\), and \(\mathcal{A}_2\) are misspecified. As a result, \(\boldsymbol{\Gamma}_{\sf{CAR,adj}}\) is nonzero, and ignoring it, as in the conventional Wald test, produces inflated standard error estimates and, consequently, conservative type~I error rates across all working models. By contrast, our model-robust procedure does not depend on the correctness of the working model. Its performance is therefore far less affected by misspecification: type~I errors remain close to the nominal level, and power is consistently higher than that of the conventional Wald test. These findings are confirmed by the empirical results reported in Tables~\ref{tb:sim2_conv} and~\ref{tb:sim2_logRR}; for instance, under \(\mathcal{A}_0\), power increases by at least 10\% across different CAR procedures. The phenomenon is further illustrated by varying \((\iota_1, \iota_2)\) in a manner similar to that in Example 1, with results summarized in Figure~\ref{fig:sim2_logRR}. Notably, in the working model \(\mathcal{A}_2\), although all covariates are included, the model is still misspecified; as a result, the model-robust tests achieve substantial power gains over their conventional counterparts. Similar observations are found with LOR and ATE, with results summarized in Figure \ref{fig:sim2_LOR} and \ref{fig:sim2_ATE}.
\begin{table}[h]
\captionsetup{font=scriptsize}
\centering
\scriptsize
\renewcommand{\arraystretch}{0.35}
\setlength{\tabcolsep}{3pt}
\setlength{\aboverulesep}{0pt}
\setlength{\belowrulesep}{0pt}
\setlength{\abovetopsep}{0pt}
\setlength{\belowbottomsep}{0pt}
\caption{$\log$RR results for Example 2, the probit model: Type I error rates and power (in $10^{-2}$) for Stage~1, Stage~2, and the combined analysis, across different working models and randomization schemes.}
\begin{tabular}{lll *{6}{c}}
\toprule
\textbf{} & \textbf{Procedure} & \textbf{Test} &
\multicolumn{3}{c}{\textbf{Type I}} & \multicolumn{3}{c}{\textbf{Power}} \\
\cmidrule(lr){4-6}\cmidrule(lr){7-9}
& & & \textbf{Stage 1} & \textbf{Stage 2} & \textbf{All} & \textbf{Stage 1} & \textbf{Stage 2} & \textbf{All} \\
\midrule
\multicolumn{3}{r}{ } & \multicolumn{3}{c}{$(\iota_1,\iota_2)=(0,0)$} & \multicolumn{3}{c}{$(\iota_1,\iota_2)=(0.2,0.3)$} \\
\midrule

\multirow{7}{*}{$\mathcal{A}_0$}
& CR &  & 4.92 & 5.16 & 5.13 & 26.68 & 37.38 & 48.64 \\

& STRPB & conv & 2.59 & 3.16 & 2.47 & 23.92 & 36.03 & 46.90 \\
& & robust & 4.82 & 4.80 & 5.07 & 31.90 & 43.50 & 56.95 \\

& HH & conv & 2.68 & 3.18 & 2.58 & 22.99 & 35.74 & 46.81 \\
& & robust & 4.66 & 4.89 & 4.85 & 31.13 & 42.55 & 56.06 \\

& PS & conv & 3.63 & 3.78 & 3.72 & 25.29 & 36.89 & 48.24 \\
& & robust & 4.86 & 5.21 & 4.95 & 29.08 & 41.35 & 53.33 \\
\midrule

\multirow{7}{*}{$\mathcal{A}_1$}
& CR &  & 5.30 & 5.16 & 5.54 & 30.28 & 41.43 & 54.43 \\

& STRPB & conv & 3.75 & 4.15 & 3.85 & 29.17 & 41.73 & 53.76 \\
& & robust & 5.11 & 4.98 & 5.14 & 33.58 & 45.44 & 58.95 \\

& HH & conv & 3.67 & 3.81 & 3.64 & 27.83 & 41.32 & 53.88 \\
& & robust & 4.96 & 4.83 & 4.88 & 32.44 & 44.91 & 58.72 \\

& PS & conv & 5.21 & 4.97 & 4.84 & 30.71 & 43.01 & 55.71 \\
& & robust & 5.10 & 4.99 & 4.84 & 30.39 & 42.95 & 55.49 \\
\midrule

\multirow{7}{*}{$\mathcal{A}_2$}
& CR &  & 5.29 & 5.31 & 5.55 & 32.33 & 43.47 & 56.41 \\

& STRPB & conv & 3.51 & 4.10 & 3.62 & 30.21 & 44.20 & 56.41 \\
& & robust & 4.88 & 5.03 & 4.93 & 35.49 & 48.26 & 61.75 \\

& HH & conv & 3.79 & 3.90 & 3.76 & 29.77 & 43.75 & 56.62 \\
& & robust & 5.02 & 4.88 & 4.97 & 34.78 & 47.58 & 61.68 \\

& PS & conv & 5.15 & 4.93 & 4.96 & 32.11 & 45.11 & 57.89 \\
& & robust & 5.06 & 4.96 & 4.89 & 31.85 & 45.14 & 57.64 \\
\bottomrule
\end{tabular}
\label{tb:sim2_logRR}
\end{table}

\begin{table}[htbp]
\captionsetup{font=scriptsize}
\centering
\scriptsize
\renewcommand{\arraystretch}{0.35}
\setlength{\tabcolsep}{3pt}
\setlength{\aboverulesep}{0pt}
\setlength{\belowrulesep}{0pt}
\setlength{\abovetopsep}{0pt}
\setlength{\belowbottomsep}{0pt}
\caption{$\log$RR results for the hypothetical trial: Type I error rates and power (in $10^{-2}$) for Stage~1, Stage~2, and the combined analysis, across different working models and randomization schemes.}
\begin{tabular}{lll *{6}{c}}
\toprule
\textbf{} & \textbf{Procedure} & \textbf{Test} &
\multicolumn{3}{c}{\textbf{Type I}} & \multicolumn{3}{c}{\textbf{Power}} \\
\cmidrule(lr){4-6}\cmidrule(lr){7-9}
& & & \textbf{Stage 1} & \textbf{Stage 2} & \textbf{All} & \textbf{Stage 1} & \textbf{Stage 2} & \textbf{All} \\
\midrule
\multicolumn{3}{r}{ } & \multicolumn{3}{c}{$(\iota_1,\iota_2)=(0,0)$} & \multicolumn{3}{c}{$(\iota_1,\iota_2)=(0.2,0.4)$} \\
\midrule

\multirow{7}{*}{$\mathcal{A}_0$}
& CR &  & 4.90 & 4.86 & 4.76 & 28.07 & 41.90 & 52.12 \\

& STRPB & conv & 3.99 & 3.81 & 3.33 & 26.08 & 41.90 & 51.67 \\
& & robust & 5.00 & 5.00 & 4.93 & 29.83 & 44.89 & 56.04 \\

& HH & conv & 3.89 & 4.30 & 4.11 & 26.27 & 41.34 & 51.32 \\
& & robust & 4.68 & 4.88 & 4.68 & 30.33 & 44.45 & 55.50 \\

& PS & conv & 4.23 & 4.10 & 3.98 & 26.99 & 41.99 & 51.58 \\
& & robust & 5.21 & 4.86 & 4.93 & 30.24 & 44.73 & 55.55 \\
\midrule

\multirow{7}{*}{$\mathcal{A}_1$}
& CR &  & 4.73 & 5.12 & 4.87 & 28.70 & 43.97 & 53.75 \\

& STRPB & conv & 4.36 & 4.11 & 3.76 & 27.87 & 43.40 & 53.86 \\
& & robust & 5.00 & 4.50 & 4.43 & 30.39 & 45.47 & 56.70 \\

& HH & conv & 4.07 & 4.74 & 4.54 & 28.07 & 42.99 & 53.59 \\
& & robust & 4.74 & 5.30 & 5.19 & 30.69 & 45.14 & 56.24 \\

& PS & conv & 4.69 & 4.46 & 4.43 & 28.90 & 43.36 & 53.71 \\
& & robust & 5.30 & 4.94 & 5.02 & 30.84 & 44.97 & 56.30 \\
\midrule

\multirow{7}{*}{$\mathcal{A}_2$}
& CR &  & 4.74 & 5.18 & 5.09 & 30.19 & 45.44 & 56.45 \\

& STRPB & conv & 5.05 & 4.79 & 4.81 & 30.15 & 45.62 & 56.72 \\
& & robust & 5.09 & 4.57 & 4.44 & 30.50 & 45.66 & 56.88 \\

& HH & conv & 4.83 & 4.64 & 4.90 & 30.41 & 45.09 & 55.99 \\
& & robust & 4.77 & 5.36 & 5.21 & 30.64 & 45.12 & 56.15 \\

& PS & conv & 4.67 & 4.96 & 4.68 & 30.94 & 45.18 & 56.62 \\
& & robust & 5.25 & 4.93 & 5.01 & 31.13 & 45.18 & 56.78 \\
\bottomrule
\end{tabular}
\label{tb:realdata_RR}
\end{table}



\section{Hypothetical trial example}\label{sec:realdata}
We illustrate the implications of our findings through a hypothetical trial modeled on a recent multi-arm Phase~II/III study~\citep{king2022two,king2021efficacy} that compared multiple doses of baricitinib against placebo in treating alopecia areata (AA). AA is an autoimmune disorder that causes sudden, patchy hair loss; its most severe forms, alopecia totalis (AT) and alopecia universalis (AU), involve complete scalp or body hair loss and can severely diminish patients’ quality of life. Conventional treatments such as topical corticosteroids or phototherapy may benefit patients with mild, patch-type AA, but are largely ineffective for chronic or refractory AT/AU. In contrast, oral Janus kinase (JAK) inhibitors have recently emerged as a promising therapeutic class. Baricitinib, a reversible and selective JAK1/JAK2 inhibitor already approved in over 70 countries for rheumatoid arthritis, has shown particular promise. In two Phase~II clinical trials~\citep{king2022two}, 36 weeks of baricitinib 4~mg treatment led to nearly 40\% of patients achieving a Scalp Alopecia Tool (SALT) score below 20, indicating over 85\% hair regrowth from baseline---establishing baricitinib as an effective therapy for severe AA.  

Motivated by this clinical evidence, we constructed a seamless Phase~II/III design to demonstrate the performance of our model-robust Wald tests. The trial adopts an adaptive two-stage framework, where the transition from Stage~1 to Stage~2 is guided by interim efficacy results without interrupting patient recruitment. In Stage~1, 420 patients are randomized equally to one control arm (placebo) and two treatment arms (baricitinib 2~mg and 4~mg). Based on interim outcomes, the superior treatment arm is advanced to Stage~2, where 500 additional patients are enrolled for confirmatory comparison with placebo. This integrated design efficiently combines dose selection with confirmatory evaluation, highlighting how our proposed methodology enhances decision-making in modern adaptive clinical trials.

The primary binary efficacy endpoint is defined as achieving a SALT score below 20, coded as 1 if achieved and 0 otherwise. Two clinically relevant baseline covariates are incorporated to capture patient heterogeneity. The first is the baseline SALT score, which quantifies the percentage of scalp hair loss at enrollment; values between 50 and 100 reflect the moderate-to-severe disease required for trial eligibility. The second covariate is disease duration, representing the length of the current episode. Consistent with observed distributions in chronic AA populations, we assume that 65\% of patients have a disease duration shorter than four years, while 35\% have a duration of four years or longer. Disease duration serves as an important prognostic factor, since shorter episodes are generally associated with more favorable treatment responses.  

To ensure balance across these key prognostic factors, we implement CAR procedures. Patients are stratified jointly by baseline SALT score ($<$75 vs.\ $\geq$75) and disease duration ($<$4 years vs.\ $\geq$4 years), yielding four strata. Within each stratum, randomization is carried out using one of three widely used CAR approaches: STRPB, PS, and HH. For comparison, we also consider complete randomization. This design mirrors the stratification logic used in the baricitinib Phase~II/III trials and ensures that treatment arms are comparable across clinically relevant subgroups.  

The binary outcome is modeled using logistic regression:  
\[
\text{logit}(p_i) = \beta_0 + \iota_1 T_{i}^1 + \iota_2 T_{i}^2 + \beta_1 X_{i,1} + \beta_2 X_{i,2},
\]
where $T_{i}^1$ and $T_{i}^2$ are indicators for the two treatment arms (with placebo as reference), $X_{i,1}$ denotes baseline SALT score, and $X_{i,2}$ denotes the binary indicator for disease duration ($1$ if $\geq4$ years, $0$ otherwise). The parameter vector is set to $(\beta_0,\iota_1,\iota_2,\beta_1,\beta_2)=(1.8,0.2,0.4,-0.04,-1.5)$. Here, $\beta_0=1.8$ represents the baseline intercept; $\iota_1=0.2$ and $\iota_2=0.4$ encode increasing treatment efficacy across the active arms; $\beta_1=-0.04$ reflects the negative effect of more severe baseline SALT scores; and $\beta_2=-1.5$ captures the reduced efficacy associated with longer disease duration. Each outcome $Y_i$ is generated from a Bernoulli distribution with success probability $p_i$.  

 We assess the performance of competing statistical procedures under CR, STRPB, PS, and HH, using 10,000 replicates and three working models $\mathcal{A}_0$, $\mathcal{A}_1$, and $\mathcal{A}_2$ as defined in Example~1. Table~\ref{tb:realdata_conv} reports type~I error rates from conventional Wald tests that ignore the contribution of CAR procedures. Consistent with our simulation findings, these tests are conservative under misspecified working models $\mathcal{A}_0$ and $\mathcal{A}_1$ across all three estimands: $\log$RR, LOR, and ATE. Additionally, Tables~\ref{tb:realdata_RR}, \ref{tb:realdata_LOR}, and \ref{tb:realdata_ATE} compare the conventional Wald test with the proposed model-robust Wald test for each metric. The results show a clear and consistent gain in power across all scenarios while maintaining valid type~I error control, thereby demonstrating the practical advantage of the model-robust approach.

\section{Discussion}\label{sec:discussion}

This paper develops the first unified framework for seamless Phase~II/III clinical trials that rigorously integrates covariate-adaptive randomization with generalized linear models. By combining model-robust Z-estimation with asymptotic theory under CAR, we establish valid inference for diverse outcome types and estimands, including log relative risk, log odds ratio, and average treatment effect. Through extensive simulations and a hypothetical trial example, we demonstrate that the proposed framework consistently improves power while maintaining Type~I error control, outperforming conventional Wald-type methods that ignore CAR. These contributions collectively bridge a critical methodological gap in the design of adaptive multi-stage trials with non-Gaussian outcomes, addressing both regulatory and practical needs in modern clinical research.  

Looking forward, several avenues merit further investigation. Extending the framework to survival outcomes with censoring or recurrent events would broaden its applicability in oncology and chronic disease trials. Incorporating high-dimensional covariates and machine-learning-based treatment effect estimators could improve precision while preserving rigorous error control. Another promising direction lies in developing computationally efficient algorithms for real-time implementation, enabling trialists to adaptively balance covariates while making interim decisions under complex outcome structures. Finally, future work should explore regulatory translation, including simulation-based operating characteristics tailored to FDA guidance, to further align seamless trial methodology with the demands of clinical practice.

\section*{Supplementary Material}
The Supplementary Material provides the proofs of the theoretical results and additional numerical results.


\bibhang=1.7pc
\bibsep=2pt
\fontsize{9}{14pt plus.8pt minus .6pt}\selectfont
\renewcommand\bibname{\large \bf References}
\expandafter\ifx\csname
natexlab\endcsname\relax\def\natexlab#1{#1}\fi
\expandafter\ifx\csname url\endcsname\relax
  \def\url#1{\texttt{#1}}\fi
\expandafter\ifx\csname urlprefix\endcsname\relax\def\urlprefix{URL}\fi








\noindent
Kun Yi \vspace{-1em}\\
Department of ISOM, School of Business and Management, HKUST \vspace{-1em}\\
E-mail: \texttt{kyiae@connect.ust.hk} \\
\noindent
Lucy Xia \vspace{-1em}\\
Department of ISOM, School of Business and Management, HKUST \vspace{-1em}\\
E-mail: \texttt{lucyxia@ust.hk}


	\clearpage
	
	\renewcommand{\refname}{References}
	
	\addcontentsline{toc}{section}{References}
	
	\bibliography{references}
	
	\clearpage
	\begin{appendices}
		\renewcommand{\thetable}{S\arabic{table}}
		\renewcommand{\thefigure}{S\arabic{figure}}
		\renewcommand{\thesection}{S\arabic{section}}
		\renewcommand{\theequation}{S\arabic{equation}}
		\setcounter{table}{0}
		\setcounter{figure}{0}
		\setcounter{section}{0}
		\setcounter{equation}{0}
		
		
\renewcommand{\baselinestretch}{2}

\markright{ \hbox{\footnotesize\rm Statistica Sinica: Supplement
}\hfill\\[-13pt]
\hbox{\footnotesize\rm
}\hfill }

\markboth{\hfill{\footnotesize\rm Kun Yi AND Lucy Xia} \hfill}
{\hfill {\footnotesize\rm MODEL ROBUST SEAMLESS TRIALS WITH CAR} \hfill}

\renewcommand{\thefootnote}{}
$\ $\par \fontsize{12}{14pt plus.8pt minus .6pt}\selectfont


 \centerline{\large\bf Model-robust Inference for Seamless II/III Trials}
\vspace{2pt}
 \centerline{\large\bf with Covariate Adaptive Randomization}
\vspace{2pt}
\vspace{.25cm}
 \centerline{Kun Yi and Lucy Xia}
\vspace{.4cm}
 \centerline{\it Department of ISOM, HKUST}
\vspace{.55cm}
 \centerline{\bf Supplementary Material}
\vspace{.55cm}
\fontsize{9}{11.5pt plus.8pt minus .6pt}\selectfont

\setcounter{section}{0}
\setcounter{equation}{0}
\def\theequation{S\arabic{section}.\arabic{equation}}
\def\thesection{S\arabic{section}}

\fontsize{12}{14pt plus.8pt minus .6pt}\selectfont
\section*{Notation and Definition}
\addcontentsline{toc}{section}{Notation and Definitions}

Throughout the supplementary material, we employ the following notation, not necessarily introduced in the text.

\begingroup
\renewcommand{\arraystretch}{1.5}
\setlength{\LTleft}{0pt}
\setlength{\LTright}{0pt}
\begin{longtable}{@{}p{3cm}p{\dimexpr\textwidth-3cm-4\tabcolsep}@{}}
$\|\boldsymbol{x}\|$ & For any column vector $\boldsymbol{x}$, $\|\boldsymbol{x}\| = \sqrt{\boldsymbol{x}^{\top}\boldsymbol{x}}$. \\
$\|\boldsymbol{M}\|$ & For a square matrix $\boldsymbol{M} \in \mathbb{R}^{q \times q}$, $\|\boldsymbol{M}\| = \sup \{\|\boldsymbol{M} \boldsymbol{x}\| : \boldsymbol{x} \in \mathbb{R}^{q}, \|\boldsymbol{x}\|=1\}$. \\
$\boldsymbol{X}_n = o_p(\boldsymbol{1})$ & For any sequence of real-valued vectors $\boldsymbol{X}_n$, if $\boldsymbol{X}_n \cvp \boldsymbol{0}$, then we denote $\boldsymbol{X}_n = o_p(\boldsymbol{1})$. \\
$\boldsymbol{X}_n = O_p(\boldsymbol{1})$ & For any sequence of real-valued vectors $\boldsymbol{X}_n$, if $\left\|\boldsymbol{X}_n\right\|$ is bounded in probability, then we denote $\boldsymbol{X}_n = O_p(\boldsymbol{1})$.\\
$\boldsymbol{M}_n = o_p(\boldsymbol{1})$ & For any sequence of real-valued square matrices $\boldsymbol{M}_n \in \mathbb{R}^{q \times q}$, if every element of $\boldsymbol{M}_n$ converges in probability to $0$, then we denote $\boldsymbol{M}_n = o_p(\boldsymbol{1})$. \\
$\boldsymbol{M}_n = O_p(\boldsymbol{1})$ & For any sequence of real-valued square matrices $\boldsymbol{M}_n \in \mathbb{R}^{q \times q}$, if every element of $\boldsymbol{M}_n$ is bounded in probability, then we denote $\boldsymbol{M}_n = O_p(\boldsymbol{1})$. \\
$[K]$ & $\{1, \ldots, K\}$. \\
$[s_{\mathrm{max}}]$ & $\{1, \ldots, s_{\mathrm{max}}\}$. \\
$\boldsymbol{Y}_i$ & $(Y_i(0), \ldots, Y_i(K))^\top$, where $Y_i(k)$ is the potential outcome of the $i$-th patient in treatment group $k$.\\
$\boldsymbol{V}$ & $(Y(0), \ldots, Y(K), \boldsymbol{X})$, where $Y(k)$ is the treatment outcome of patients in treatment group $k$ and $\boldsymbol{X}$ are baseline covariates .\\
$\boldsymbol{V}_i$ & $(Y_i(0), \ldots, Y_i(K), \boldsymbol{X_i})$, where $\boldsymbol{X}_i$ are baseline covariates. \\
$S$ & Stratification variable with possible values $s \in [s_\mathrm{max}]$. \\
$T_i^k$ & Treatment indicator (1 if i-th patient  is assigned to treatment $k \in 0 \cup [K]$, 0 otherwise) .\\
$\boldsymbol{T}_i$ & $(T_i^0, T_i^1, \ldots, T_i^K)$. \\
$n_k(s)$ & Number of patients assigned to treatment $k$ within stratum $s \in [s_\mathrm{max}]$. \\
$n_k$    & Number of patients assigned to treatment $k$. \\
$n(s)$  & Number of patients within stratum $s \in [s_\mathrm{max}]$.\\
$D_{n}^k(s)$ & Imbalance measure for treatment $k$ in stratum $s$. \\
$\boldsymbol{\theta}$ & Parameter vector of dimension $K+1+q$. \\
$\boldsymbol{\psi}^k(\boldsymbol{\theta})$ & Estimating equation for treatment group $k$, defined as $\boldsymbol{\psi}^k(Y(k),\bm{X}; \boldsymbol{\theta})$, where $\boldsymbol{\theta}$ is the parameter vector, $\bm{X}$ are baseline covariates and $Y(k)$ is the potential outcome under treatment $k$. \\
$\boldsymbol{\psi}_i^k(\boldsymbol{\theta})$ & The estimating equation for the $i$-th patient under treatment $k$, which is defined as $\boldsymbol{\psi}^k(Y_i(k),\bm{X}_i; \boldsymbol{\theta})$. \\
$\psi^{k,j}(\boldsymbol{\theta})$ & The $j$-th component (scalar) of the vector-valued estimating function $\boldsymbol{\psi}^k(\boldsymbol{\theta})$. \\
$\psi_{i}^{k,j}(\boldsymbol{\theta})$ & The $j$-th component (scalar) of the vector-valued estimating function $\boldsymbol{\psi}_i^k(\boldsymbol{\theta})$. \\
$\dot{\boldsymbol{\psi}}_i^k(\boldsymbol{\theta})$ & First derivative matrix with respect to $\boldsymbol{\theta}$: $\frac{\partial}{\partial \boldsymbol{\theta}} \boldsymbol{\psi}_i^k(\boldsymbol{\theta})$. \\
$\ddot{\boldsymbol{\psi}}_{i}^{k,j}(\boldsymbol{\theta})$ &Second derivative matrix for the $j$-th component with respect to $\boldsymbol{\theta}$: $\frac{\partial^2}{\partial \boldsymbol{\theta} \partial \boldsymbol{\theta}^{\top}} \psi_{i}^{k,j}(\boldsymbol{\theta})$. \\
$\boldsymbol{e}_j$ & Standard basis vector in $\mathbb{R}^{K+1}$ (1 in position $j$, 0 elsewhere). \\
$\boldsymbol{\Psi}$ & $\frac{1}{K+1}\sum_{k \in 0 \cup [K]}  \mathbb{E}\left[\left.\frac{\partial}{\partial \boldsymbol{\theta}} \boldsymbol{\psi}^k(Y(k),\boldsymbol{X}; \boldsymbol{\theta})\right|_{\boldsymbol{\theta}=\boldsymbol{\theta}^*}\right]$.\\
$\boldsymbol{h}(\boldsymbol{X};\boldsymbol{\theta})$ & $(h^0(\boldsymbol{X};\boldsymbol{\theta}),...,h^K(\boldsymbol{X};\boldsymbol{\theta}))^\top$.\\
$\boldsymbol{h}(\boldsymbol{X}_i;\boldsymbol{\theta})$ & $(h^0(\boldsymbol{X}_i;\boldsymbol{\theta}),...,h^K(\boldsymbol{X}_i;\boldsymbol{\theta}))^\top$.\\
$\widehat{\boldsymbol{\mu}}$ & $(\widehat{\mu}_0,...,\widehat{\mu}_K)^\top$, where $\widehat{\mu}_k = n^{-1}\sum_{i=1}^n h^k(\boldsymbol{X}_i,\widehat{\boldsymbol{\theta}})$ for all $k = 0,...,K$.\\
$\boldsymbol{r}(\boldsymbol{\theta})$ & $(r^0(\boldsymbol{\theta}),...,r^K(\boldsymbol{\theta}))^\top$, where $r^k(\boldsymbol{\theta}) =  Y(k) - h^k(\boldsymbol{X},\boldsymbol{\theta})$ for all $k = 0,...,K$.\\
$\boldsymbol{r}_i(\boldsymbol{\theta})$ & $(r_i^0(\boldsymbol{\theta}),...,r_i^K(\boldsymbol{\theta}))^\top$, where $r_i^k(\boldsymbol{\theta}) =  Y_i - h^k(\boldsymbol{X}_i,\boldsymbol{\theta})$ for all $k = 0,...,K$.
\end{longtable}
\endgroup

\section{Assumptions}
\begin{customass}{1}[Reproduced from Assumption \ref{Assumption 3.1} in the main text]\quad
\begin{enumerate}[leftmargin=*]
\item  \(\{ (\bm{X}_i,Y_i(0),\ldots,Y_i(K) \}_{i = 1}^n\) are independently identically distributed as \\ $(\bm{X},Y(0),\ldots, Y(K))$, with $\mathbb{E}\left[\|\bm{X}\|^2\right]<\infty$, $\max_{k\in 0\cup [K]}\mathrm{Var}[Y(k)]<\infty$.
   \item Let $\pi_{s}$ denote the proportion of observations in stratum $s$. For any \(s \in [s_{\mathrm{max}}]\), we have \(0 < \pi_{s} < 1\), and \(\sum_{s \in [s_{\mathrm{max}}]} \pi_{s} = 1\).

    \item Recall that \(\mathcal F_{i-1}=\mathcal X_{i,\mathrm{ex}}\otimes \mathcal T_{i-1}\otimes \mathcal X_{i-1}\otimes \mathcal Y_{i-1}\) and the assignment rule satisfies $\bm{\phi}_{i} = \mathbb{E}\left[\bm{T}_i|\mathcal F_{i-1}\right]$ for $i\in [n]$. Let us define the imbalance within stratum $s$ and treatment arm $k$ as $D_{n}^k(s)$ with the following decomposition 
    \begin{align*}
    D_{n}^k(s)&:=\sum_{i = 1}^{n}\left({T}_{i}^k - \frac{1}{K+1} \right)\mathbb{I}\{S_{i}=s\}\cr 
    &=\sum_{i = 1}^{n}M_{i}^k(s)+d_{n}^k(s), \quad s\in[s_{\mathrm{max}}],\quad k\in 0\cup[K].
    \end{align*}
   We assume that the CAR procedure satisfies
    \begin{enumerate}[leftmargin=*]
        \item \(\mathbb E[\lbrace d_{n}^k(s) \rbrace
    ^{2}]=o(n)\); and
        \item \(\{M_{i}^k(s)\}_{i = 1}^{n}\) is a sequence of bounded zero-mean martingale differences with respect to \(\mathcal{F}_{i - 1}\). Further with vectorized 
        \[
\bm M_i
:= \operatorname{vec}\!\Big(\big(M_i^k(s)\big)_{s\in[s_{\mathrm{max}}],\;k\in[K]}\Big)
\in \mathbb R^{s_{\mathrm{max}}(K+1)},
\]
the averaged conditional second moment converges in probability to a block covariance matrix:
\[
\frac{1}{n}\sum_{i=1}^{n}
\mathbb E\!\left[\bm M_i\,\bm M_i^\top \,\big|\, \mathcal F_{i-1}\right]
\ \cvp\ 
\boldsymbol{\Sigma}^{\sf{CAR}}
\ \in\ \mathbb R^{s_{\mathrm{max}}(K+1)\times s_{\mathrm{max}}(K+1)}.
\]

Entrywise, if we index $\boldsymbol{\Sigma}^{\sf{CAR}}$ by $(s,s')\in [s_{\mathrm{max}}]\times [s_{\mathrm{max}}]$ as a block matrix:
\[
\boldsymbol{\Sigma}^{\sf{CAR}}
=\big[\,\boldsymbol{\Sigma}^{\sf{CAR}}(s,s')\,\big]_{s,s'\in [s_{\mathrm{max}}]},
\qquad
\boldsymbol{\Sigma}^{\sf{CAR}}(s,s') \in \mathbb R^{(K+1)\times (K+1)},
\]
and defined the $(k,k')$-th element in $\boldsymbol{\Sigma}^{\sf{CAR}}(s,s')$ as $\bm{\Sigma}^{\sf{CAR}}_{kk'}(s,s')$, then 

        \begin{align*}
        &\frac{1}{n}\sum_{i = 1}^{n}\mathbb E[M_{i}^k(s)M_{i}^{k'}(s') \mid \mathcal{F}_{i - 1}] \cvp \bm{\Sigma}^{\sf{CAR}}_{kk'}(s,s').
        \end{align*}
    \end{enumerate}
\end{enumerate}
\label{supp:Assumption 3.1}
\end{customass}

\begin{customass}{S2}[Regularity conditions for the $Z$-estimator]\quad
\begin{enumerate}[leftmargin=*,nosep]
    \item Compactness of the parameter space: $\bm{\theta} \in \bm{\Theta} \subset \mathbb{R}^{K+1+q}$, with $\bm{\Theta}$ compact.
    \item Existence of a unique solution and the invertibility condition: There exists a unique $\bm{\theta}^* \in \mathrm{int}(\bm{\Theta})$ such that:
\[ \frac{1}{K+1}\sum_{k=0}^K  \mathbb{E}[\bm{\psi}^k(\bm{\theta}^*)] = \bm{0}, \]
    and $\bm{\Psi} = \sum_{k=0}^K  \mathbb{E}[\dot{\bm{\psi}}^k(\bm{\theta}^*)]/(K+1)$ is invertible. 
    \item Bounded second moment:
    \[
        \sup_{\bm{\theta} \in \bm{\Theta}, k \in 0\cup[K]} \mathbb{E}\left[\|\bm{\psi}^k(\bm{\theta})\|^2\right] < \infty, \quad \sup_{ k \in 0\cup [K]} \mathbb{E}\left[\|\dot{\bm{\psi}}^k(\bm{\theta}^*)\|^2\right] < \infty.
    \]

    \item Differentiability and bounded second derivatives: For each $k \in 0\cup [K]$,
    \begin{itemize}[nosep,leftmargin=*]
        \item $\bm{\psi}^k(\bm{\theta})$ is twice continuously differentiable for every $(y,\bm{x})$ in the support of $( Y(k),\bm{X})$ and is dominated by an integrable function $\bm{u}(Y(k),\bm{X})$.
        \item Bounded second derivatives: $\exists \:C > 0$ and integrable $v(Y(k),\bm{X})$ s.t. $\forall j \in \{1,...,K+1+q\}$:
    \[ \|\ddot{\bm{\psi}}^{k,j}(\bm{\theta})\| < v( Y(k),\bm{X}) \ \mathrm{for} \ \|\bm{\theta}-\bm{\theta}^*\| < C. \]
    \end{itemize}
    \item Conditions for the estimating equation:  $\widehat{\bm{\theta}}$ satisfies
\[
\sum_{k=0}^K\sum_{i=1}^n T_i^k \bm{\psi}^k(Y_i(k), \bm{X}_i; \widehat{\bm{\theta}})= o_p(n^{-1/2}).
\]
    \end{enumerate}

\label{Assump:Z-estimation}
\end{customass}

\section{Proof of Theoretical Results}

\subsection{Proof of Proposition 1}

\begin{custompro}{1}[Reproduced from Proposition 1 in the main text]
    \label{supp:prop:consistency}
Under Assumptions \ref{supp:Assumption 3.1} and \ref{Assump:Z-estimation}, 
\[
\widehat{\bm{\theta}} - \bm{\theta}^* =O_p(\frac{1}{\sqrt{n}}),\qquad \widehat{\bm{\mu}}-\bm{\mu} =O_p(\frac{1}{\sqrt{n}}).
\]
\end{custompro}

\subsubsection{Lemmas}

\begin{customlm}{1}\label{lem:stratum}
	Under Assumption \ref{supp:Assumption 3.1}, for each \(k \in 0 \cup [K]\) and \(s \in [s_\mathrm{max}]\),
		\[
		\frac{n_k(s)}{n} = \frac{1}{K+1} \pi_{s} + o_p(1) \quad \mathrm{and} \quad D_n^k(s) = O_p(\sqrt{n}).
		\]
\end{customlm}

\begin{customlm}{2}\label{lem:uniformconverge}
	Under Assumption \ref{supp:Assumption 3.1}, let \(Z_{i}(\boldsymbol{\theta}) := Z(\boldsymbol{V}_{i} ; \boldsymbol{\theta})\), $i=1,\cdots,n$, be i.i.d. as \( Z(\boldsymbol{\theta}):=Z(\boldsymbol{V} ; \boldsymbol{\theta})\) for some measurable real-valued function \(Z\), and \(\boldsymbol{\theta} \in \boldsymbol{\Theta}\). Further, for each \(s \in [s_\mathrm{max}]\), let \(Z^{(s)}\) denote the conditional process of \(Z = \{Z(\boldsymbol{\theta}): \boldsymbol{\theta} \in \boldsymbol{\Theta}\}\) given \(S = s\). 
    Then, if \(\sup_{\boldsymbol{\theta} \in \boldsymbol{\Theta}}\left|\mathbb{\mathbb{E}}\left[Z^{(s)}(\boldsymbol{\theta})\right]\right| < \infty\) and \(Z^{(s)}\) is \(P\)-Glivenko-Cantelli for each \(s \in [s_\mathrm{max}]\), we have 
    \[
    \sup_{\boldsymbol{\theta} \in \boldsymbol{\Theta}} \left|\frac{1}{n} \sum_{i = 1}^{n} T_i^{k} Z_{i}(\boldsymbol{\theta}) - \frac{1}{K+1} \mathbb{\mathbb{E}}[Z(\boldsymbol{\theta})] \right| \cvp 0 ,\ \ \mathrm{for \ all}\ \  k \in 0 \cup [K].
    \]
\end{customlm}

\begin{customlm}{3}
    For each $k \in 0 \cup [K]$ and $s \in [s_\mathrm{max}]$, (\ref{eq:lem2:5}) holds.
    \label{lem:3}
\end{customlm}

\begin{customlm}{4}[Consistency of $\widehat{\boldsymbol{\theta}}$]\label{lemma:consistency}
Under Assumptions \ref{supp:Assumption 3.1} and \ref{Assump:Z-estimation}, the Z-estimator $\widehat{\boldsymbol{\theta}}$ is consistent.
\label{lem:theta:Consistency}
\end{customlm}

\begin{customlm}{5}
Under Assumptions \ref{supp:Assumption 3.1} and \ref{Assump:Z-estimation}, we have:
\begin{align*}
     \frac{1}{\sqrt{n}} \sum_{i=1}^{n} \sum_{k \in 0 \cup [K]} T_i^k\boldsymbol{\psi} _{i}^{k}(\boldsymbol{\theta}^*) \cvd N(\boldsymbol{0},  \boldsymbol{\Lambda} ),
\end{align*}
where
\begin{equation*}
\begin{aligned}
\boldsymbol{\Lambda} =& \mathrm{Var}\left[\mathbb{E}\left(\frac{1}{K+1}\sum_{k \in 0 \cup [K]}\boldsymbol{\psi}^k(\boldsymbol{\theta}^*)\mid S\right)\right]+\frac{1}{K+1}\sum_{k \in 0 \cup [K]}\mathbb{E}\left[\mathrm{Var}\left[\boldsymbol{\psi}^k(\boldsymbol{\theta}^*)\mid S\right]\right]\\
+&\sum_{k,k'\in 0 \cup [K]}\sum_{s,s' \in [s_\mathrm{max}]}\boldsymbol{\Sigma}_{kk'}^{\sf{CAR}}(s,s')\mathbb{E} \left[ \boldsymbol{\psi}^{k}(\boldsymbol{\theta}^*) \mid S = s \right]\mathbb{E} \left[ \boldsymbol{\psi}^{k'}(\boldsymbol{\theta}^*) \mid S = s' \right]^{\top}.
\end{aligned}
\end{equation*}
\label{lem:psi:normality}
\end{customlm}

\begin{customlm}{6}
Under Assumptions \ref{supp:Assumption 3.1} and \ref{Assump:Z-estimation}, the estimator $\widehat{\boldsymbol{\theta}}$ is asymptotically linear. Specifically, we have the following result:
\begin{align}
    \sqrt{n}(\widehat{\boldsymbol{\theta}} - \boldsymbol{\theta}^*) = -\frac{1}{\sqrt{n}}\sum_{i = 1}^n\boldsymbol{\Psi}^{-1}\boldsymbol{\psi}(\boldsymbol{T}_i,\boldsymbol{X}_i,Y_i;\boldsymbol{\theta}^*) + o_{p}(\boldsymbol{1}).
    \label{eq:theta:linearity}
\end{align}
\label{lem:theta:linearity}
\end{customlm}

\begin{customlm}{7}
    Under Assumptions \ref{supp:Assumption 3.1} and \ref{Assump:Z-estimation}, we have:
\begin{align*}
    \sqrt{n} (\widehat{\boldsymbol{\theta}} - \boldsymbol{\theta}^*) \cvd N(\boldsymbol{0}, \boldsymbol{\Psi}^{-1} \boldsymbol{\Lambda} \lbrace\boldsymbol{\Psi}^{-1} \rbrace^\top).
\end{align*}
\label{lem:theta:normality}
\end{customlm}

\begin{customlm}{8}
Under Assumptions \ref{supp:Assumption 3.1} and  \ref{Assump:Z-estimation}, $\widehat{\boldsymbol{\mu}}$ is consistent.
 \label{lem:mu:consistency}
\end{customlm}

\begin{customlm}{9}
\label{lem:mu_linearity}
Under Assumptions \ref{supp:Assumption 3.1} and \ref{Assump:Z-estimation}, the population mean estimator $\widehat{\boldsymbol{\mu}}$ have the asymptotic linearity:
\small{
\begin{equation*}
\sqrt{n}(\widehat{\boldsymbol{\mu}}-\boldsymbol{\mu}) = \frac{1}{\sqrt{n}}\sum_{i=1}^n\lbrace\boldsymbol{h}(\boldsymbol{X}_i;\boldsymbol{\theta}^*)-\boldsymbol{\mu}\rbrace-\frac{K+1}{\sqrt{n}}\sum_{i = 1}^n\sum_{k \in 0 \cup [K]}T_i^k\lbrace h^k(\boldsymbol{X_i};\boldsymbol{\theta}^*)-Y_i\rbrace\boldsymbol{e}_k+o_p(\bm{1}).
\end{equation*}
}
\end{customlm}

\begin{customlm}{10}
\label{lem:mu_normality}
Under Assumptions \ref{supp:Assumption 3.1} and \ref{Assump:Z-estimation}, we have:
\begin{equation}
\sqrt{n}(\widehat{\bm{\mu}}-\bm{\mu})\cvd N(\bm{0},\bm{\Gamma}),
\end{equation}
where
\begin{align*}
    &\bm{\Gamma}=(K+1)\mathrm{diag}\left \{  \mathrm{Var}\left[r^0(\boldsymbol{\theta}^*)\right],\dots, \mathrm{Var}\left[r^K(\boldsymbol{\theta}^*)\right]\right \} +\mathrm{Var}\left[\boldsymbol{Y}\right]-\mathrm{Var}\left[\boldsymbol{r}\right] \\
    & \quad-(K+1)^2\sum_{k,k'\in 0 \cup [K]}\sum_{s,s'\in [s_\mathrm{max}]}\lbrace\boldsymbol{\Sigma}_{kk'}^{\sf{CR}}(s,s')-\boldsymbol{\Sigma}_{kk'}^{\sf{CAR}}(s,s')\rbrace \boldsymbol{L}(s,k)\boldsymbol{L}(s',k')^{\top}, \\
     &\boldsymbol{L}(s,k)= \mathbb E\left[r^{k}(\boldsymbol{\theta}^*)\mid S = s\right]\boldsymbol{e}_k.
\end{align*}
\end{customlm}

\subsubsection{Main proof of Proposition \ref{supp:prop:consistency}}
\begin{proof}
The result follows immediately from Lemmas \ref{lem:theta:normality}, \ref{lem:mu_normality}.
\end{proof}

\subsubsection{Proof of lemmas}

\begin{proof}[Proof of Lemma \ref{lem:stratum}]
    The proof follows similarly from Proposition 1.1 in \citep{liu2023impacts}
\end{proof}

\begin{proof}[Proof of Lemma \ref{lem:uniformconverge}]
The proof contains three steps. In the first step, we rearrange the samples and transform the sample function into a sorted sample function. In the second step, we prove that the sample function is equivalent to the sum of the sample functions at each stratum. In the third step, we prove that the sample function of each stratum satisfies uniform convergence by using Lemma \ref{lem:3}. Then, by combining the conclusions of the first and second steps, we prove the uniform convergence of the sample function.\\
Step 1: Let \((\widetilde{\boldsymbol{T}}_1, \widetilde{\boldsymbol{V}}_1, \ldots, \widetilde{\boldsymbol{T}}_1, \widetilde{\boldsymbol{V}}_n)\) be a permutation of  \((\boldsymbol{T}_1, \boldsymbol{V}_1, \ldots, \boldsymbol{T}_n, \boldsymbol{V}_n)\) such that each \((\boldsymbol{T}_i, \boldsymbol{V}_i)\) is first ordered by strata (starting with those having \(S_i = 1\), then \(S_i = 2\), and so on) and then by treatment group (with \(T_i^0 = 1\) coming first) within each stratum. By construction, the process \(\left\{\frac{1}{n}\sum_{i = 1}^n T_i^k Z_i(\boldsymbol{\theta}): \boldsymbol{\theta} \in \boldsymbol{\Theta}\right\}\) is identical to \(\left\{\frac{1}{n}\sum_{i = 1}^n \widetilde{T}_i^k Z(\widetilde{\boldsymbol{V}}_i; \boldsymbol{\theta}): \boldsymbol{\theta} \in \boldsymbol{\Theta}\right\}\). Thus, with probability \(1\), we have
{\small
\begin{align}
    & \sup_{\boldsymbol{\theta} \in \boldsymbol{\Theta}} \left|\frac{1}{n}\sum_{i = 1}^n T_i^k Z_i(\boldsymbol{\theta}) - \frac{1}{K+1} \mathbb{\mathbb{E}}[Z(\boldsymbol{\theta})]\right| = \sup_{\boldsymbol{\theta} \in \boldsymbol{\Theta}} \left|\frac{1}{n}\sum_{i = 1}^n \widetilde{T}_i^k Z(\widetilde{\boldsymbol{V}}_i; \boldsymbol{\theta}) - \frac{1}{K+1} \mathbb{\mathbb{E}}[Z(\boldsymbol{\theta})]\right|.
    \label{eq:lem2:1}
\end{align}
}
Step 2: For each \(s \in [s_{\mathrm{max}}]\), let \(\boldsymbol{V}_1^{(s)}, \ldots, \boldsymbol{V}_n^{(s)}\) be i.i.d.\ copies of \(\boldsymbol{V}^{(s)}\), independent of \((\widetilde{\boldsymbol{T}}_1, \ldots, \widetilde{\boldsymbol{T}}_n)\) and \((\widetilde{\boldsymbol{V}}_1, \ldots, \widetilde{\boldsymbol{V}}_n)\). We use the notation \(\boldsymbol{V}_i^{(S_i)}\) to denote \(\sum_{s \in [s_\mathrm{max}]} \mathbb{I}\{S_i = s\} \boldsymbol{V}_i^{(s)}\). Let \(\widetilde{S}_i\) denote the stratum variable encoded in \(\widetilde{\boldsymbol{V}}_i\). Then, we have the following derivation:
\begin{align*}
    \ \ \mathbb{P}(\widetilde{\boldsymbol{V}}_1, \ldots, \widetilde{\boldsymbol{V}}_n | \widetilde{\boldsymbol{T}}_1, \widetilde{S}_1, \ldots, \widetilde{\boldsymbol{T}}_1, \widetilde{S}_n) &= \mathbb{P}(\widetilde{\boldsymbol{V}}_1, \ldots, \widetilde{\boldsymbol{V}}_n | \widetilde{S}_1, \ldots, \widetilde{S}_n)\\
    &= \prod_{i = 1}^n \mathbb{P}(\widetilde{\boldsymbol{V}}_i | \widetilde{S}_i) = \prod_{i = 1}^n \mathbb{P}(\boldsymbol{V}_i^{( \widetilde{S}_i)})\\
    &= \mathbb{P}(\boldsymbol{V}_1^{(\widetilde{S}_1)}, \ldots, \boldsymbol{V}_n^{(\widetilde{S}_n)} | \widetilde{\boldsymbol{T}}_1, \widetilde{S}_1, \ldots, \widetilde{\boldsymbol{T}}_1, \widetilde{S}_n).
\end{align*}


By the chain rule of joint probability, we obtain the following equality of the joint distributions: 
\begin{align*}
	\mathbb{P}(\widetilde{\boldsymbol{T}}_1, \widetilde{\boldsymbol{V}}_1, \ldots, \widetilde{\boldsymbol{T}}_1, \widetilde{\boldsymbol{V}}_n) = \mathbb{P}(\widetilde{\boldsymbol{T}}_1, \boldsymbol{V}_1^{(\widetilde{S}_1)}, \ldots, \widetilde{\boldsymbol{T}}_1, \boldsymbol{V}_n^{(\widetilde{S}_n)})
\end{align*}
unconditionally. This implies that, for any \(\varepsilon > 0\),
\begin{align}
    &\mathbb{P}\left(\sup_{\boldsymbol{\theta} \in \boldsymbol{\Theta}} \left|\frac{1}{n}\sum_{i = 1}^n \widetilde{T}_i^k Z(\widetilde{\boldsymbol{V}}_i; \boldsymbol{\theta}) - \frac{1}{K+1} \mathbb{\mathbb{E}}[Z(\boldsymbol{\theta})]\right| > \varepsilon\right)\nonumber\\
    =& \mathbb{P}\left(\sup_{\boldsymbol{\theta} \in \boldsymbol{\Theta}} \left|\frac{1}{n}\sum_{i = 1}^n \widetilde{T}_i^k Z(\boldsymbol{V}_i^{(\widetilde{S}_i)}; \boldsymbol{\theta}) - \frac{1}{K+1} \mathbb{\mathbb{E}}[Z(\boldsymbol{\theta})]\right| > \varepsilon\right).
    \label{eq:lem2:2}
\end{align}\\
Step 3: Based on our construction of \((\widetilde{\boldsymbol{T}}_i, \widetilde{\boldsymbol{V}}_i)\), we have
\begin{align}
\sum_{i = 1}^n \widetilde{T}_i^k Z(\boldsymbol{V}_i^{(s = \widetilde{S}_i)}; \boldsymbol{\theta}) = &\sum_{s \in [s_\mathrm{max}]}  \sum_{i = N(s) + \sum_{j = 0}^{k - 1}n_j(s)+1}^{N(s) + \sum_{j = 0}^kn_j(s)} Z(\boldsymbol{V}_i^{(s)}; \boldsymbol{\theta}) \nonumber \\
	= &\sum_{s \in [s_\mathrm{max}]}  \sum_{i = N(s)+\sum_{j = 0}^{k - 1}n_j(s)+1}^{N(s) + \sum_{j = 0}^kn_j(s)} Z_i^{(s)}(\boldsymbol{\theta}),\label{eq:lem2:3}
\end{align}
where \(N(s) = \sum_{i = 1}^n \mathbb{I}\{S_i < s\}\), \(n_k(s) = \sum_{i = 1}^n T_i^k \mathbb{I}\{S_i = s\}\) and \(Z_i^{(s)}(\boldsymbol{\theta}) = Z(\boldsymbol{V}_i^{(s)}; \boldsymbol{\theta})\). 


The fact that the corresponding joint distributions are identical implies:
\begin{equation*}
\mathbb{P}(\boldsymbol{V}_{N(s) + \sum_{j = 0}^{k - 1}n_j(s)+1}^{(s)}, \ldots, \boldsymbol{V}_{N(s) + \sum_{j = 0}^{k - 1}n_j(s)+n_k(s)}^{(s)}) = \mathbb{P}(\boldsymbol{V}_1^{(s)}, \ldots, \boldsymbol{V}_{n_k(s)}^{(s)}).
\end{equation*}

Therefore, for any \(\varepsilon > 0\),
\begin{align}
    &\mathbb{P}\left(\sup_{\boldsymbol{\theta} \in \boldsymbol{\Theta}} \left|\frac{1}{n}\sum_{i = N(s) +\sum_{j = 0}^{k - 1}n_j(s)+ 1}^{N(s) + \sum_{j = 0}^{k}n_j(s)} Z_i^{(s)}(\boldsymbol{\theta}) - \frac{1}{K+1} \pi_{s} \mathbb{E}[Z^{(s)}(\boldsymbol{\theta})]\right| > \varepsilon\right) \nonumber\\
    =& \mathbb{P}\left(\sup_{\boldsymbol{\theta} \in \boldsymbol{\Theta}} \left|\frac{1}{n}\sum_{i = 1}^{n_k(s)} Z_i^{(s)}(\boldsymbol{\theta}) - \frac{1}{K+1} \pi_{s} \mathbb{E}[Z^{(s)}(\boldsymbol{\theta})]\right| > \varepsilon\right).
    \label{eq:lem2:4}
\end{align}
By Lemma \ref{lem:3}, we know: 
\begin{align}
	\sup_{\boldsymbol{\theta} \in \boldsymbol{\Theta}} \left|\frac{1}{n}\sum_{i = 1}^{n_k(s)} Z_i^{(s)}(\boldsymbol{\theta}) - \frac{1}{K+1} \pi_{s} \mathbb{E}[Z^{(s)}(\boldsymbol{\theta})]\right| \cvp 0.
    \label{eq:lem2:5}
\end{align}
Combing (\ref{eq:lem2:1}), (\ref{eq:lem2:2}), (\ref{eq:lem2:3}), (\ref{eq:lem2:4}) and (\ref{eq:lem2:5}), we finish this proof.
\end{proof}

\begin{proof}[Proof of Lemma \ref{lem:3}]
By Lemma \ref{lem:stratum}, we know that \(\frac{n_k(s)}{n} \cvp \frac{\pi_{s}}{K+1} \). Then
\begin{align*}
&\sup_{\boldsymbol{\theta} \in \boldsymbol{\Theta}}\left|\frac{1}{n}\sum_{i = 1}^{n_k(s)}Z_i^{(s)}(\boldsymbol{\theta})-\frac{1}{K+1} \pi_{s}\mathbb{E}[Z^{(s)}(\boldsymbol{\theta})]\right|\\
=&\frac{n_k(s)}{n}\sup_{\boldsymbol{\theta} \in \boldsymbol{\Theta}}\left|\frac{1}{n_k(s)}\sum_{i = 1}^{n_k(s)}Z_i^{(s)}(\boldsymbol{\theta})-\frac{n \pi_{s}}{(K+1)n_k(s)}\mathbb{E}[Z^{(s)}(\boldsymbol{\theta})]\right|\\
\leq&\frac{n_k(s)}{n}\sup_{\boldsymbol{\theta} \in \boldsymbol{\Theta}}\left|\frac{1}{n_k(s)}\sum_{i = 1}^{n_k(s)}Z_i^{(s)}(\boldsymbol{\theta})-\mathbb{E}[Z^{(s)}(\boldsymbol{\theta})]\right| \\ 
&+\frac{n_k(s)}{n}\sup_{\boldsymbol{\theta} \in \boldsymbol{\Theta}}\left|\left(\frac{n\frac{1}{K+1} \pi_{s}}{n_k(s)}-1\right)\mathbb{E}[Z^{(s)}(\boldsymbol{\theta})]\right|\\
=&\bigg(\frac{\pi_{s}}{K+1} +o_p(1)\bigg)\sup_{\boldsymbol{\theta} \in \boldsymbol{\Theta}}\left|\frac{1}{n_k(s)}\sum_{i = 1}^{n_k(s)}Z_i^{(s)}(\boldsymbol{\theta})-\mathbb{E}[Z^{(s)}(\boldsymbol{\theta})]\right| \\ &+o_p(1)\sup_{\boldsymbol{\theta} \in \boldsymbol{\Theta}}|\mathbb{E}[Z^{(s)}(\boldsymbol{\theta})]|.
\end{align*}

Since \(\sup_{\boldsymbol{\theta} \in \boldsymbol{\Theta}}|\mathbb{E}[Z^{(s)}(\boldsymbol{\theta})]|<\infty\) by assumption, it suffices to show that for each $k \in 0 \cup [K]$ and \(s\in [s_\mathrm{max}]\), 
\begin{align}
    \sup_{\boldsymbol{\theta} \in \boldsymbol{\Theta}}\Biggl|\frac{1}{n_k(s)}\sum_{i = 1}^{n_k(s)}Z_i^{(s)}(\boldsymbol{\theta}) - \mathbb{E}[Z^{(s)}(\boldsymbol{\theta})]\Biggr|\cvp0.
    \label{eq:lem3:1}
\end{align}
To this end, by the Skorokhod's representation theorem, we can construct \(\frac{\tilde{n}_k(s)}{n}\) such that \(\frac{\tilde{n}_k(s)}{n}\) has the same distribution as \(\frac{n_k(s)}{n}\), \(\frac{\tilde{n}_k(s)}{n}\to\frac{\pi_{s}}{K+1} \) almost surely and \(\frac{\tilde{n}_k(s)}{n}\) is independent of \(\{Z_1^{(s)},\dots,Z_n^{(s)}\}\). Since \((\boldsymbol{T}_1,\dots,\boldsymbol{T}_n)\) and \((S_1,\dots,S_n)\) are independent of \(\{Z_1^{(s)},\dots,Z_n^{(s)}\}\), we have, for any \(\varepsilon>0\) and \(n > 0\),
\begin{align}
&\mathbb{P}\left(\sup_{\boldsymbol{\theta} \in \boldsymbol{\Theta}} \left|\frac{1}{n_k(s)}\sum_{i = 1}^{n_k(s)} Z_i^{(s)}(\boldsymbol{\theta}) - \mathbb{E}[Z^{(s)}(\boldsymbol{\theta})]\right| > \varepsilon\right) \nonumber\\
=&\mathbb{P}\left(\sup_{\boldsymbol{\theta} \in \boldsymbol{\Theta}} \left|\frac{1}{n\frac{\tilde{n}_k(s)}{n}}\sum_{i = 1}^{n\frac{\tilde{n}_k(s)}{n}} Z_i^{(s)}(\boldsymbol{\theta}) - \mathbb{E}[Z^{(s)}(\boldsymbol{\theta})]\right| > \varepsilon\right) \nonumber\\
 \leq& \mathbb{E}\left[\mathbb{P}\left(\sup_{\boldsymbol{\theta} \in \boldsymbol{\Theta}} \left|\frac{1}{n\frac{\tilde{n}_k(s)}{n}}\sum_{i = 1}^{n\frac{\tilde{n}_k(s)}{n}} Z_i^{(s)}(\boldsymbol{\theta}) - \mathbb{E}[Z^{(s)}(\boldsymbol{\theta})]\right| > \varepsilon \Bigg\vert \frac{\tilde{n}_k(s)}{n} \right)\right].
 \label{eq:lem3:2}
\end{align}
Recall that by assumption, $Z^{(s)}$ is a $P$-Glivenko--Cantelli process. This implies that for any sequence $\{n_k\}$ such that $n_k \to \infty$ as $k \to \infty$, we have
\begin{align*}
    \sup_{\boldsymbol{\theta} \in \boldsymbol{\Theta}} \left|\frac{1}{n_k}\sum_{i = 1}^{n_k} Z_i^{(s)}(\boldsymbol{\theta}) - \mathbb{E}[Z^{(s)}(\boldsymbol{\theta})]\right| \cvp 0.
\end{align*}
Since the almost sure convergence of $\tilde{n}_k(s)$ to infinity and the independence of $\frac{\tilde{n}_k(s)}{n}$ from the sequence $\{Z_1^{(s)}, \dots, Z_n^{(s)}\}$, we have:
\begin{equation}
\mathbb{P}\left(\sup_{\boldsymbol{\theta} \in \boldsymbol{\Theta}} \left|\frac{1}{\tilde{n}_k(s)}\sum_{i = 1}^{\tilde{n}_k(s)} Z_i^{(s)}(\boldsymbol{\theta}) - \mathbb{E}[Z^{(s)}(\boldsymbol{\theta})]\right| > \varepsilon \Bigg\vert \frac{\tilde{n}_k(s)}{n} \right) \to 0 \quad a.s.,
\label{G-C Lemma}
\end{equation}
then by (\ref{eq:lem3:2}), (\ref{G-C Lemma}) and the dominated convergence theorem, we finish the proof of (\ref{eq:lem3:1}). This concludes the proof of the entire lemma.
\end{proof}

\begin{proof}[Proof of Lemma \ref{lem:theta:Consistency}]
The proof leverages the decomposition of the estimating function:
\begin{equation}\label{eq:decomposition}
\frac{1}{\sqrt{n}} \sum_{i=1}^{n} \boldsymbol{\psi}_{i}(\boldsymbol{\theta}) = \frac{1}{\sqrt{n}} \sum_{i=1}^{n} \sum_{k \in 0 \cup [K]} T_i^k \boldsymbol{\psi}_{i}^{k}(\boldsymbol{\theta}).
\end{equation}

By Assumption \ref{Assump:Z-estimation}.3 and \ref{Assump:Z-estimation}.4, we know $\lbrace\boldsymbol{\psi}^k(\boldsymbol{\theta}):\boldsymbol{\theta} \in \boldsymbol{\Theta}\rbrace$ is~P-Glivenko--Cantelli class, then by Lemma \ref{lem:uniformconverge}, we establish the uniform convergence:
\begin{equation*}\label{eq:uniform_conv}
\sup_{\boldsymbol{\theta} \in \boldsymbol{\Theta}} \left\| \frac{1}{n} \sum_{i=1}^n \sum_{k \in 0 \cup [K]} T_i^k \boldsymbol{\psi}_i^{k}(\boldsymbol{\theta}) - \frac{1}{K+1}\sum_{k \in 0 \cup [K]} \mathbb{E}\left[\boldsymbol{\psi}^{k}(\boldsymbol{\theta})\right] \right\| \cvp 0.
\end{equation*}

The result follows since $\widehat{\boldsymbol{\theta}}$ satisfies the empirical version of the estimating equation:
\begin{equation*}\label{eq:estimator_eq}
\frac{1}{n} \sum_{i=1}^n \sum_{k \in 0 \cup [K]} T_i^k \boldsymbol{\psi}_i^{k}(\widehat{\boldsymbol{\theta}}) = o_p(n^{-1/2}),
\end{equation*}
combined with the uniqueness of $\boldsymbol{\theta}^*$ and Theorem 5.9 of \cite{van2000asymptotic}.
\end{proof}

\begin{proof}[Proof of Lemma \ref{lem:psi:normality}]
    we rewrite $n^{-1/2} \sum_{i=1}^{n} \sum_{k \in 0 \cup [K]} T_i^k\boldsymbol{\psi} _{i}^{k}(\boldsymbol{\theta}^*)$ as
\[
\begin{aligned}
& \frac{1}{\sqrt{n}} \sum_{i=1}^{n} \sum_{k \in 0 \cup [K]} T_i^k\boldsymbol{\psi} _{i}^{k}(\boldsymbol{\theta}^*) \\
=& \frac{1}{\sqrt{n}} \sum_{i=1}^{n} \sum_{k \in 0 \cup [K]} (T_i^k-\frac{1}{K+1})\boldsymbol{\psi} _{i}^{k}(\boldsymbol{\theta}^*)  + \frac{1}{\sqrt{n}(K+1)} \sum_{i=1}^{n} \sum_{k \in 0 \cup [K]}\boldsymbol{\psi} _{i}^{k}(\boldsymbol{\theta}^*)\\
=& \frac{1}{\sqrt{n}} \sum_{i=1}^{n} \sum_{k \in 0 \cup [K]}\frac{1}{K+1}\boldsymbol{\psi} _{i}^{k}(\boldsymbol{\theta}^*) \\
&+\frac{1}{\sqrt{n}} \sum_{i=1}^{n} \sum_{k \in 0 \cup [K]} (T_i^k-\frac{1}{K+1})\left[\boldsymbol{\psi} _{i}^{k}(\boldsymbol{\theta}^*) - \mathbb{E}\left [\boldsymbol{\psi} _{i}^{k}(\boldsymbol{\theta}^*)\mid S_i  \right ] \right] \\
&+\frac{1}{\sqrt{n}} \sum_{s \in [s_\mathrm{max}]} \sum_{k \in 0 \cup [K]}D_n^k(s)\mathbb{E}\left [\boldsymbol{\psi} ^{k}(\boldsymbol{\theta}^*)\mid S=s  \right ]\\
:=& \mathbb{G}_{n,1}^*(\boldsymbol{\psi}) + \mathbb{G}_{n,2}^*(\boldsymbol{\psi}) + \mathbb{G}_{n,3}(\boldsymbol{\psi}), 
\end{aligned}
\]
where 
\begin{align*}
\mathbb{G}_{n,1}^*(\boldsymbol{\psi})&=\frac{1}{\sqrt{n}(K+1)} \sum_{i = 1}^{n} \sum_{k \in 0 \cup [K]}\boldsymbol{\psi} _{i}^{k}(\boldsymbol{\theta}^*),\\
\mathbb{G}_{n,2}^*(\boldsymbol{\psi})&=\frac{1}{\sqrt{n}} \sum_{i = 1}^{n} \sum_{k \in 0 \cup [K]} (T_i^k-\frac{1}{K+1})\left[\boldsymbol{\psi} _{i}^{k}(\boldsymbol{\theta}^*) - \mathbb{E}\left [\boldsymbol{\psi} _{i}^{k}(\boldsymbol{\theta}^*)\mid S_i  \right ] \right],\\
\mathbb{G}_{n,3}^*(\boldsymbol{\psi})&=\frac{1}{\sqrt{n}} \sum_{s \in [s_\mathrm{max}]} \sum_{k \in 0 \cup [K]}D_n^k(s)\mathbb{E}\left [\boldsymbol{\psi} _{i}^{k}(\boldsymbol{\theta}^*)\mid S_i=s  \right ].
\end{align*}
By Assumption \ref{supp:Assumption 3.1}.3, we can rewrite
\begin{align*}
\frac{1}{\sqrt{n}} D_n^k(s) \mathbb{E} \left[ \boldsymbol{\psi}^{k}(\boldsymbol{\theta}^*) \mid S = s \right] 
= \frac{1}{\sqrt{n}} \sum_{i=1}^{n} M_i^k(s) \mathbb{E} \left[ \boldsymbol{\psi}^{k}(\boldsymbol{\theta}^*) \mid S = s \right] + o_p(\boldsymbol{1}).
\end{align*}

Define a vector of martingale differences:
\begin{align*}
\Delta G_{i}(\boldsymbol{\psi}) = \bigl(&(\Delta G_{i,1}(\boldsymbol{\psi}))^\top,\ 
(\Delta G_{i,2}(\boldsymbol{\psi}))^\top,\ (\Delta G_{i,3}(\boldsymbol{\psi}))^\top,\ 
(\Delta G_{i,4}(\boldsymbol{\psi}))^\top \bigr)^\top,
\end{align*}
where
\begin{align*}
\Delta G_{i,1}(\boldsymbol{\psi}) &=\frac{1}{K+1} \sum_{k \in 0 \cup [K]}\left[\boldsymbol{\psi} _{i}^{k}(\boldsymbol{\theta}^*)-\mathbb{E}\left [\boldsymbol{\psi} _{i}^{k}(\boldsymbol{\theta}^*)\mid S_i  \right ]\right], \\
\Delta G_{i,3}(\boldsymbol{\psi}) &= \sum_{k \in 0 \cup [K]} (T_i^k-\frac{1}{K+1})\left[\boldsymbol{\psi} _{i}^{k}(\boldsymbol{\theta}^*) -\mathbb{E}\left [\boldsymbol{\psi} _{i}^{k}(\boldsymbol{\theta}^*)\mid S_i  \right ] \right], \\
\Delta G_{i,4}(\boldsymbol{\psi}) &= \sum_{k \in 0 \cup [K]}\sum_{s \in [s_\mathrm{max}]} M_i^k(s) \mathbb{E} \left[ \boldsymbol{\psi}_{i}^{k}(\boldsymbol{\theta}^*) \mid S_i = s \right],
\end{align*}

and $\Delta G_{i,2}(\boldsymbol{\psi}) = 
\begin{cases}
\frac{1}{K+1}\sum_{k \in 0 \cup [K]}\mathbb{E}\left[ \boldsymbol{\psi} _{i + 1}^{k}(\boldsymbol{\theta}^*)\mid S_{i + 1} \right ], & \mathrm{for }\ 1\leq i\leq n - 1 \\
0, & \mathrm{for } \ i = n
\end{cases}$
with this construction, we have
\begin{align*}
& \mathbb{G}_{n,1}^*(\boldsymbol{\psi}) = \frac{1}{\sqrt{n}}\sum_{i=1}^n(\Delta G_{i,1}(\boldsymbol{\psi})+\Delta G_{i,2}(\boldsymbol{\psi}))+o_p(\boldsymbol{1}), \\
& \mathbb{G}_{n,2}^*(\boldsymbol{\psi}) = \frac{1}{\sqrt{n}}\sum_{i=1}^n\Delta G_{i,3}(\boldsymbol{\psi}), \\
& \mathbb{G}_{n,3}^*(\boldsymbol{\psi}) = \frac{1}{\sqrt{n}}\sum_{i=1}^n\Delta G_{i,4}(\boldsymbol{\psi})+o_{p}(\boldsymbol{1}).
\end{align*}

Since Assumption \ref{supp:Assumption 3.1}.2 implies that $\boldsymbol{T}_i \perp \boldsymbol{V}_i$ given $S_i$, we have
\[
\mathbb{E}[\Delta G_i(\boldsymbol{\psi}) \mid \mathcal{F}_{i-1}] = 0,
\]
so $\{\Delta G_i(\boldsymbol{\psi})\}_{i=1}^n$ is a sequence of zero-mean martingale differences with respect to the filtration $\mathcal{F}_{i-1}$.

Additionally, since $\mathbb{E} \|\boldsymbol{\psi}^{k}(\boldsymbol{\theta}^*)\|^2 < \infty$ for all $k \in \{0, \ldots, K\}$ and Assumption \ref{supp:Assumption 3.1}.4, we obtain
\[
\mathbb{E} \left [ \sum_{i=1}^n \| \Delta G_i(\boldsymbol{\psi}) \| \right ]  = O(n).
\]

Therefore, the conditional Lindeberg condition \cite[Corollary 3.1]{hall2014martingale} holds for $\lbrace \Delta G_i(\boldsymbol{\psi})\rbrace_{i=1}^n$. To derive the normality, it suffices to calculate the conditional covariance
matrices for \(\{\Delta G_i(\boldsymbol{\psi})\}_{i=1}^n\). Let \(\boldsymbol{\psi}(\boldsymbol{\theta})=(K+1)^{-1}\sum_{k \in 0 \cup [K]}\boldsymbol{\psi}^k(\boldsymbol{\theta})\). The following results hold:
{\small
\begin{align*}
    \frac{1}{n}\sum_{i = 1}^nE[\Delta G_{i,1}(\boldsymbol{\psi})(\Delta G_{i,1}(\boldsymbol{\psi}))^{\top}|\mathcal{F}_{i - 1}]=&\mathbb{E}\left[\mathrm{Cov}\left[\boldsymbol{\psi}(\boldsymbol{\theta}^*),\boldsymbol{\psi}(\boldsymbol{\theta}^*)^{\top}\mid S\right]\right]+o_p(\boldsymbol{1}),\\
    \frac{1}{n}\sum_{i = 1}^nE[\Delta G_{i,2}(\boldsymbol{\psi})(\Delta G_{i,2}(\boldsymbol{\psi}))^{\top}|\mathcal{F}_{i - 1}] =& \mathrm{Cov}\left[\mathbb{E}\left[\boldsymbol{\psi}(\boldsymbol{\theta}^*)\mid S\right],\mathbb{E}\left[\boldsymbol{\psi}(\boldsymbol{\theta}^*)^{\top}\mid S\right]\right]+o_p(\boldsymbol{1}),\\
    \frac{1}{n}\sum_{i = 1}^nE[\Delta G_{i,3}(\boldsymbol{\psi})(\Delta G_{i,3}(\boldsymbol{\psi}))^{\top}|\mathcal{F}_{i - 1}]=&\frac{1}{K+1}\sum_{k \in 0 \cup [K]} \mathbb{E}\left[\mathrm{Cov}\left[\boldsymbol{\psi}^k(\boldsymbol{\theta}^*),\boldsymbol{\psi}^k(\boldsymbol{\theta}^*)^{\top}\mid S\right]\right]\\
    -& \mathbb{E}\left[\mathrm{Cov}\left[\boldsymbol{\psi}(\boldsymbol{\theta}^*),\boldsymbol{\psi}(\boldsymbol{\theta}^*)^{\top}\mid S\right]\right]+o_p(\boldsymbol{1}).
\end{align*}
}
In addition, we have
\begin{align*}
    \mathbb{E}[\Delta G_{i,1}(\boldsymbol{\psi})(\Delta G_{i,4}(\boldsymbol{\psi}))^{\top}|\mathcal{F}_{i - 1}]&=\mathbb{E}[\Delta G_{i,1}(\boldsymbol{\psi})(\Delta G_{i,2}(\boldsymbol{\psi}))^{\top}|\mathcal{F}_{i - 1}]\\
    &=\mathbb{E}[\Delta G_{i,3}(\boldsymbol{\psi})(\Delta G_{i,2}(\boldsymbol{\psi}))^{\top}|\mathcal{F}_{i - 1}]\\
    &=\mathbb{E}[\Delta G_{i,4}(\boldsymbol{\psi})(\Delta G_{i,2}(\boldsymbol{\psi}))^{\top}|\mathcal{F}_{i - 1}]\\
    &=\mathbb{E}[\Delta G_{i,3}(\boldsymbol{\psi})(\Delta G_{i,4}(\boldsymbol{\psi}))^{\top}|\mathcal{F}_{i - 1}]=0.
\end{align*}

Moreover, 
\begin{align*}
    &\frac{1}{n}\sum_{i = 1}^{n}\mathbb{E}[\Delta G_{i,3}(\boldsymbol{\psi})(\Delta G_{i,1}(\boldsymbol{\psi}))^{\top}|\mathcal{F}_{i - 1}]
    \\=&\frac{1}{n(K+1)}\sum_{s \in [s_\mathrm{max}]}\sum_{k,k' \in 0 \cup [K]}\mathbb{I}(S_i = s)\mathbb{E}[(T_i^k-\frac{1}{K+1})|\mathcal{F}_{i - 1}] \\ \times &\mathbb{E}\left[\left[\boldsymbol{\psi}_{i}^{k}(\boldsymbol{\theta}^*)-\mathbb{E}\left[\boldsymbol{\psi}_{i}^{k}(\boldsymbol{\theta}^*)\mid S_i\right]\right]\right. \times\left.\left[\boldsymbol{\psi}_{i}^{k}(\boldsymbol{\theta}^*)-\mathbb{E}\left[\boldsymbol{\psi}_{i}^{k}(\boldsymbol{\theta}^*)\mid S_i\right]\right]^{\top}|S_i=s\right]  .
\end{align*}
And,
\begin{align*}
    D_n^k(s)-\sum_{i = 1}^{n}\mathbb{I}(S_i = s)\mathbb{E}[(T_i^k-\frac{1}{K+1})|\mathcal{F}_{i - 1}]&=\sum_{i = 1}^n\mathbb{I}(S_i = s)(T_i^k - \mathbb{E}[T_i^k|\mathcal{F}_{i - 1}]).
\end{align*}
Since \(\lbrace\mathbb{I}(S_i = s)(T_i^k - \mathbb{E}[T_i^k|\mathcal{F}_{i - 1}])\rbrace\) forms a sequence of bounded martingale differences, by the martingale convergence theorem, we have
\[
\frac{1}{n}\lbrace D_n^k(s)-\sum_{i = 1}^{n}\mathbb{I}(S_i = s)\mathbb{E}[(T_i^k-\frac{1}{K+1})|\mathcal{F}_{i - 1}]\rbrace = o_p(1).
\]
By Lemma \ref{lem:stratum}, \(n^{-1}D_n^k(s)=o_p(1)\), which implies
\[
\frac{1}{n}\sum_{i = 1}^{n}\mathbb{I}(S_i = s)\mathbb{E}[(T_i^k-\frac{1}{K+1})|\mathcal{F}_{i - 1}]=o_p(1).
\]
Consequently,
\[
\frac{1}{n}\sum_{i = 1}^{n}\mathbb{E}[\Delta G_{i,3}(\boldsymbol{\psi})(\Delta G_{i,1}(\boldsymbol{\psi}))^{\top}|\mathcal{F}_{i - 1}]=o_p(\boldsymbol{1}).
\]

Assumptions \ref{supp:Assumption 3.1}.2 and \ref{supp:Assumption 3.1}.3 further imply that
\begin{align*}
    &\frac{1}{n}\sum_{i = 1}^{n}\mathbb{E}[\Delta G_{i,4}(\boldsymbol{\psi})(\Delta G_{i,4}(\boldsymbol{\psi}))^{\top}|\mathcal{F}_{i - 1}]\\
    &=\frac{1}{n}\sum_{i = 1}^nE\left[\sum_{k \in 0 \cup [K]}\sum_{s \in [s_\mathrm{max}] }M_i^k(s)\mathbb{E}\left[\boldsymbol{\psi}_{i}^{k}(\boldsymbol{\theta}^*)\mid S = s\right]\right.\\
    &\quad\times\left.\sum_{k' \in 0 \cup [K]}\sum_{s' \in [s_\mathrm{max}]}M_i^{k'}(s')\mathbb{E}\left[\boldsymbol{\psi}_{i}^{k'}(\boldsymbol{\theta}^*)\mid S = s'\right]^{\top}|\mathcal{F}_{i - 1}\right]+o_p(\boldsymbol{1})\\
    &=\sum_{k,k'\in 0 \cup [K]}\sum_{s,s' \in [s_\mathrm{max}]}\boldsymbol{\Sigma}_{kk'}^{\sf{CAR}}(s,s') \mathbb{E}\left[\boldsymbol{\psi}^{k}(\boldsymbol{\theta}^*)\mid S = s\right]\mathbb{E}\left[\boldsymbol{\psi}^{k'}(\boldsymbol{\theta}^*)\mid S = s'\right]^{\top}+o_p(\boldsymbol{1}).
\end{align*}

Combining the above results, we obtain
{\small
\begin{align*}
    &\frac{1}{n}\sum_{i = 1}^{n}\mathbb{E}\left[\sum_{j = 1}^4\Delta G_{i,j}(\boldsymbol{\psi})\sum_{j = 1}^4(\Delta G_{i,j}(\boldsymbol{\psi}))^{\top}\middle|\mathcal{F}_{i - 1}\right] \nonumber\\
    &=\frac{1}{n}\sum_{j = 1}^4\sum_{i = 1}^{n}\mathbb{E}\left[\Delta G_{i,j}(\boldsymbol{\psi})(\Delta G_{i,j}(\boldsymbol{\psi}))^{\top}\middle|\mathcal{F}_{i - 1}\right] \nonumber\\
    &=\mathbb{E}\left[\mathrm{Cov}\left[\boldsymbol{\psi}(\boldsymbol{\theta}^*),\boldsymbol{\psi}(\boldsymbol{\theta}^*)^{\top}\mid S\right]\right]+\mathrm{Cov}\left[\mathbb{E}\left[\boldsymbol{\psi}(\boldsymbol{\theta}^*)\mid S_i\right],\mathbb{E}\left[\boldsymbol{\psi}(\boldsymbol{\theta}^*)^{\top}\mid S\right]\right] \nonumber\\
    &\quad+\frac{1}{K+1}\sum_{k \in 0 \cup [K]} \mathbb{E}\left[\mathrm{Cov}\left[\boldsymbol{\psi}^k(\boldsymbol{\theta}^*),\boldsymbol{\psi}^k(\boldsymbol{\theta}^*)^{\top}\mid S\right]\right]-\mathbb{E}\left[\mathrm{Cov}\left[\boldsymbol{\psi}(\boldsymbol{\theta}^*),\boldsymbol{\psi}(\boldsymbol{\theta}^*)^{\top}\mid S\right]\right]\\
    &\quad+\sum_{k,k'\in 0 \cup [K]}\sum_{s,s'\in [s_\mathrm{max}]}\boldsymbol{\Sigma}_{kk'}^{\sf{CAR}}(s,s')\mathbb{E}\left[\boldsymbol{\psi}^{k}(\boldsymbol{\theta}^*)\mid S = s\right]\mathbb{E}\left[\boldsymbol{\psi}^{k'}(\boldsymbol{\theta}^*)\mid S = s'\right]^{\top}+o_p(\boldsymbol{1})\\
    &=\mathrm{Var}\left[\mathbb{E}\left(\boldsymbol{\psi}(\boldsymbol{\theta}^*)\mid S\right)\right]+\frac{1}{K+1}\sum_{k \in 0 \cup [K]} \mathbb{E}\left[\mathrm{Var}\left[\boldsymbol{\psi}^k(\boldsymbol{\theta}^*)\mid S\right]\right] \nonumber\\
    &\quad+\sum_{k,k'\in 0 \cup [K]}\sum_{s,s'\in [s_\mathrm{max}]}\boldsymbol{\Sigma}_{kk'}^{\sf{CAR}}(s,s')\mathbb{E}\left[\boldsymbol{\psi}^{k}(\boldsymbol{\theta}^*)\mid S = s\right]\mathbb{E}\left[\boldsymbol{\psi}^{k'}(\boldsymbol{\theta}^*)\mid S = s'\right]^{\top}+o_p(\boldsymbol{1}) \nonumber\\
    &=\boldsymbol{\Lambda} + o_p(\boldsymbol{1}).
\end{align*}}

By the martingale central limit theorem \citep{hall2014martingale} and Slutsky’s theorem, we finish the proof of this lemma.
\end{proof}

\begin{proof}[Proof of Lemma \ref{lem:theta:linearity}]
By Assumption \ref{Assump:Z-estimation}.5 and performing a multivariate Taylor expansion of the function $\sum_{i = 1}^{n}\sum_{k \in 0 \cup [K]} T_i^k \boldsymbol{\psi}_{i}^{k}(\boldsymbol{\theta})$ around $\boldsymbol{\theta}=\boldsymbol{\theta}^*$, we obtain
\[
\begin{aligned}
& \frac{1}{n} \sum_{i = 1}^{n}\sum_{k \in 0 \cup [K]} T_i^k \boldsymbol{\psi}_{i}^{k}(\widehat{\boldsymbol{\theta}}) \\
=& \frac{1}{n} \sum_{i = 1}^{n}\sum_{k \in 0 \cup [K]} T_i^k \boldsymbol{\psi}_{i}^{k}(\boldsymbol{\theta}^*)+\frac{1}{n} \sum_{i = 1}^{n}\sum_{k \in 0 \cup [K]} T_i^k \dot{\boldsymbol{\psi}}_{i}^{k}(\boldsymbol{\theta}^*) (\widehat{\boldsymbol{\theta}} - \boldsymbol{\theta}^*)\\
&+\sum_{j = 1}^{K+1+q} \frac{1}{2}(\widehat{\boldsymbol{\theta}} - \boldsymbol{\theta}^*)^{\top} \biggl(\frac{1}{n}\sum_{i = 1}^{n}\sum_{k \in 0 \cup [K]} T_i^k \ddot{\boldsymbol{\psi}}_{i}^{k,j}(\widetilde{\boldsymbol{\theta}})\biggr)(\widehat{\boldsymbol{\theta}} - \boldsymbol{\theta}^*) \boldsymbol{e}_{j}=o_p(n^{-1/2}).
\end{aligned}
\]
From Assumption \ref{Assump:Z-estimation}.4, there exists a ball $Q$ centered at $\boldsymbol{\theta}^*$ such that $\|\ddot{\boldsymbol{\psi}}_{i}^{k,j}(\boldsymbol{\theta})\|$ is dominated by an integrable function $v(Y_{i}(k),\bm{X}_{i})$ for each $j\in [K+1+q]$ and $\boldsymbol{\theta}\in Q$. Since $\widetilde{\boldsymbol{\theta}}$ lies on the line segment between $\widehat{\boldsymbol{\theta}}$ and $\boldsymbol{\theta}^*$ and $\mathbb{P}(\widetilde{\boldsymbol{\theta}} \in Q) \rightarrow 1$, we have 
\begin{equation}
    \Biggl\|\frac{1}{n} \sum_{i = 1}^{n}\sum_{k \in 0 \cup [K]} T_i^k \ddot{\boldsymbol{\psi}}_{i}^{k,j}(\widetilde{\boldsymbol{\theta}})\Biggr\|=O_{p}(1),
    \label{eq:Second derivative}
\end{equation}
and thus  the last term in the Taylor expansion is $O_{p}(\boldsymbol{1})\|\widehat{\boldsymbol{\theta}} - \boldsymbol{\theta}^*\|^{2}$.

By Lemma \ref{lem:uniformconverge} and Assumption \ref{Assump:Z-estimation}.2, we have 
\begin{align*}
\frac{1}{n} \sum_{i = 1}^{n}\sum_{k \in 0 \cup [K]} T_i^k \dot{\boldsymbol{\psi}}_{i}^{k}(\boldsymbol{\theta}^*)\cvp \frac{1}{K+1}\sum_{k \in 0 \cup [K]} \mathbb{E}\left[\dot{\boldsymbol{\psi}}^{k}(\boldsymbol{\theta}^*)\right]=\boldsymbol{\Psi}.
\end{align*}
Combining the above, we get the equation
\begin{equation}
    \frac{1}{n} \sum_{i = 1}^{n}\sum_{k \in 0 \cup [K]} T_i^k \boldsymbol{\psi}_{i}^{k}(\boldsymbol{\theta}^*)=-\boldsymbol{\Psi}(\widehat{\boldsymbol{\theta}} - \boldsymbol{\theta}^*)-O_{p}(\boldsymbol{1})\|\widehat{\boldsymbol{\theta}} - \boldsymbol{\theta}^*\|^{2}+o_{p}(1)(\widehat{\boldsymbol{\theta}} - \boldsymbol{\theta}^*).
    \label{eq:Taylor expansion1}
\end{equation}
Multiplying  both sides of  (\ref{eq:Taylor expansion1}) by $\sqrt{n}$ and by Lemma  \ref{lem:psi:normality}, we have:
\begin{align*}
-\sqrt{n}\boldsymbol{\Psi}(\widehat{\boldsymbol{\theta}} - \boldsymbol{\theta}^*)-\sqrt{n}\ o_p(\boldsymbol{1}) \|\widehat{\boldsymbol{\theta}} - \boldsymbol{\theta}^*\|+\sqrt{n}\ o_p(1)(\widehat{\boldsymbol{\theta}} - \boldsymbol{\theta}^*)=O_p(\bm{1}).
\end{align*}
Since $\boldsymbol{\Psi}$ is invertible, we get
\begin{align*}
\sqrt{n}\|\widehat{\boldsymbol{\theta}} - \boldsymbol{\theta}^*\|&\leq\|\boldsymbol{\Psi}^{-1}\|\sqrt{n}\|\boldsymbol{\Psi}(\widehat{\boldsymbol{\theta}} - \boldsymbol{\theta}^*)\| = O_p(1)+o_p(1)\sqrt{n}\|\widehat{\boldsymbol{\theta}} - \boldsymbol{\theta}^*\|,
\end{align*}
which implies $\sqrt{n}(\widehat{\boldsymbol{\theta}} - \boldsymbol{\theta}^*)=O_p(\bm{1})$, then the asymptotic linearity of $\widehat{\boldsymbol{\theta}}$ follows from (\ref{eq:Taylor expansion1}).
\end{proof}

\begin{proof}[Proof of Lemma \ref{lem:theta:normality}]
The result follows immediately from Lemmas \ref{lem:psi:normality}, \ref{lem:theta:linearity} and Slutsky’s theorem.
\end{proof}

\begin{proof}[Proof of Lemma \ref{lem:mu:consistency}]
    Note that for all $k \in 0 \cup [K]$, the definition of $\bm{\psi}^k(Y_i(k), \bm{X}_i; \bm{\theta})$ implies that $\mathbb{E}[Y(k)] = \mathbb{E}[h^k(\boldsymbol{X};\boldsymbol{\theta}^*)]$; and thus
\begin{equation}
    \frac{1}{n}\sum_{i=1}^nh^k(\boldsymbol{X}_i,\boldsymbol{\theta}^*) = \mathbb{E}[Y(k)]+o_p(1).
    \label{eq:h:consistency}
\end{equation}
Next, by the multivariate Taylor's expansion of $\sum_{i=1}^n \boldsymbol{h}(\boldsymbol{X}_i;\widehat{\boldsymbol{\theta}})$ around the point $\boldsymbol{\theta}=\boldsymbol{\theta}^*$, there exists a $\widetilde{\boldsymbol{\theta}}$ (which depends on $\widehat{\boldsymbol{\theta}}$) on the line segment between $\widehat{\boldsymbol{\theta}}$ and $\boldsymbol{\theta}^*$ such that
\begin{align*}
    &\frac{1}{n}\sum_{i=1}^n\left\{ \boldsymbol{h}(\boldsymbol{X}_i;\widehat{\boldsymbol{\theta}}) -\boldsymbol{h}(\boldsymbol{X}_i;\boldsymbol{\theta}^*)\right\}\\
    &=\left\{\frac{1}{n}\sum_{i=1}^n\frac{\partial}{\partial \boldsymbol{\theta}} \boldsymbol{h}(\boldsymbol{X}_i;\boldsymbol{\theta}) \Big|_{\boldsymbol{\theta} = \boldsymbol{\theta}^*}\right\}(\widehat{\boldsymbol{\theta}}-\boldsymbol{\theta})\\
    &+\sum_{k \in 0 \cup [K]} \frac{1}{2}(\widehat{\boldsymbol{\theta}} - \boldsymbol{\theta}^*)^{\top} \left(\frac{1}{n}\sum_{i = 1}^{n}\frac{\partial^2}{\partial \boldsymbol{\theta} \partial \boldsymbol{\theta}^{\top}} h^k(\boldsymbol{X}_i;\widetilde{\boldsymbol{\theta}})\right)(\widehat{\boldsymbol{\theta}} - \boldsymbol{\theta}^*) \boldsymbol{e}_{k}.
\end{align*}
From  Assumption \ref{Assump:Z-estimation}.4 and (\ref{eq:Second derivative}), we know $\left\|\sum_{i = 1}^{n}\frac{\partial^2}{\partial \boldsymbol{\theta} \partial \boldsymbol{\theta}^{\top}} h^k(\boldsymbol{X}_i;\widetilde{\boldsymbol{\theta}})/n\right\|=O_{p}(1)$, and thus  the last term in the Taylor expansion is $O_{p}(\boldsymbol{1})\|\widehat{\boldsymbol{\theta}} - \boldsymbol{\theta}^*\|^{2}$, By Lemma \ref{lem:theta:Consistency}, we have
\begin{align*}
    \frac{1}{n}\sum_{i=1}^n\left\{ \boldsymbol{h}(\boldsymbol{X}_i;\widehat{\boldsymbol{\theta}}) -\boldsymbol{h}(\boldsymbol{X}_i;\boldsymbol{\theta}^*)\right\} = o_p(\boldsymbol{1}).
\end{align*}
Combined with (\ref{eq:h:consistency}), we complete the proof of this lemma.
\end{proof}

\begin{proof}[Proof of Lemma \ref{lem:mu_linearity}]

Denote $\bm{C} = \left((K+1)\boldsymbol{I}_{(K+1)\times(K+1)},\boldsymbol{0}_{(K+1)\times q}\right)$, from the definition of $\boldsymbol{\Psi}$, it follows that
{\small
\begin{align*}
    \boldsymbol{\Psi} &= \frac{1}{K+1}\sum_{k \in 0 \cup [K]}  \mathbb{E}\left[\left.\frac{\partial}{\partial \boldsymbol{\theta}} \boldsymbol{\psi}^k(\boldsymbol{\theta})\right|_{\boldsymbol{\theta}=\boldsymbol{\theta}^*}\right] = \frac{1}{K+1}\begin{bmatrix}\mathbb{E}\left[
\frac{\partial}{\partial \boldsymbol{\theta}} h^0(\boldsymbol{X};\boldsymbol{\theta})|_{\boldsymbol{\theta}=\boldsymbol{\theta}^*}\right] \\
\mathbb{E}\left[\frac{\partial}{\partial \boldsymbol{\theta}} h^1(\boldsymbol{X};\boldsymbol{\theta})|_{\boldsymbol{\theta}=\boldsymbol{\theta}^*}\right] \\
\vdots \\
\mathbb{E}\left[\frac{\partial}{\partial \boldsymbol{\theta}} h^K(\boldsymbol{X};\boldsymbol{\theta})|_{\boldsymbol{\theta}=\boldsymbol{\theta}^*}\right] \\
\sum_{k \in 0 \cup [K]}\mathbb{E}\left[\frac{\partial}{\partial \boldsymbol{\theta}}\boldsymbol{\psi}_1^k(\boldsymbol{X}, Y; \boldsymbol{\theta}) |_{\boldsymbol{\theta}=\boldsymbol{\theta}^*}\right]
\end{bmatrix} \\
&=\frac{1}{K+1} \begin{bmatrix}
\mathbb{E}\left[\frac{\partial}{\partial \boldsymbol{\theta}} \boldsymbol{h}(\boldsymbol{X};\boldsymbol{\theta})|_{\boldsymbol{\theta}=\boldsymbol{\theta}^*}\right] \\
\sum_{k \in 0 \cup [K]}\mathbb{E}\left[\frac{\partial}{\partial \boldsymbol{\theta}}\boldsymbol{\xi}^k(\boldsymbol{X}, Y; \boldsymbol{\theta}) |_{\boldsymbol{\theta}=\boldsymbol{\theta}^*}\right]
\end{bmatrix}.
\end{align*}
}
Moreover, since $\bm{\Psi}\bm{\Psi}^{-1} = \bm{I}_{(K+1+q)\times (K+1+q)}$, we have
\begin{align*}
    \boldsymbol{\Psi}\boldsymbol{\Psi}^{-1} = \frac{1}{K+1} \begin{bmatrix}
\mathbb{E}\left[\frac{\partial}{\partial \boldsymbol{\theta}} \boldsymbol{h}(\boldsymbol{X};\boldsymbol{\theta})|_{\boldsymbol{\theta}=\boldsymbol{\theta}^*}\right]\\
\sum_{k \in 0 \cup [K]}\mathbb{E}\left[\frac{\partial}{\partial \boldsymbol{\theta}}\boldsymbol{\xi}^k(\boldsymbol{X}, Y; \boldsymbol{\theta}) |_{\boldsymbol{\theta}=\boldsymbol{\theta}^*}\right]
\end{bmatrix} \times \boldsymbol{\Psi}^{-1} = \boldsymbol{I}_{(K+1+q)\times (K+1+q)}.
\end{align*}
Then,
\begin{align*}
    \mathbb{E}\left[\frac{\partial}{\partial \boldsymbol{\theta}} \boldsymbol{h}(\boldsymbol{X};\boldsymbol{\theta})|_{\boldsymbol{\theta}=\boldsymbol{\theta}^*}\right]\times \boldsymbol{\Psi}^{-1}  = (K+1)\bm{C}.
\end{align*}
By the multivariate Taylor's expansion and Lemma \ref{lem:theta:linearity}, we have:
\begin{align*}
    &\frac{1}{\sqrt{n}}\sum_{i=1}^n\left\{ \boldsymbol{h}(\boldsymbol{X}_i;\widehat{\boldsymbol{\theta}}) -\boldsymbol{h}(\boldsymbol{X}_i;\boldsymbol{\theta}^*)\right\}\\
    &=\left\{\frac{1}{n}\sum_{i=1}^n\frac{\partial}{\partial \boldsymbol{\theta}} \boldsymbol{h}(\boldsymbol{X}_i;\boldsymbol{\theta}) \Big|_{\boldsymbol{\theta} = \boldsymbol{\theta}^*}\right\}\sqrt{n}(\widehat{\boldsymbol{\theta}}-\boldsymbol{\theta})+\sqrt{n}\left\|\widehat{\boldsymbol{\theta}}-\boldsymbol{\theta}\right\|^2 O_p(\boldsymbol{1})\\
    &=-\frac{1}{\sqrt{n}}\left \{\mathbb{E}\left[\frac{\partial}{\partial \boldsymbol{\theta}} \boldsymbol{h}(\boldsymbol{X};\boldsymbol{\theta}) \Big|_{\boldsymbol{\theta} = \boldsymbol{\theta}^*} \right] +o_p{(\boldsymbol{1}})\right\}\sum_{i = 1}^n\boldsymbol{\Psi}^{-1}\boldsymbol{\psi}(\boldsymbol{T}_i,\boldsymbol{X}_i,Y_i;\boldsymbol{\theta}^*)+o_p(\boldsymbol{1})\\
    & = -\frac{1}{\sqrt{n}}\sum_{i = 1}^n\bm{C}\boldsymbol{\psi}(\boldsymbol{T}_i,\boldsymbol{X}_i,Y_i;\boldsymbol{\theta}^*)+o_p(\boldsymbol{1}) \\
    &= -\frac{K+1}{\sqrt{n}}\sum_{i = 1}^n\sum_{k \in 0 \cup [K]}T_i^k\lbrace h^k(\boldsymbol{X_i};\boldsymbol{\theta}^*)-Y_i\rbrace\boldsymbol{e}_k+o_p(\boldsymbol{1}).
\end{align*}
Combined with (\ref{eq:h:consistency}), we finish the proof of this lemma.
\end{proof}

\begin{proof}[Proof of Lemma \ref{lem:mu_normality}]
By simple calculations, we have the variance-covariance matrix of imbalance measure under complete randomization, denoted as 
\begin{align*}
    \boldsymbol{\Sigma}^{\sf{CR}} =\begin{bmatrix}
  \boldsymbol{\Sigma}_{D}(1) & \bm{0}      & \cdots  & \bm{0}      \\
  \bm{0}      & \boldsymbol{\Sigma}_{D}(2) & \cdots  & \bm{0}      \\
  \vdots          & \vdots          & \ddots & \vdots          \\
  \bm{0}      & \bm{0}      & \cdots  & \boldsymbol{\Sigma}_{D}(s_{\mathrm{max}})
\end{bmatrix},
\end{align*}
where,
\begin{align*}
    \boldsymbol{\Sigma}_{D}(s) = \pi_s \left[ 
    \mathrm{diag}\left(\frac{\bm{1}_K}{K+1}\right) - \frac{1}{(K+1)^2} \bm{1}_K \bm{1}_K^\top 
\right].
\end{align*}
 By Lemma \ref{lem:mu_linearity}, we have:
{\small
   \begin{align*}
       &\sqrt{n}(\widehat{\boldsymbol{\mu}}-\boldsymbol{\mu}) = \frac{1}{\sqrt{n}}\sum_{i=1}^n\lbrace\boldsymbol{h}(\boldsymbol{X}_i;\boldsymbol{\theta}^*)-\boldsymbol{\mu}\rbrace-\frac{K+1}{\sqrt{n}}\sum_{i = 1}^n\sum_{k \in 0 \cup [K]}T_i^kr_i^k(\boldsymbol{\theta}^*)\boldsymbol{e}_k+o_p(\boldsymbol{1})\\
       &=\frac{1}{\sqrt{n}} \sum_{i=1}^{n} \lbrace\boldsymbol{h}(\boldsymbol{X}_i;\boldsymbol{\theta}^*)-\boldsymbol{\mu}-\sum_{k \in 0 \cup [K]}r_i^k(\boldsymbol{\theta}^*)\boldsymbol{e}_k\rbrace \\
       &-\frac{K+1}{\sqrt{n}} \sum_{s \in [s_\mathrm{max}]} \sum_{k \in 0 \cup [K]}D_n^k(s)\mathbb{E}\left [r_i^k(\boldsymbol{\theta}^*)\boldsymbol{e}_k| S=s  \right ]\\
    &-\frac{K+1}{\sqrt{n}} \sum_{i=1}^{n} \sum_{k \in 0 \cup [K]} (T_i^k-\frac{1}{K+1})\left[r_i^k(\boldsymbol{\theta}^*)\boldsymbol{e}_k - \mathbb{E}\left [r_i^k(\boldsymbol{\theta}^*)\boldsymbol{e}_k\mid S_i  \right ] \right]+o_p(\boldsymbol{1})\\
    &=\frac{1}{\sqrt{n}} \sum_{i=1}^{n}\lbrace\boldsymbol{Y}_i-\boldsymbol{\mu}\rbrace-\frac{K+1}{\sqrt{n}} \sum_{s \in [s_\mathrm{max}]} \sum_{k \in 0 \cup [K]}D_n^k(s)\mathbb{E}\left [r_i^k(\boldsymbol{\theta}^*)| S=s  \right ]\boldsymbol{e}_k\\
    &-\frac{K+1}{\sqrt{n}} \sum_{i=1}^{n} \sum_{k \in 0 \cup [K]} (T_i^k-\frac{1}{K+1})\left[r_i^k(\boldsymbol{\theta}^*) - \mathbb{E}\left [r_i^k(\boldsymbol{\theta}^*)\mid S_i  \right ] \right]\boldsymbol{e}_k+o_p(\boldsymbol{1}).
   \end{align*}
   }
Following a similar derivation of Lemma \ref{lem:psi:normality}, we know
\begin{align*}
    \sqrt{n}(\widehat{\bm{\mu}}-\bm{\mu})\cvd N(\bm{0},\bm{\Gamma}),
\end{align*}
{\small
\begin{align*}
    &\boldsymbol{\Gamma} = \mathrm{Var}\left[\boldsymbol{Y}\right]+(K+1)\sum_{k \in 0 \cup [K]} \mathbb{E}\left[\mathrm{Var}\left[r^k(\boldsymbol{\theta}^*)\mid S\right]\right]\boldsymbol{e}_k\boldsymbol{e}_k^\top- \mathbb{E}\left[\mathrm{Var}\left[\boldsymbol{r}\mid S\right]\right]\\
    &+(K+1)^2\sum_{k,k'\in 0 \cup [K]}\sum_{s,s'\in [s_\mathrm{max}]}\boldsymbol{\Sigma}_{kk'}^{\sf{CAR}}(s,s')\boldsymbol{e}_kE\left[r^k(\boldsymbol{\theta}^*)\mid S = s\right]\mathbb{E}\left[r^{k'}(\boldsymbol{\theta}^*)\mid S = s'\right]\boldsymbol{e}_{k'}^\top\\
   &=\mathrm{Var}\left[\boldsymbol{Y}\right]-\mathrm{Var}\left[\boldsymbol{r}\right]+\lbrace\mathrm{Var}\left[\boldsymbol{r}\right]- \mathbb{E}\left[\mathrm{Var}\left[\boldsymbol{r}\mid S\right]\right] +(K+1)\sum_{k \in 0 \cup [K]} \mathbb{E}\left[\mathrm{Var}\left[r^k(\boldsymbol{\theta}^*)\mid S\right]\right]\boldsymbol{e}_k\boldsymbol{e}_k^\top\\
    &+(K+1)^2\sum_{k,k'\in 0 \cup [K]}\sum_{s,s'\in [s_\mathrm{max}]}\boldsymbol{\Sigma}_{kk'}^{\sf{CAR}}(s,s')\boldsymbol{e}_kE\left[r^k(\boldsymbol{\theta}^*)\mid S = s\right]\mathbb{E}\left[r^{k'}(\boldsymbol{\theta}^*)\mid S = s'\right]\boldsymbol{e}_{k'}^\top\rbrace \\
    &=\mathrm{Var}\left[\boldsymbol{Y}\right]-\mathrm{Var}\left[\boldsymbol{r}\right]+(K+1)\sum_{k \in 0 \cup [K]}\mathbb{E}\left[(r^k(\boldsymbol{\theta}^*))^2\right]\boldsymbol{e}_k\boldsymbol{e}_k^\top-(K+1)\sum_{k \in 0 \cup [K]} \mathbb{E}\left[\mathbb{E}\left[r^k(\boldsymbol{\theta}^*)\mid S\right]^2\right]\boldsymbol{e}_k\boldsymbol{e}_k^\top \\
    &+\mathrm{Var}\left[\mathbb{E}\left[\boldsymbol{r}\mid S\right]\right]+(K+1)^2\sum_{k,k'\in 0 \cup [K]}\sum_{s,s'\in [s_\mathrm{max}]}\boldsymbol{\Sigma}_{kk'}^{\sf{CAR}}(s,s')\mathbb{E}\left[r^k(\boldsymbol{\theta}^*)\mid S = s\right]\mathbb{E}\left[r^{k'}(\boldsymbol{\theta}^*)\mid S = s'\right]\boldsymbol{e}_k\boldsymbol{e}_{k'}^\top\\
    &=(K+1)\mathrm{diag}\left \{  \mathrm{Var}\left[r^0(\boldsymbol{\theta}^*)\right],\dots, \mathrm{Var}\left[r^K(\boldsymbol{\theta}^*)\right]\right \} +\mathrm{Var}\left[\boldsymbol{Y}\right]-\mathrm{Var}\left[\boldsymbol{r}\right] \\
    &-(K+1)^2\sum_{k,k'\in 0 \cup [K]}\sum_{s,s'\in [s_\mathrm{max}]}\lbrace\boldsymbol{\Sigma}_{kk'}^{\sf{CR}}(s,s')-\boldsymbol{\Sigma}_{kk'}^{\sf{CAR}}(s,s')\rbrace \boldsymbol{L}(s,k)\boldsymbol{L}(s',k')^{\top}.
\end{align*}}
\end{proof}
\subsection{Proof of Theorem 2 and Corollary 3}

\begin{customth}{2}[Reproduced from Theorem 2 in the main text]
\label{supp:thm:mu_normality}
 Under Assumptions \ref{supp:Assumption 3.1} and \ref{Assump:Z-estimation}, $\widehat{\bm{\mu}}$ in (\ref{eq:mu_hat}) satisfies:
\begin{equation}
\sqrt{n}(\widehat{\bm{\mu}}-\bm{\mu})\cvd N(\bm{0},\bm{\Gamma}),
\end{equation}
with $\boldsymbol{\Gamma} = \boldsymbol{\Gamma}_{\sf{base}}-\boldsymbol{\Gamma}_{\sf{adj}}-\boldsymbol{\Gamma}_{\sf{CAR,adj}}$. Here,
\begin{itemize}
\item $\boldsymbol{\Gamma}_{\sf{base}} = (K+1)\mathrm{diag}\left \{  \mathrm{Var}\left[Y(0)\right],\dots,\mathrm{Var}\left[Y(K)\right]\right \}$ represents the baseline covariance matrix that does not account for covariate adjustment or the CAR procedures. 
\item With $r^k(\boldsymbol{\theta}^*) = Y(k) - h^k(\boldsymbol{X},\boldsymbol{\theta}^*)$ and $\boldsymbol{r} = (r^0(\boldsymbol{\theta}^*), \ldots, r^K(\boldsymbol{\theta}^*))^\top$,
\begin{align*}
\boldsymbol{\Gamma}_{\sf{adj}} &=(K+1)\mathrm{diag}\left \{ \mathrm{Var}\left[Y(0)\right]-\mathrm{Var}\left[r^0(\boldsymbol{\theta}^*)\right],\dots,\mathrm{Var}\left[Y(K)\right]-\mathrm{Var}\left[r^K(\boldsymbol{\theta}^*)\right]\right \}\\
&\quad +\mathrm{Var}\left[\boldsymbol{r}\right]-\mathrm{Var}\left[\boldsymbol{Y}\right]
\end{align*}
captures the contribution from covariate adjustment in Z-estimation.
\item With $\boldsymbol{L}(s,k) = \mathbb E\left[r^{k}(\boldsymbol{\theta}^*)\mid S = s\right]\boldsymbol{e}_k$, 
\[
\boldsymbol{\Gamma}_{\sf{CAR,adj}} = (K+1)^2\sum_{k,k'\in 0 \cup [K]}\sum_{s,s'\in [s_{\mathrm{max}}]}\lbrace\boldsymbol{\Sigma}^{\sf{CR}}_{kk'}(s,s')-\boldsymbol{\Sigma}^{\sf{CAR}}_{kk'}(s,s')\rbrace \boldsymbol{L}(s,k)\boldsymbol{L}(s',k')^{\top}
\]
depicts the interplay between covariate adjustment in Z-estimation and the CAR procedures.
\end{itemize}
\end{customth}

\subsubsection{Corollary}

\begin{customcor}{3}[Reproduced from Corollary 3  in the main text]
    \begin{enumerate} \item Under CR, Assumption \ref{supp:Assumption 3.1} holds with $\boldsymbol{\Sigma}_{kk'}^{\sf{CR}}(s,s') = \boldsymbol{\Sigma}_{kk'}^{\sf{CAR}}(s,s')$ for $s,s'\in [s_{\mathrm{max}}]$ and $k,k'\in 0 \cup [K]$, and $\boldsymbol{\Gamma}_{\sf{CAR,adj}} = 0$.
\item When $q = 0$ in the Z-estimation, $\boldsymbol{\Gamma}_{\sf{adj}} = 0$.
\item Under  or STRPB,  $\boldsymbol{\Gamma}_{\sf{CAR,adj}}$ is positive definite because $\boldsymbol{\Sigma}^{\sf{CAR}}_{kk'}(s,s') = 0$ for all $k, k' \in 0 \cup[K]$ and $s,s' \in [s_\mathrm{max}]$.
\end{enumerate}
\label{cor:specialcase}
\end{customcor}

\subsubsection{Main proof of Theorem \ref{supp:thm:mu_normality}}

\begin{proof}
By Lemma \ref{lem:mu_normality}, we decompose $\bm{\Gamma}$ into three parts:
\begin{align*}
     \bm{\Gamma} &=(K+1)\mathrm{diag}\left \{  \mathrm{Var}\left[r^0(\boldsymbol{\theta}^*)\right],\dots, \mathrm{Var}\left[r^K(\boldsymbol{\theta}^*)\right]\right \} +\mathrm{Var}\left[\boldsymbol{Y}\right]-\mathrm{Var}\left[\boldsymbol{r}\right] \\
    &-(K+1)^2\sum_{k,k'\in 0 \cup [K]}\sum_{s,s'\in [s_\mathrm{max}]}\lbrace\boldsymbol{\Sigma}_{kk'}^{\sf{CR}}(s,s')-\boldsymbol{\Sigma}_{kk'}^{\sf{CAR}}(s,s')\rbrace \boldsymbol{L}(s,k)\boldsymbol{L}(s',k')^{\top} \\
    &=(K+1)\mathrm{diag}\left \{  \mathrm{Var}\left[Y(0)\right],\dots,\mathrm{Var}\left[Y(K)\right]\right \} +\mathrm{Var}\left[\boldsymbol{Y}\right]-\mathrm{Var}\left[\boldsymbol{r}\right] \\
    &+ (K+1)\mathrm{diag}\left \{  \mathrm{Var}\left[r^0(\boldsymbol{\theta}^*)\right]-\mathrm{Var}\left[Y(0)\right],\dots,\mathrm{Var}\left[r^K(\boldsymbol{\theta}^*)\right]-\mathrm{Var}\left[Y(K)\right]\right \}\\
    &-(K+1)^2\sum_{k,k'\in 0 \cup [K]}\sum_{s,s'\in [s_\mathrm{max}]}\lbrace\boldsymbol{\Sigma}_{\sf{CR}}^{k,k'}(s,s')-\boldsymbol{\Sigma}_{kk'}^{\sf{CAR}}(s,s')\rbrace \boldsymbol{L}(s,k)\boldsymbol{L}(s',k')^{\top}\\
    &:= \boldsymbol{\Gamma}_{\sf{base}}-\boldsymbol{\Gamma}_{\sf{adj}}-\boldsymbol{\Gamma}_{\sf{CAR,adj}}.
\end{align*}

Furthermore, to facilitate estimation, we expand $\boldsymbol{\Gamma}_{\sf{adj}}$ as follows:
\begin{align*}
\boldsymbol{\Gamma}_{\sf{adj}} 
&= (K+1) \cdot \mathrm{diag} \left\{ \mathrm{Var}[Y(0)] - \mathrm{Var}[r^0(\boldsymbol{\theta}^*)], \dots, \mathrm{Var}[Y(K)] - \mathrm{Var}[r^K(\boldsymbol{\theta}^*)] \right\} \\
&\quad + \mathrm{Var}[\boldsymbol{r}] - \mathrm{Var}[\boldsymbol{Y}] \\
&= (K+1) \cdot \mathrm{diag} \left\{ \mathrm{Var}[Y(0)] - \mathrm{Var}[r^0(\boldsymbol{\theta}^*)], \dots, \mathrm{Var}[Y(K)] - \mathrm{Var}[r^K(\boldsymbol{\theta}^*)] \right\} \\
&\quad + \mathrm{Var}[\boldsymbol{h}(\boldsymbol{X};\boldsymbol{\theta}^*)] - \mathrm{Cov}[\boldsymbol{h}(\boldsymbol{X};\boldsymbol{\theta}^*), \boldsymbol{Y}] - \mathrm{Cov}[\boldsymbol{h}(\boldsymbol{X};\boldsymbol{\theta}^*), \boldsymbol{Y}]^\top,
\end{align*}
where the second to fourth lines are obtained by substituting $\boldsymbol{r} = \boldsymbol{h}(\boldsymbol{X};\boldsymbol{\theta}^*)$ and expanding the expression using the relationship between covariance and variance, thereby yielding a form more amenable to estimation.
\end{proof}

\subsubsection{Proof of Corollary \ref{cor:specialcase}}

\begin{proof}
For the first claim, since the complete randomization procedure satisfies Assumption~\ref{supp:Assumption 3.1}, the result follows directly from Theorem~\ref{supp:thm:mu_normality}. 

For the second claim, when $q=0$, i.e., when no covariate is included, we have $h^k(\bm{X},\bm{\theta}^*) = h^k(\bm{\theta}^*)$, which is a constant. Consequently, $\mathrm{Var}[r^k(\bm{\theta}^*)] = \mathrm{Var}[Y(k)]$ and $\mathrm{Var}[\bm{r}] = \mathrm{Var}[\bm{Y}]$.

For the third claim, it is straightforward to verify that $D_n^k(s) = O_p(1)$ for $k \in {0} \cup [K]$ and $s \in [s_{\mathrm{max}}]$ under the STRPB procedure. Under the HH procedure, where $w_s > 0$ as noted in \citet{hu2023multi}, and by Theorem 3.2 in \citet{hu2023multi}, we also have $D_n^k(s) = O_p(1)$ for $k \in {0} \cup [K]$ and $s \in [s_{\mathrm{max}}]$. Consequently, $\boldsymbol{\Sigma}^{\sf CAR}_{kk'}(s,s') = 0$ for all $k, k' \in {0} \cup [K]$ and for all $s, s' \in [s_{\mathrm{max}}]$ under both the STRPB and HH procedures.

\end{proof}

\subsection{Proof of Theorem \ref{supp:thm:contrast_normality}}\label{supp:asymp_delta}

\begin{customth}{4}[Reproduced from Theorem 4 in the main text]
\label{supp:thm:contrast_normality}
Consider the parameter of interest \(\bm{\delta} = (g(\mu_1) - g(\mu_0), \ldots, g(\mu_K) - g(\mu_0))^\top\), and denote by \(\widehat{\bm{\delta}}\) its estimator obtained by replacing \(\bm{\mu}\) with \(\widehat{\bm{\mu}}\).
 Then under Assumptions \ref{supp:Assumption 3.1} and \ref{Assump:Z-estimation},
\begin{equation}
\sqrt{n}(\widehat{\bm{\delta}} - \bm{\delta}) \cvd N(\bm{0}, \bm{G}\bm{\bm{\Gamma}}\bm{G}^\top),
\end{equation}
where $\bm{G} = \frac{\partial \bm{\delta}}{\partial \bm{\mu}}$ is the $K \times (K+1)$ Jacobian matrix of $\bm{\delta}$ evaluated at $\bm{\mu}$, and $\bm{\Gamma}$ is defined in Theorem~\ref{supp:thm:mu_normality}.
\end{customth}


\subsubsection{Main proof of Theorem \ref{supp:thm:contrast_normality}}
\begin{proof}
    The result follows immediately from Theorem \ref{supp:thm:mu_normality} 
    and the delta method \cite[Theorem 5.5.28]{casella2002statistical}.
\end{proof}

\subsection{Proof of Proposition \ref{pro:Gamma_consistency}}
\begin{custompro}{5}[Reproduced from Proposition 5 in the main text]
    Under Assumptions \ref{Assumption 3.1} and \ref{Assump:Z-estimation}, $\widehat{\bm{\Gamma}}_{\sf{conv}} \cvp \bm{\Gamma}_{\sf{conv}}$ and $\widehat{\bm{\Gamma}} \cvp \bm{\Gamma}$.
    \label{pro:Gamma_consistency}
\end{custompro}

\subsubsection{Lemmas}

\begin{customlm}{11}
\label{lem:Gamma_consistency}
Under Assumptions \ref{supp:Assumption 3.1} and \ref{Assump:Z-estimation}, the estimator $\widehat{\boldsymbol{\Gamma}} := \widehat{\boldsymbol{\Gamma}}_{\sf{base}} - \widehat{\boldsymbol{\Gamma}}_{\sf{adj}}-\widehat{\boldsymbol{\Gamma}}_{\sf{CAR,adj}}$ are consistent, where:
\begin{align*}
    & \widehat{\boldsymbol{\Gamma}}_{\sf{base}} := \mathrm{diag}\left \{ \frac{1}{\widehat{\pi}(0)}\widehat{\mathrm{Var}}\left[Y(0)\right],\cdots,\frac{1}{\widehat{\pi}(K)}\left[Y(K)\right]\right \}, \\
    & \widehat{\boldsymbol{\Gamma}}_{\sf{adj}} := \widehat{\mathrm{Var}}\left[\boldsymbol{h}(\boldsymbol{X};\boldsymbol{\theta}^*)\right]-\widehat{\mathrm{Cov}}\left[\boldsymbol{h}(\boldsymbol{X};\boldsymbol{\theta}^*),\boldsymbol{Y}\right]-\widehat{\mathrm{Cov}}\left[\boldsymbol{h}(\boldsymbol{X};\boldsymbol{\theta}^*),\boldsymbol{Y}\right]^\top\\
    & + \mathrm{diag}\left \{ \frac{1}{\widehat{\pi}(0)}\lbrace\widehat{\rm{Var}}\left[Y(0)\right]-\widehat{\mathrm{Var}}\left[r^0(\boldsymbol{\theta}^*)\right]\rbrace,\cdots,\frac{1}{\widehat{\pi}(K)}\lbrace\widehat{\rm{Var}}\left[Y(K)\right]-\widehat{\mathrm{Var}}\left[r^K(\boldsymbol{\theta}^*)\right]\rbrace\right \} ,\\
    & \widehat{\boldsymbol{\Gamma}}_{\sf{CAR,adj}} := \widehat{\boldsymbol{\Gamma}}_{\sf{CR}}-\widehat{\boldsymbol{\Gamma}}_{\sf{CAR}}, \ \mathrm{with} \\ 
    & \widehat{\boldsymbol{\Gamma}}_{\sf{CR}} := \sum_{k,k'\in0 \cup [K]} \sum_{s,s'\in[s_\mathrm{max}]}\frac{1}{\widehat{\pi}(k)} \frac{1}{\widehat{\pi}(k')}  \widehat{\boldsymbol{\Sigma}}_{kk'}^{\sf{CR}}(s,s')\widehat{\boldsymbol{L}}(s,k)\widehat{\boldsymbol{L}}(s',k')^{\top}, \\
& \widehat{\boldsymbol{\Gamma}}_{\sf{CAR}} := \sum_{k,k'\in0 \cup [K]} \sum_{s,s'\in[s_\mathrm{max}]}\frac{1}{\widehat{\pi}(k)}\frac{1}{\widehat{\pi}(k')}  \widehat{\boldsymbol{\Sigma}}_{kk'}^{\mathrm{CAR}}(s,s')\widehat{\boldsymbol{L}}(s,k)\widehat{\boldsymbol{L}}(s',k')^{\top}. \\
    & \widehat{\mathrm{Var}}\left[Y(k)\right] := \frac{1}{n_k}\sum_{i=1}^n T_i^k[Y_i-\overline{Y}(k)]^2, \quad \overline{Y}(k)  := \frac{1}{n_k}\sum_{i=1}^n T_i^kY_i ,\\
    & \widehat{\mathrm{Cov}}\left[\boldsymbol{h}(\boldsymbol{X};\boldsymbol{\theta}^*),\boldsymbol{Y}\right]:= \begin{bmatrix} \frac{1}{n_0}\sum_{i=1}^n T_i^0\lbrace Y_i-\overline{Y}(0)\rbrace \boldsymbol{h}(\boldsymbol{X}_i;\widehat{\boldsymbol{\theta}})\rbrace ^\top\\
\frac{1}{n_1}\sum_{i=1}^n T_i^1\lbrace Y_i-\overline{Y}(1)\rbrace \boldsymbol{h}(\boldsymbol{X}_i;\widehat{\boldsymbol{\theta}}) ^\top\\
\vdots \\
\frac{1}{n_K}\sum_{i=1}^n T_i^K\lbrace Y_i-\overline{Y}(K)\rbrace \boldsymbol{h}(\boldsymbol{X}_i;\widehat{\boldsymbol{\theta}})^\top
\end{bmatrix} , \\
&\widehat{\mathrm{Var}}\left[r^k(\boldsymbol{\theta}^*)\right] := \frac{1}{\widehat{\pi}(k)}\sum_{i=1}^n 
T_i^k(r_i^k(\widehat{\boldsymbol{\theta}}))^2 ,\\
&\widehat{\pi}_s := \frac{n(s)}{n}, \quad  \widehat{\pi}(k) := \frac{1}{n}\sum_{i=1}^n T_i^k, \quad
\widehat{\boldsymbol{L}}(s,k) := \frac{1}{n_k(s)}\sum_{i=1}^n T_i^k r_i^k(\widehat{\boldsymbol{\theta}}) \mathbb{I}(S_i = s)\bm{e}_k.
\end{align*}
First, by substituting $\widehat{\pi}_s$ and $\widehat{\pi}(k)$, we can compute $\widehat{\boldsymbol{\Sigma}}_{kk'}^{\sf{CR}}(s,s')$. 
Next, we estimate $\widehat{\boldsymbol{\Gamma}}_{\sf{CAR}}$ using a bootstrap-based approach. 
Let $B$ denote the number of bootstrap replicates, for each $1 \leq b \leq B$, generate $\{\{S_{i,b}\}_{i=1}^n\}_{b=1}^B$ from the empirical distribution of $\{S_i\}_{i=1}^n$. 
Then, generate $\{\{\bm{T}_{i,b}^*\}_{i=1}^n\}_{b=1}^B$ according to the same CAR procedure as used in the trial, and evaluate the imbalance 
$\{D_{n,b}^{*k}(s), 1 \leq s \leq s_{\mathrm{max}}, 0 \leq k \leq K\}_{b=1}^B$. 
Subsequently, each element of $\widehat{\boldsymbol{\Sigma}}^{\sf{CAR}}$ can be estimated as
\begin{align*}
    \widehat{\boldsymbol{\Sigma}}_{kk'}^{\sf{CAR}}(s,s') 
    = \frac{1}{n} \left \{  \frac{1}{B}\sum_{b=1}^B D_{n,b}^{*k}(s) D_{n,b}^{*k'}(s')
    - \left(\frac{1}{B}\sum_{b=1}^B D_{n,b}^{*k}(s)\right)
      \left(\frac{1}{B}\sum_{b=1}^B D_{n,b}^{*k'}(s')\right) \right \} .
\end{align*}

By plugging in $\widehat{\boldsymbol{L}}(s,k)$ and $\widehat{\boldsymbol{\Sigma}}_{kk'}^{\mathsf{CAR}}(s,s')$, we obtain 
$\widehat{\boldsymbol{\Gamma}}_{\mathsf{CAR}}$. 
Furthermore, when the dimension of $\widehat{\boldsymbol{\Sigma}}^{\mathsf{CAR}}$ is large and memory efficiency is a concern, 
we can instead compute
\[
    \bm{O}_b =   \sum_{k=0}^K\sum_{s=1}^{s_{\mathrm{max}}} D_{n,b}^{*k}(s)\, \widehat{\boldsymbol{L}}(s, k),
    \quad 1 \leq b \leq B,
\]
and directly estimate $\widehat{\boldsymbol{\Gamma}}_{\mathrm{CAR}}$ using the sample covariance of $\{\bm{O}_b\}_{b=1}^B$, that is,
\begin{align*}
   \widehat{\boldsymbol{\Gamma}}_{\mathrm{CAR}} 
   = \frac{1}{nB}\sum_{b=1}^B \bm{O}_b \bm{O}_b^\top
   - \frac{1}{n}\left(\frac{1}{B}\sum_{b=1}^B \bm{O}_b\right)
     \left(\frac{1}{B}\sum_{b=1}^B \bm{O}_b\right)^\top.
\end{align*}
It is straightforward to verify that both methods yield equivalent estimates of $\widehat{\boldsymbol{\Gamma}}_{\mathrm{CAR}}$.
\end{customlm}

\begin{customlm}{12}
    Under Assumptions \ref{supp:Assumption 3.1} and \ref{Assump:Z-estimation}, $\widehat{\mathrm{Cov}}\left[\boldsymbol{h}(\boldsymbol{X};\boldsymbol{\theta}^*),\boldsymbol{Y}\right]$ is consistent.
    \label{Lemma S4.1}
\end{customlm}

\begin{customlm}{13}
    Under Assumption \ref{supp:Assumption 3.1}, $\widehat{\boldsymbol{\Sigma}}^{\sf{CAR}}$ is consistent.
    \label{lemma S4.2}
\end{customlm}

\begin{customlm}{14}
For any $\tilde{\boldsymbol{\pi}} \in \tilde{\boldsymbol{\Pi}}^0(\boldsymbol{\kappa})$, where the latter is defined in the proof of Lemma 13, \eqref{Bootstrap1} holds.
\label{lemma:bootstrap}
\end{customlm}

\subsubsection{Main proof of Proposition \ref{pro:Gamma_consistency}}

\begin{proof}
    The result follows immediately from Lemma \ref{lem:Gamma_consistency}.
\end{proof}

\subsubsection{Proof of lemmas}
\begin{proof}[Proof of Lemma \ref{lem:Gamma_consistency}]
By Lemma \ref{Lemma S4.1}, we know 
\[
\widehat{\mathrm{Cov}}\left[\boldsymbol{h}(\boldsymbol{X};\boldsymbol{\theta}^*),\boldsymbol{Y}\right] \cvp \mathrm{Cov}\left[\boldsymbol{h}(\boldsymbol{X};\boldsymbol{\theta}^*),\boldsymbol{Y}\right].\]
Similarly, we can prove the consistency of $\widehat{\boldsymbol{\Gamma}}_{\sf{base}}$,  $\widehat{\boldsymbol{\Gamma}}_{\sf{adj}}$ and $r_s^k(\widehat{\boldsymbol{\theta}})$. By Lemma \ref{lemma S4.2} and Slutsky’s theorem, we can prove the consistency of $\widehat{\boldsymbol{\Gamma}}_{\sf{CAR}}$ and $\widehat{\boldsymbol{\Gamma}}_{\sf{CR}}$, then we finish the proof of this theorem. 
\end{proof}

\begin{proof}[Proof of Lemma \ref{Lemma S4.1}]
    By the multivariate Taylor's expansion of $\sum_{i=1}^n T_i^kY_i\boldsymbol{h}(\boldsymbol{X}_i;\widehat{\boldsymbol{\theta}})$ around the point $\boldsymbol{\theta}=\boldsymbol{\theta}^*$, there exists a $\widetilde{\boldsymbol{\theta}}$ (which depends on $\widehat{\boldsymbol{\theta}}$) on the line segment between $\widehat{\boldsymbol{\theta}}$ and $\boldsymbol{\theta}^*$ such that
\begin{align*}
    &\frac{1}{n_k}\sum_{i=1}^nT_i^kY_i\left\{\boldsymbol{h}(\boldsymbol{X}_i;\widehat{\boldsymbol{\theta}}) -\boldsymbol{h}(\boldsymbol{X}_i;\boldsymbol{\theta}^*)\right\}=\left\{\frac{1}{n_k}\sum_{i=1}^nT_i^kY_i\frac{\partial}{\partial \boldsymbol{\theta}} \boldsymbol{h}(\boldsymbol{X}_i;\boldsymbol{\theta}) \Big|_{\boldsymbol{\theta} = \boldsymbol{\theta}^*}\right\}(\widehat{\boldsymbol{\theta}}-\boldsymbol{\theta})\\
    &+\sum_{k' \in 0 \cup [K]} \frac{1}{2}(\widehat{\boldsymbol{\theta}} - \boldsymbol{\theta}^*)^{\top} \left(\frac{1}{n_k}\sum_{i = 1}^{n}T_i^kY_i\frac{\partial^2}{\partial \boldsymbol{\theta} \partial \boldsymbol{\theta}^{\top}} h^{k'}(\boldsymbol{X}_i;\widetilde{\boldsymbol{\theta}})\right)(\widehat{\boldsymbol{\theta}} - \boldsymbol{\theta}^*) \boldsymbol{e}_{k'}.
\end{align*}
From  Assumptions \ref{supp:Assumption 3.1}.1, \ref{Assump:Z-estimation} and (\ref{eq:Second derivative}), we know $\left\| \sum_{i = 1}^{n}T_i^kY_i\frac{\partial^2}{\partial \boldsymbol{\theta} \partial \boldsymbol{\theta}^{\top}} h^{k'}(\boldsymbol{X}_i;\widetilde{\boldsymbol{\theta}})/n_k\right\|=O_{p}(1)$, and thus  the last term in the Taylor expansion is $O_{p}(\boldsymbol{1})\|\widehat{\boldsymbol{\theta}} - \boldsymbol{\theta}^*\|^{2}$. By Lemma \ref{lem:theta:Consistency}, we have
\begin{align*}
    \frac{1}{n_k}\sum_{i=1}^nT_i^kY_i\left\{\boldsymbol{h}(\boldsymbol{X}_i;\widehat{\boldsymbol{\theta}}) -\boldsymbol{h}(\boldsymbol{X}_i;\boldsymbol{\theta}^*)\right\} = o_p(\boldsymbol{1}).
\end{align*}
Then, by Assumption \ref{supp:Assumption 3.1}.1 and Lemma \ref{lem:mu:consistency}, we have
\begin{align*}
    \frac{1}{n_k}\sum_{i=1}^nT_i^kY_i\lbrace\boldsymbol{h}(\boldsymbol{X}_i;\widehat{\boldsymbol{\theta}})-\widehat{\boldsymbol{\mu}}\rbrace &= \frac{1}{n_k}\sum_{i=1}^nT_i^kY_i\lbrace\boldsymbol{h}(\boldsymbol{X}_i;\boldsymbol{\theta}^*)-\boldsymbol{\mu}\rbrace+o_p(\boldsymbol{1}) \\
    &= \mathbb{E}\left[Y(k)(\boldsymbol{h}(\boldsymbol{X}_i;\boldsymbol{\theta}^*)-\boldsymbol{\mu})\right]+o_p(\boldsymbol{1}),
\end{align*}
so we finish the proof of this lemma.
\end{proof}

\begin{proof}[Proof of Lemma \ref{lemma S4.2}]
    Let $\boldsymbol{D}_n^{\mathcal{K}}(\mathcal{S}) \in \mathbb{R}^{(K+1)\times s_\mathrm{max}}$  denote the vectors of the observed within-stratum imbalance $\{D_n^k(s),k\in 0 \cup [K],s\in [s_\mathrm{max}]\}$. Denote the vector of the probabilities of observed strata $\{\pi_s, s\in [s_\mathrm{max}]\}$ as $\boldsymbol{\pi}^0$, and the vector of the corresponding empirical probabilities $\{\pi_s^n = n^{-1}n(s), s\in [s_\mathrm{max}]\}$ as $\boldsymbol{\pi}^n$. Let $\mathcal{P}_{\boldsymbol{\pi}^0}$ and $\mathcal{P}_{\boldsymbol{\pi}^n}$ be the probability measures associated with the parameters $\boldsymbol{\pi}^0$ and $\boldsymbol{\pi}^n$, respectively. Then, according to Assumption \ref{supp:Assumption 3.1}, there exists a variance - covariance matrix $\boldsymbol{\Xi}_D(\boldsymbol{\pi}^0)$, whose value may depend on $\boldsymbol{\pi}^0$, such that
\begin{equation}
\frac{\boldsymbol{D}_n^{\mathcal{K}}(\mathcal{S})}{\sqrt{n}} \stackrel{d}{\underset{\mathcal{P}_{\boldsymbol{\pi}^0}}{\longrightarrow}} \mathcal{N}\left(\bm{0}_{|\mathcal{V}|}, \boldsymbol{\Xi}_D(\boldsymbol{\pi}^0)\right),
\label{eq:Bootstrap}
\end{equation}
where $\mathcal{P}_{\boldsymbol{\pi}^0}$ beneath the arrow indicates that the convergence is with respect to the probability measure of $\mathcal{P}_{\boldsymbol{\pi}^0}$, and $|\mathcal{V}|=(K + 1)\times s_{\mathrm{max}}$. Since \eqref{eq:Bootstrap} is presupposed, the theoretical framework for the bootstrap based on local asymptotic normality \citep{Beran1997} can be utilized to establish the desired results.

The proof strategy is as follows. To apply the local asymptotic normality framework, we define $\tilde{\boldsymbol{\Pi}}^0(\boldsymbol{\kappa})=\{\tilde{\boldsymbol{\pi}}: \tilde{\boldsymbol{\pi}}=\boldsymbol{\pi}^0 + n^{-1/2}\boldsymbol{\kappa}\}$, where $\boldsymbol{\kappa}\in\mathbb{R}^{s_{\mathrm{max}}}$ represents a local parameter. For any $\tilde{\boldsymbol{\pi}}\in\tilde{\boldsymbol{\Pi}}(\boldsymbol{\kappa})$, we will leverage Le Cam's third lemma (see \citet[Example 6.7 on page 90]{van2000asymptotic}) to demonstrate that, under $\mathcal{P}_{\tilde{\boldsymbol{\pi}}}$, $n^{-1/2}\boldsymbol{D}_n^{\mathcal{K}}(\mathcal{S})$ has the same distribution as in \eqref{eq:Bootstrap}, i.e.,
\begin{equation}
\frac{\boldsymbol{D}_n^{\mathcal{K}}(\mathcal{S})}{\sqrt{n}} \stackrel{d}{\underset{\mathcal{P}_{\tilde{\boldsymbol{\pi}}}}{\longrightarrow}} \mathcal{N}\left(\bm{0}_{|\mathcal{V}|}, \boldsymbol{\Xi}_D(\boldsymbol{\pi}^0)\right).
\label{Bootstrap1}
\end{equation}

Consequently, $\widehat{\boldsymbol{\Xi}}_D(\tilde{\boldsymbol{\pi}})$, the sample variance-covariance matrix computed using $B$ random samples $\{\boldsymbol{D}_{n,b}^{\mathcal{K}}(\mathcal{S})\}_{b = 1}^B$ from $\boldsymbol{D}_n^{\mathcal{K}}(\mathcal{S})$ with respect to the probability measure $\mathcal{P}_{\tilde{\boldsymbol{\pi}}}$, converges to $\boldsymbol{\Xi}_D(\boldsymbol{\pi}^0)$ by virtue of equation \eqref{Bootstrap1} and the Law of Large Numbers (LLN). Since equation \eqref{Bootstrap1} is independent of the choice of $\boldsymbol{\kappa}$, it holds for any $\boldsymbol{\kappa}\in\mathbb{R}^{|\mathcal{S}|}$. Moreover, by the Central Limit Theorem, $n^{1/2}(\boldsymbol{\pi}^n - \boldsymbol{\pi}^0)=O_p(\bm{1})$, which implies that $\boldsymbol{\pi}^n$ is an element of $\tilde{\boldsymbol{\Pi}}^0(\boldsymbol{\kappa})$ for some $\boldsymbol{\kappa}\in\mathbb{R}^{|\mathcal{S}|}$. Then, according to the extended continuous mapping theorem (see page 67 of \citet[Theorem 1.11.1]{VanDerVaartWellner1996}), $\widehat{\boldsymbol{\Xi}}_D(\boldsymbol{\pi}^n)$ converges to $\boldsymbol{\Xi}_D(\boldsymbol{\pi}^0)$ under $\mathcal{P}_{\boldsymbol{\pi}^n}$. Notice that $\{\boldsymbol{D}_{n,b}^{*\mathcal{K}}(\mathcal{S})\}_{b = 1}^B$ considered in Lemma \ref{lem:Gamma_consistency} is equivalent to $\{\boldsymbol{D}_{n,b}^{\mathcal{K}}(\mathcal{S})\}_{b = 1}^B$ with respect to the probability measure $\mathcal{P}_{\boldsymbol{\pi}^n}$. With the above results,
\begin{align*}
\widehat{\boldsymbol{\Sigma}}_{kk'}^{\sf{CAR}}(s,s') = \boldsymbol{\Sigma}_{kk'}^{\sf{CAR}}(s,s')+o_{p}(1).
\end{align*}
\end{proof}
\begin{proof}[Proof of Lemma \ref{lemma:bootstrap}]
For notational simplicity, let $\pi(\tilde{\boldsymbol{x}}_i)$ denote the probability mass function of $\tilde{\boldsymbol{x}}_i$ with parameter $\boldsymbol{\pi}$, i.e., 
\[
\pi(\tilde{\boldsymbol{x}}_i)=\prod_{s \in \mathcal{S}} [\pi_s]^{\mathbb{I}\{\tilde{\boldsymbol{x}}_i = \boldsymbol{x}(s)\}},
\]
and let $\tilde{\pi}(\tilde{\boldsymbol{x}}_i)$ and $\pi^0(\tilde{\boldsymbol{x}}_i)$ denote the analogous functions with parameters $\tilde{\boldsymbol{\pi}}$ and $\boldsymbol{\pi}^0$, respectively. To apply Le Cam's third lemma, we consider the likelihood ratio of $\tilde{\boldsymbol{\pi}}$ and $\boldsymbol{\pi}^0$, denoted as $l_n(\tilde{\boldsymbol{\pi}}, \boldsymbol{\pi}^0)$. Let $\kappa_s$ be the $s$-th element of $\boldsymbol{\kappa}$. By a Taylor expansion of $\log(1 + x)$, we have
{\small
\begin{align*}
l_n(\tilde{\boldsymbol{\pi}}, \boldsymbol{\pi}^0) & = \sum_{i = 1}^{n} \log\{\tilde{\pi}(\tilde{\boldsymbol{x}}_i)/\pi^0(\tilde{\boldsymbol{x}}_i)\} = \sum_{i = 1}^{n} \log\left\{\prod_{s \in [s_\mathrm{max}]} \left(\frac{\tilde{\pi}_s}{\pi_s^0}\right)^{\mathbb{I}\{\tilde{\boldsymbol{x}}_i = \boldsymbol{x}(s)\}}\right\}\\
& = \sum_{s \in [s_\mathrm{max}]} \sum_{i = 1}^{n} \mathbb{I}\{\tilde{\boldsymbol{x}}_i = \boldsymbol{x}(s)\} \log\left\{1 + \frac{\kappa_s}{\sqrt{n}\pi_s^0}\right\}\\
& = \sum_{i = 1}^{n} \sum_{s \in [s_\mathrm{max}]} \mathbb{I}\{\tilde{\boldsymbol{x}}_i = \boldsymbol{x}(s)\} \frac{\kappa_s}{\sqrt{n}\pi_s^0} - \frac{1}{2n} \sum_{s \in [s_\mathrm{max}]} \sum_{i = 1}^{n} \mathbb{I}\{\tilde{\boldsymbol{x}}_i = \boldsymbol{x}(s)\} \left(\frac{\kappa_s}{\pi_s^0}\right)^2 + O_p(n^{-1/2})\\
& = \frac{1}{\sqrt{n}} \sum_{i = 1}^{n} \sum_{s \in [s_\mathrm{max}]} (\mathbb{I}\{\tilde{\boldsymbol{x}}_i = \boldsymbol{x}(s)\} - \pi_s^0) \frac{\kappa_s}{\pi_s^0} - \frac{1}{2} \sum_{s \in [s_\mathrm{max}]} \frac{\kappa_s^2}{\pi_s^0} + o_p(\boldsymbol{1}),
\end{align*}
}
provided that $\sum_{s \in \mathcal{S}} \kappa_s = \sqrt{n} \sum_{s \in \mathcal{S}} (\pi_s - \pi_s^0) = 0$.

Next, we establish the joint normality of $(n^{-1/2}\boldsymbol{D}_n^{\mathcal{K}}(\mathcal{S})^{\top}, l_n(\tilde{\boldsymbol{\pi}}, \boldsymbol{\pi}^0))^{\top}$. Define $\Delta l_i(\tilde{\boldsymbol{\pi}}, \boldsymbol{\pi}) = \mathbb{E}[l_n(\tilde{\boldsymbol{\pi}}, \boldsymbol{\pi})|\mathcal{F}_i] - \mathbb{E}[l_n(\tilde{\boldsymbol{\pi}}, \boldsymbol{\pi})|\mathcal{F}_{i - 1}] = \sum_{s \in \mathcal{S}} (\mathbb{I}\{\tilde{\boldsymbol{x}}_{i + 1} = \boldsymbol{x}(s)\} - \pi_s^0)(\pi_s^0)^{-1}\kappa_s$, and let $\widetilde{M}_i = (M_i^{\mathcal{K}}(\mathcal{S})^{\top}, \Delta l_i(\tilde{\boldsymbol{\pi}}, \boldsymbol{\pi}))^{\top}$, where $M_i^{\mathcal{K}}(\mathcal{S})$ represents the $|\mathcal{V}|$-vector of martingale difference $\{M_i^k(s), k \in 0 \cup [K], s \in [s_\mathrm{max}]\}$. Since the last term of $\widetilde{M}_i$ is bounded, it is straightforward to show that $\widetilde{M}_i$ is a sequence of $(|\mathcal{V}| + 1)$-dimensional zero - mean martingale differences satisfying the conditional Lindeberg's condition. Moreover, the conditional variance - covariance matrix satisfies the following:
\begin{align*}
&\frac{1}{n}\sum_{i = 1}^{n} \mathbb{E}[M_i^{\mathcal{K}}(\mathcal{S})M_i^{\mathcal{K}}(\mathcal{S})^{\top}|\mathcal{F}_{i - 1}] \stackrel{\mathrm{P}}{\underset{\mathcal{P}_{\boldsymbol{\pi}^0}}{\longrightarrow}} \boldsymbol{\Xi}_D(\boldsymbol{\pi}^0),\\
&\sum_{i = 1}^{n} \mathbb{E}[M_i^{\mathcal{K}}(\mathcal{S})\Delta l_i(\tilde{\boldsymbol{\pi}}, \boldsymbol{\pi})|\mathcal{F}_{i - 1}]\\
&=\sum_{i = 1}^{n} \mathbb{E}[M_i^{\mathcal{K}}(\mathcal{S})|\mathcal{F}_{i - 1}] \mathbb{E}\left[\sum_{s \in \mathcal{S}} \left(\mathbb{I}\{\tilde{\boldsymbol{x}}_{i + 1} = \boldsymbol{x}(s)\} - \pi_s^0\right)|\mathcal{F}_{i - 1}\right] \frac{\kappa_s}{\pi_s^0} = \bm{0}_{|\mathcal{V}|},
\end{align*}
and
\begin{align*}
\frac{1}{n}\sum_{i = 1}^{n} \mathbb{E}[\Delta l_i(\tilde{\boldsymbol{\pi}}, \boldsymbol{\pi})^2|\mathcal{F}_{i - 1}]&=\sum_{s \in \mathcal{S}} \left(\frac{\kappa_s}{\pi_s^0}\right)^2 \pi_s^0(1 - \pi_s^0) - \sum_{s \neq s'} \kappa_s \kappa_{s'}\\
&=\sum_{s \in \mathcal{S}} \frac{\kappa_s^2}{\pi_s^0} - \left(\sum_{s \in \mathcal{S}} \kappa_s\right)^2 = \sum_{s \in \mathcal{S}} \frac{\kappa_s^2}{\pi_s^0},
\end{align*}
by the conditions of Assumption \ref{supp:Assumption 3.1}. Then, by the martingale central limit theorem and the Cramér-Wold device, we obtain
\[
\begin{pmatrix}
n^{-1/2}\boldsymbol{D}_n^{\mathcal{K}}(\mathcal{S}) \\
l_n(\tilde{\boldsymbol{\pi}}, \boldsymbol{\pi}^0)
\end{pmatrix}
\stackrel{d}{\underset{\mathcal{P}_{\boldsymbol{\pi}^0}}{\longrightarrow}}
\mathcal{N}\left(
\begin{pmatrix}
\bm{0}_{|\mathcal{V}|} \\
-\frac{\nu^2}{2}
\end{pmatrix},
\begin{pmatrix}
\boldsymbol{\Xi}_D(\boldsymbol{\pi}^0) & \bm{0}_{|\mathcal{V}|}^{\top} \\
\bm{0}_{|\mathcal{V}|} & \nu^2
\end{pmatrix}
\right),
\]
where $\nu^2 = \sum_{s \in \mathcal{S}} (\pi_s^0)^{-1} \kappa_s^2$. Thus, by Le Cam's third lemma, \eqref{Bootstrap1} holds. 
\end{proof}
\subsection{Proof of Theorem \ref{supp:thm:multiple_testing} and Theorem \ref{supp:thm:comparison}}
\begin{customth}{6}[Reproduced from Theorem 6 in the main text]
\label{supp:thm:multiple_testing}
Under Assumptions~\ref{Assumption 3.1} and~\ref{Assump:Z-estimation}, we consider the individual hypothesis test $H_{0}: \delta_k = 0$ versus $H_{1}: \delta_k > 0$. Define the individual conventional and model-robust Wald test statistics for $k \in [K]$ as
\begin{equation*}
W^{\sf{conv}}_k = \frac{\sqrt{n}\,\widehat{\delta}_k}{\sqrt{[\widehat{\bm{G}}\,\widehat{\bm{\Gamma}}_{\sf{conv}}\,\widehat{\bm{G}}^\top]_{(k,k)}}}, 
\quad \text{and} \quad 
W^{\sf{robust}}_k = \frac{\sqrt{n}\,\widehat{\delta}_k}{\sqrt{[\widehat{\bm{G}}\,\widehat{\bm{\Gamma}}\,\widehat{\bm{G}}^\top]_{(k,k)}}}.
\end{equation*}
$W^{\sf{robust}}_k$ always produces a valid type~I error by Theorem \ref{thm:contrast_normality} and Proposition \ref{prop:consistency}. Because $\boldsymbol{\Gamma}_{\sf conv} - \boldsymbol{\Gamma}$ is positive definite under HH and STRPB, the conventional statistic $W^{\sf conv}_k$ exhibits a reduced type~I error under these procedures.
\end{customth}

\begin{customth}{7}[Reproduced from Theorem 7 in the main text]
\label{supp:thm:comparison}
  Under Assumptions \ref{Assumption 3.1} and \ref{Assump:Z-estimation},  we aim to conduct the following hypothesis test
\begin{align*}
\vspace{-10pt}
H^1_{0}&: \delta_1 = \delta_2=\cdots =\delta_K= 0,\\
H^1_{1}&: \exists k\in[K], \delta_k > 0.
\end{align*}
In particular, $(W^{\sf{robust}}_1,\ldots,W^{\sf{robust}}_K)^\top \cvd N(\bm{0}, \bm{R})$ with 
\begin{equation*}
\bm{R}_{(k,k')} = \frac{[\bm{G}\bm{\Gamma}\bm{G}^\top]_{(k,k')}}{\sqrt{[\bm{G}\bm{\Gamma}\bm{G}^\top]_{(k,k)}[\bm{G}\bm{\Gamma}\bm{G}^\top]_{(k',k')}}}, \quad k, k'\in[K].
\end{equation*}
\end{customth}

\subsubsection{Main proof of Theorem \ref{supp:thm:multiple_testing}}
\begin{proof}
By Theorem~\ref{thm:contrast_normality} and Proposition~\ref{pro:Gamma_consistency}, the robust statistic $W^{\sf robust}_k$ satisfies $W^{\sf robust}_k \cvd N(0,1)$, and therefore controls the type~I error at the nominal level. In contrast, Corollary~\ref{cor:specialcase} implies that $\boldsymbol{\Gamma}_{\sf conv} - \boldsymbol{\Gamma}$ is positive definite under the HH and STRPB procedures. Consequently, the conventional statistic $W^{\sf conv}_k$ converges in distribution to $N(0,\varrho)$ for some $0<\varrho<1$, leading to a reduced type~I error under these randomization schemes.
\end{proof}
\subsubsection{Main proof of Theorem \ref{supp:thm:comparison}}
\begin{proof}
For any fixed vector $\bm{a}\in\mathbb{R}^K$,  
consider the linear combination 
\begin{align*}
    \bm{a}^\top (W^{\sf robust}_1,\ldots,W^{\sf robust}_K)^\top,
\end{align*} since $\widehat{\bm{\delta}}$ is asymptotically normal and $\widehat{\bm{G}}\,\widehat{\bm{\Gamma}}\,\widehat{\bm{G}}^\top$ is consistent, Slutsky’s theorem implies that this
linear combination converges in distribution to a univariate normal random
variable. Because this holds for every $\bm{a}$, the Cramér--Wold theorem ensures
that $(W^{\sf robust}_1,\ldots,W^{\sf robust}_K)^\top$ is jointly asymptotically
multivariate normal. Moreover, a direct computation verifies that its limiting
covariance matrix is
\[
\bm{R}_{(k,k')}
= \frac{[\bm{G}\bm{\Gamma}\bm{G}^\top]_{(k,k')}}%
{\sqrt{[\bm{G}\bm{\Gamma}\bm{G}^\top]_{(k,k)}[\bm{G}\bm{\Gamma}\bm{G}^\top]_{(k',k')}}},
\]
which matches the correlation matrix stated in the theorem.
\end{proof}
\newpage

\section{Additional Numerical Results}

\subsection{Example 1}
\begin{figure}[htbp] 
  \centering
  \includegraphics[width=\textwidth]{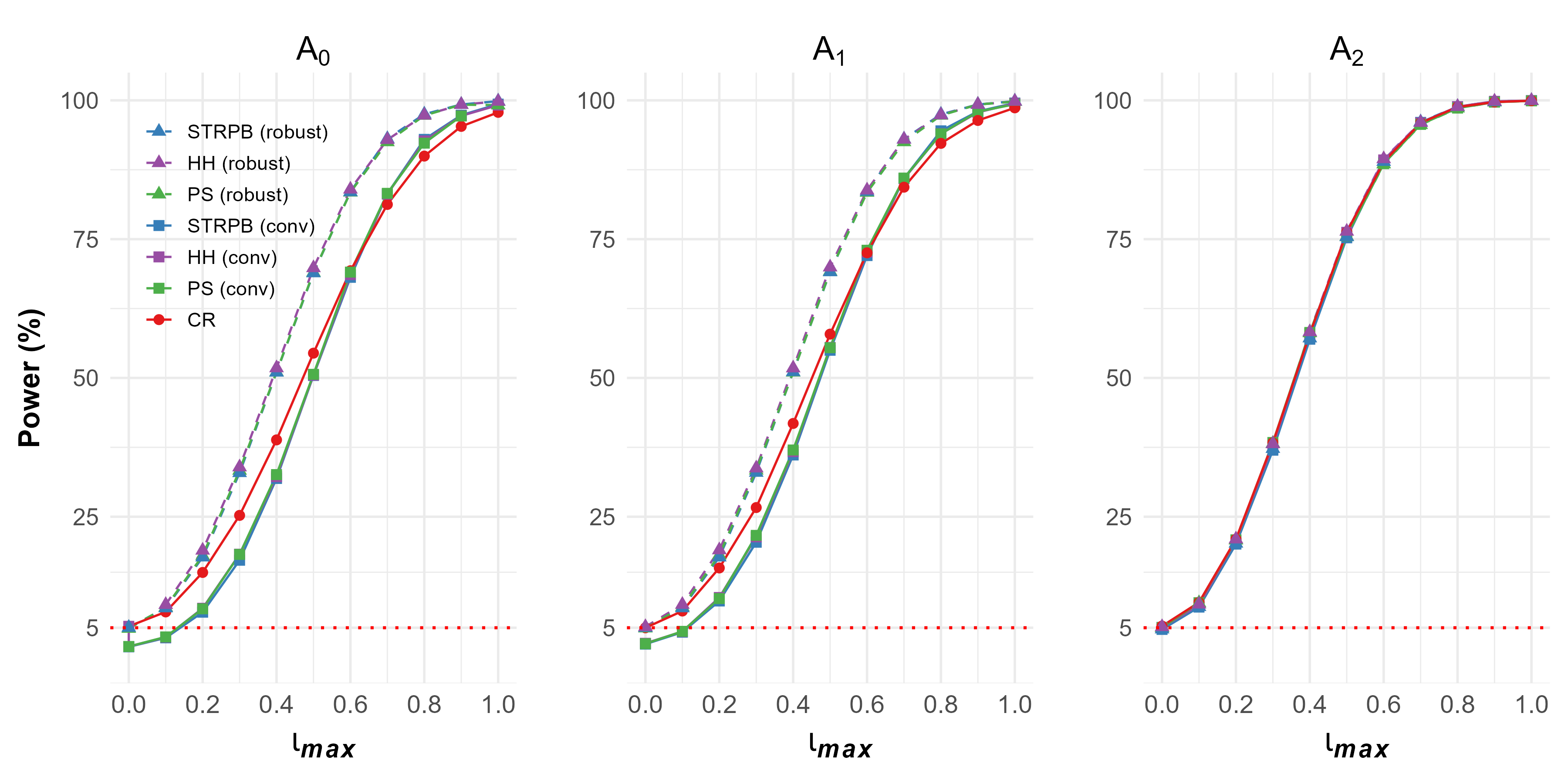}
    \caption{
   LOR for Example~1, power comparison of conventional and model-robust tests 
under $(\mathcal{A}_0, \mathcal{A}_1, \mathcal{A}_2)$ across $\iota_{\max}$; 
the red dashed line denotes $\alpha=0.05$ under the null hypothesis.
}
\label{fig:sim1_LOR}
\end{figure}

\begin{figure}[H] 
  \centering
  \includegraphics[width=\textwidth]{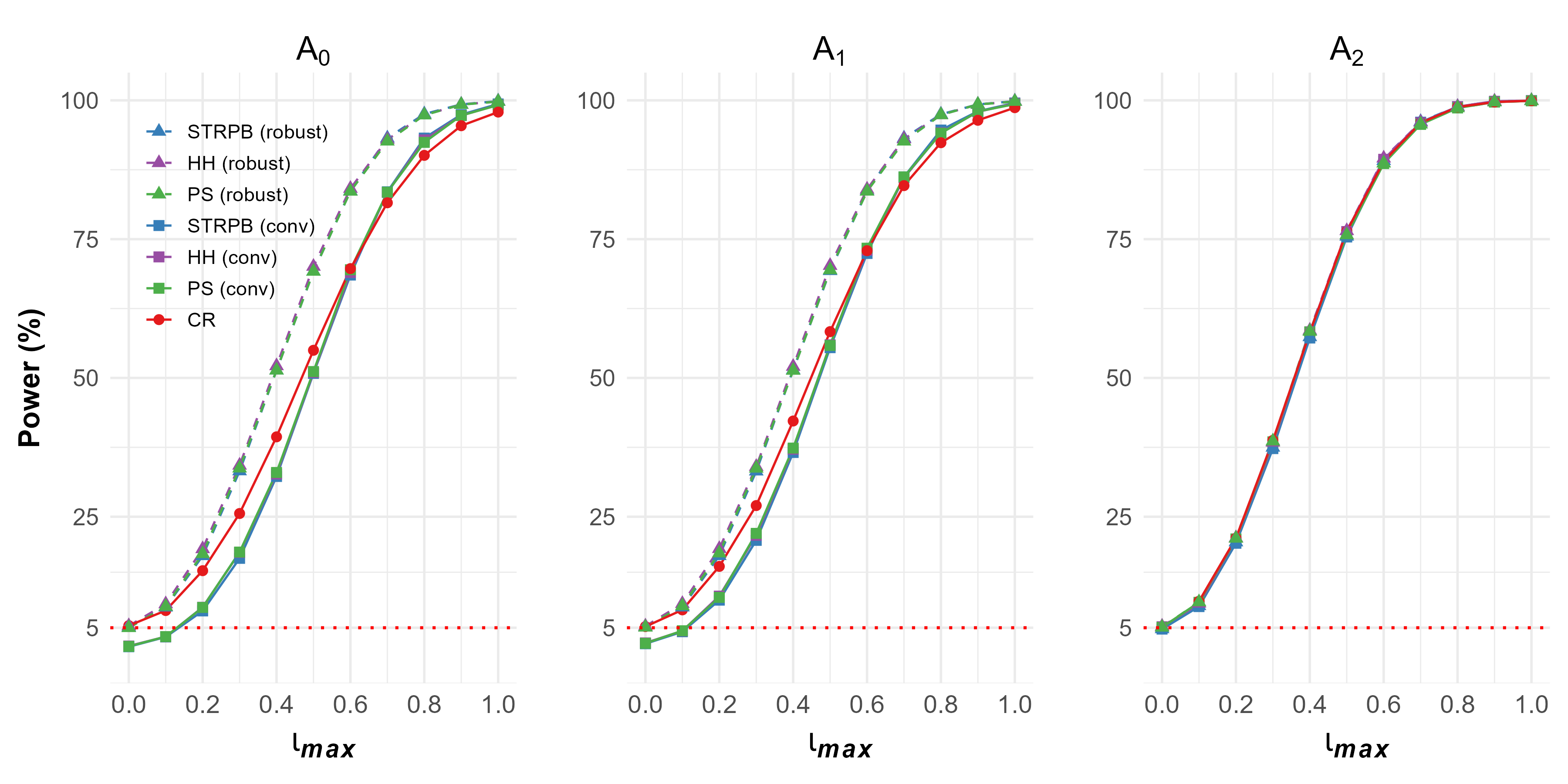}
    \caption{
   ATE for Example~1, power comparison of conventional and model-robust tests 
under $(\mathcal{A}_0, \mathcal{A}_1, \mathcal{A}_2)$ across $\iota_{\max}$; 
the red dashed line denotes $\alpha=0.05$ under the null hypothesis.
}
\label{fig:sim1_ATE}
\end{figure}

In this section, Tables \ref{tb:sim1_conv}, \ref{tb:sim1_LOR}, \ref{tb:sim1_ATE} and Figures \ref{fig:sim1_LOR}, \ref{fig:sim1_ATE} present additional simulation results from Example 1. More specifically, Table~\ref{tb:sim1_conv} reports the results of conventional Wald tests across different estimands and randomization procedures; Tables \ref{tb:sim1_LOR} and \ref{tb:sim1_ATE} compare the conventional Wald test with the proposed model-robust version for the LOR and ATE metrics, respectively; Figures \ref{fig:sim1_LOR} and \ref{fig:sim1_ATE} display the statistical power of the conventional and model-robust tests under different randomization schemes as $\iota_{\max}$ increases for LOR and ATE, respectively.

\begin{table}[H]
\centering
\footnotesize
\captionsetup{font=scriptsize} 
\setlength{\tabcolsep}{2pt}    
\renewcommand{\arraystretch}{0.4} 
\caption{Results from conventional Wald tests for Example 1, the logistic model: test statistics, SD ($\times \sqrt{n}$), Type I error rates (in \%) for Stage~1, Stage~2, and the overall Type I error, evaluated across different working models and randomization schemes: $(\iota_1,\iota_2)=(0,0)$.}
\begin{tabular}{llllcccccc}
\toprule
\textbf{Metric} & \textbf{Procedure} & \textbf{} & 
& \multicolumn{2}{c}{\textbf{Stage 1}} & \multicolumn{2}{c}{\textbf{Stage 2}} & \textbf{Overall} \\
\cmidrule(lr){5-6} \cmidrule(lr){7-8} \cmidrule(lr){9-9}
& & & & \textbf{Mean (SD)} & \textbf{Type I} & \textbf{Mean (SD)} & \textbf{Type I} & \textbf{Type I} \\
\midrule
\multirow{12}{*}{$\log$RR}
& \multirow{3}{*}{CR} & $\mathcal{A}_0$ & & 0.0664 (2.9292) & 5.15 & -0.0004 (2.6664) & 5.02 & 5.11 \\
& & $\mathcal{A}_1$ & & 0.0631 (2.8222) & 5.13 & -0.0004 (2.5550) & 5.05 & 4.85 \\
& & $\mathcal{A}_2$ & & 0.0511 (2.2592) & 5.49 & -0.0273 (0.0908) & 5.22 & 5.14 \\
\cmidrule{2-9}
& \multirow{3}{*}{STRPB} & $\mathcal{A}_0$ & & 0.0502 (2.4190) & 1.76 & 0.0009 (2.1713) & 2.41 & 1.53 \\
& & $\mathcal{A}_1$ & & 0.0502 (2.4189) & 2.23 & 0.0009 (2.1702) & 2.84 & 2.01 \\
& & $\mathcal{A}_2$ & & 0.0460 (2.2209) & 4.84 & 0.0091 (0.0884) & 4.70 & 4.62 \\
\cmidrule{2-9}
& \multirow{3}{*}{HH} & $\mathcal{A}_0$ & & 0.0532 (2.3964) & 1.62 & 0.0004 (2.2353) & 2.67 & 1.60 \\
& & $\mathcal{A}_1$ & & 0.0531 (2.3946) & 2.17 & 0.0004 (2.2336) & 3.21 & 2.15 \\
& & $\mathcal{A}_2$ & & 0.0491 (2.1902) & 5.00 & 0.0021 (0.0912) & 5.23 & 4.96 \\
\cmidrule{2-9}
& \multirow{3}{*}{PS} & $\mathcal{A}_0$ & & 0.0529 (2.4672) & 1.76 & 0.0009 (2.2249) & 2.23 & 1.61 \\
& & $\mathcal{A}_1$ & & 0.0528 (2.4663) & 2.37 & 0.0008 (2.2233) & 2.88 & 2.07 \\
& & $\mathcal{A}_2$ & & 0.0477 (2.2283) & 4.87 & 0.0107 (0.0908) & 4.97 & 4.95 \\
\midrule
\multirow{12}{*}{LOR}
& \multirow{3}{*}{CR} & $\mathcal{A}_0$ & & 0.1040 (4.6484) & 5.23 & -0.0007 (4.1738) & 5.05 & 5.26 \\
& & $\mathcal{A}_1$ & & 0.0989 (4.4781) & 5.13 & -0.0006 (3.9998) & 5.11 & 4.99 \\
& & $\mathcal{A}_2$ & & 0.0801 (3.5755) & 5.51 & -0.0426 (0.1422) & 5.25 & 5.30 \\
\cmidrule{2-9}
& \multirow{3}{*}{STRPB} & $\mathcal{A}_0$ & & 0.0786 (3.8335) & 1.78 & 0.0013 (3.4012) & 2.43 & 1.56 \\
& & $\mathcal{A}_1$ & & 0.0785 (3.8333) & 2.27 & 0.0013 (3.3995) & 2.88 & 2.05 \\
& & $\mathcal{A}_2$ & & 0.0720 (3.5149) & 4.88 & 0.0141 (0.1385) & 4.77 & 4.70 \\
\cmidrule{2-9}
& \multirow{3}{*}{HH} & $\mathcal{A}_0$ & & 0.0833 (3.7977) & 1.65 & 0.0006 (3.4996) & 2.69 & 1.61 \\
& & $\mathcal{A}_1$ & & 0.0832 (3.7949) & 2.22 & 0.0006 (3.4969) & 3.23 & 2.20 \\
& & $\mathcal{A}_2$ & & 0.0770 (3.4677) & 5.05 & 0.0021 (0.1427) & 5.28 & 5.07 \\
\cmidrule{2-9}
& \multirow{3}{*}{PS} & $\mathcal{A}_0$ & & 0.0828 (3.9043) & 1.82 & 0.0013 (3.4826) & 2.28 & 1.62 \\
& & $\mathcal{A}_1$ & & 0.0826 (3.9028) & 2.42 & 0.0013 (3.4801) & 2.89 & 2.12 \\
& & $\mathcal{A}_2$ & & 0.0747 (3.5217) & 4.93 & 0.0171 (0.1420) & 5.04 & 5.02 \\
\midrule
\multirow{12}{*}{ATE}
& \multirow{3}{*}{CR} & $\mathcal{A}_0$ & & 0.0239 (1.0774) & 5.33 & -0.0002 (0.9599) & 5.10 & 5.38 \\
& & $\mathcal{A}_1$ & & 0.0227 (1.0379) & 5.28 & -0.0002 (0.9201) & 5.14 & 5.21 \\
& & $\mathcal{A}_2$ & & 0.0184 (0.8285) & 5.60 & -0.0097 (0.0327) & 5.28 & 5.38 \\
\cmidrule{2-9}
& \multirow{3}{*}{STRPB} & $\mathcal{A}_0$ & & 0.0180 (0.8888) & 1.81 & 0.0003 (0.7831) & 2.46 & 1.58 \\
& & $\mathcal{A}_1$ & & 0.0180 (0.8887) & 2.33 & 0.0003 (0.7827) & 2.88 & 2.12 \\
& & $\mathcal{A}_2$ & & 0.0165 (0.8145) & 4.92 & 0.0032 (0.0319) & 4.81 & 4.76 \\
\cmidrule{2-9}
& \multirow{3}{*}{HH} & $\mathcal{A}_0$ & & 0.0191 (0.8805) & 1.67 & 0.0001 (0.8054) & 2.74 & 1.71 \\
& & $\mathcal{A}_1$ & & 0.0191 (0.8799) & 2.27 & 0.0001 (0.8048) & 3.29 & 2.27 \\
& & $\mathcal{A}_2$ & & 0.0177 (0.8039) & 5.17 & 0.0003 (0.0328) & 5.35 & 5.18 \\
\cmidrule{2-9}
& \multirow{3}{*}{PS} & $\mathcal{A}_0$ & & 0.0190 (0.9043) & 1.89 & 0.0003 (0.8013) & 2.38 & 1.64 \\
& & $\mathcal{A}_1$ & & 0.0190 (0.9039) & 2.51 & 0.0003 (0.8008) & 2.92 & 2.19 \\
& & $\mathcal{A}_2$ & & 0.0172 (0.8154) & 4.99 & 0.0040 (0.0327) & 5.06 & 5.07 \\
\bottomrule
\end{tabular}
\label{tb:sim1_conv}
\end{table}

\begin{table}[htbp]
\captionsetup{font=footnotesize}
\centering
\footnotesize
\renewcommand{\arraystretch}{0.5} 
\setlength{\tabcolsep}{4pt} 
\caption{LOR results for Example 1, the logistic model: Type I error rates and power (in \%) for Stage~1, Stage~2, and the combined analysis, evaluated across different working models and randomization schemes.}
\begin{tabular}{lll *{6}{c}}
\toprule
\textbf{} & \textbf{Procedure} & \textbf{Test} &
\multicolumn{3}{c}{\textbf{Type I}} & \multicolumn{3}{c}{\textbf{Power}} \\
\cmidrule(lr){4-6}\cmidrule(lr){7-9}
& & & \textbf{Stage 1} & \textbf{Stage 2} & \textbf{All} & \textbf{Stage 1} & \textbf{Stage 2} & \textbf{All} \\
\midrule
\multicolumn{3}{r}{ } & \multicolumn{3}{c}{$(\iota_1,\iota_2)=(0,0)$} & \multicolumn{3}{c}{$(\iota_1,\iota_2)=(0.3,0.4)$} \\
\midrule

\multirow{7}{*}{$\mathcal{A}_0$}
& CR &  & 5.23 & 5.05 & 5.26 & 21.77 & 29.72 & 38.84 \\

& STRPB & conv & 1.78 & 2.43 & 1.56 & 14.76 & 26.48 & 31.89 \\
& & robust & 4.89 & 4.92 & 5.04 & 28.42 & 39.29 & 51.01 \\

& HH & conv & 1.65 & 2.69 & 1.61 & 15.90 & 26.56 & 32.22 \\
& & robust & 5.08 & 5.47 & 5.26 & 29.69 & 39.63 & 51.79 \\

& PS & conv & 1.82 & 2.28 & 1.62 & 15.46 & 26.69 & 32.59 \\
& & robust & 4.90 & 4.90 & 4.86 & 28.50 & 39.31 & 51.07 \\
\midrule

\multirow{7}{*}{$\mathcal{A}_1$}
& CR &  & 5.13 & 5.11 & 4.99 & 23.76 & 31.94 & 41.80 \\

& STRPB & conv & 2.27 & 2.88 & 2.05 & 17.69 & 29.34 & 36.12 \\
& & robust & 4.90 & 5.02 & 5.04 & 28.45 & 39.35 & 51.10 \\

& HH & conv & 2.22 & 3.23 & 2.20 & 18.57 & 29.58 & 36.74 \\
& & robust & 5.11 & 5.45 & 5.17 & 29.67 & 39.63 & 51.77 \\

& PS & conv & 2.42 & 2.89 & 2.12 & 18.25 & 29.80 & 37.02 \\
& & robust & 4.93 & 4.88 & 4.93 & 28.58 & 39.39 & 51.02 \\
\midrule

\multirow{7}{*}{$\mathcal{A}_2$}
& CR &  & 5.51 & 5.25 & 5.30 & 33.18 & 43.80 & 57.91 \\

& STRPB & conv & 4.88 & 4.77 & 4.70 & 31.59 & 44.55 & 56.97 \\
& & robust & 4.95 & 4.78 & 4.75 & 31.95 & 44.66 & 57.17 \\

& HH & conv & 5.05 & 5.28 & 5.07 & 32.37 & 44.15 & 58.00 \\
& & robust & 5.21 & 5.30 & 5.16 & 32.64 & 44.18 & 58.29 \\

& PS & conv & 4.93 & 5.04 & 5.02 & 31.92 & 43.98 & 58.11 \\
& & robust & 4.98 & 5.05 & 5.04 & 32.07 & 43.98 & 58.19 \\
\bottomrule
\end{tabular}
\label{tb:sim1_LOR}
\end{table}

\begin{table}[htbp]
\captionsetup{font=footnotesize}
\centering
\footnotesize
\renewcommand{\arraystretch}{0.5} 
\setlength{\tabcolsep}{4pt} 
\caption{ATE results for Example 1, the logistic model: Type I error rates and power (in \%) for Stage~1, Stage~2, and the combined analysis, evaluated across different working models and randomization schemes.}
\begin{tabular}{lll *{6}{c}}
\toprule
\textbf{} & \textbf{Procedure} & \textbf{Test} &
\multicolumn{3}{c}{\textbf{Type I}} & \multicolumn{3}{c}{\textbf{Power}} \\
\cmidrule(lr){4-6}\cmidrule(lr){7-9}
& & & \textbf{Stage 1} & \textbf{Stage 2} & \textbf{All} & \textbf{Stage 1} & \textbf{Stage 2} & \textbf{All} \\
\midrule
\multicolumn{3}{r}{ } & \multicolumn{3}{c}{$(\iota_1,\iota_2)=(0,0)$} & \multicolumn{3}{c}{$(\iota_1,\iota_2)=(0.3,0.4)$} \\
\midrule

\multirow{7}{*}{$\mathcal{A}_0$}
& CR &  & 5.33 & 5.10 & 5.38 & 22.27 & 30.02 & 39.39 \\

& STRPB & conv & 1.81 & 2.46 & 1.58 & 15.10 & 26.70 & 32.25 \\
& & robust & 4.96 & 4.95 & 5.10 & 28.65 & 39.40 & 51.37 \\

& HH & conv & 1.67 & 2.74 & 1.71 & 16.25 & 26.87 & 32.69 \\
& & robust & 5.16 & 5.52 & 5.33 & 29.89 & 39.83 & 52.16 \\

& PS & conv & 1.89 & 2.38 & 1.64 & 15.87 & 27.01 & 32.97 \\
& & robust & 4.97 & 4.98 & 4.99 & 28.73 & 39.48 & 51.38 \\
\midrule

\multirow{7}{*}{$\mathcal{A}_1$}
& CR &  & 5.28 & 5.14 & 5.21 & 24.18 & 32.18 & 42.24 \\

& STRPB & conv & 2.33 & 2.88 & 2.12 & 18.00 & 29.58 & 36.57 \\
& & robust & 4.99 & 5.03 & 5.11 & 28.71 & 39.56 & 51.45 \\

& HH & conv & 2.27 & 3.29 & 2.27 & 18.87 & 29.78 & 37.16 \\
& & robust & 5.17 & 5.51 & 5.28 & 30.04 & 39.79 & 52.05 \\

& PS & conv & 2.51 & 2.92 & 2.19 & 18.64 & 30.06 & 37.36 \\
& & robust & 5.05 & 4.99 & 5.08 & 28.86 & 39.59 & 51.34 \\
\midrule

\multirow{7}{*}{$\mathcal{A}_2$}
& CR &  & 5.60 & 5.28 & 5.38 & 33.43 & 43.99 & 58.16 \\

& STRPB & conv & 4.92 & 4.81 & 4.76 & 31.83 & 44.69 & 57.19 \\
& & robust & 5.00 & 4.82 & 4.79 & 32.22 & 44.79 & 57.39 \\

& HH & conv & 5.17 & 5.35 & 5.18 & 32.59 & 44.21 & 58.29 \\
& & robust & 5.28 & 5.37 & 5.25 & 32.87 & 44.31 & 58.56 \\

& PS & conv & 4.99 & 5.06 & 5.07 & 32.20 & 44.08 & 58.29 \\
& & robust & 5.03 & 5.06 & 5.07 & 32.35 & 44.11 & 58.37 \\
\bottomrule
\end{tabular}
\label{tb:sim1_ATE}
\end{table}

\subsection{Example 2}
\begin{figure}[htbp] 
  \centering
  \includegraphics[width=\textwidth]{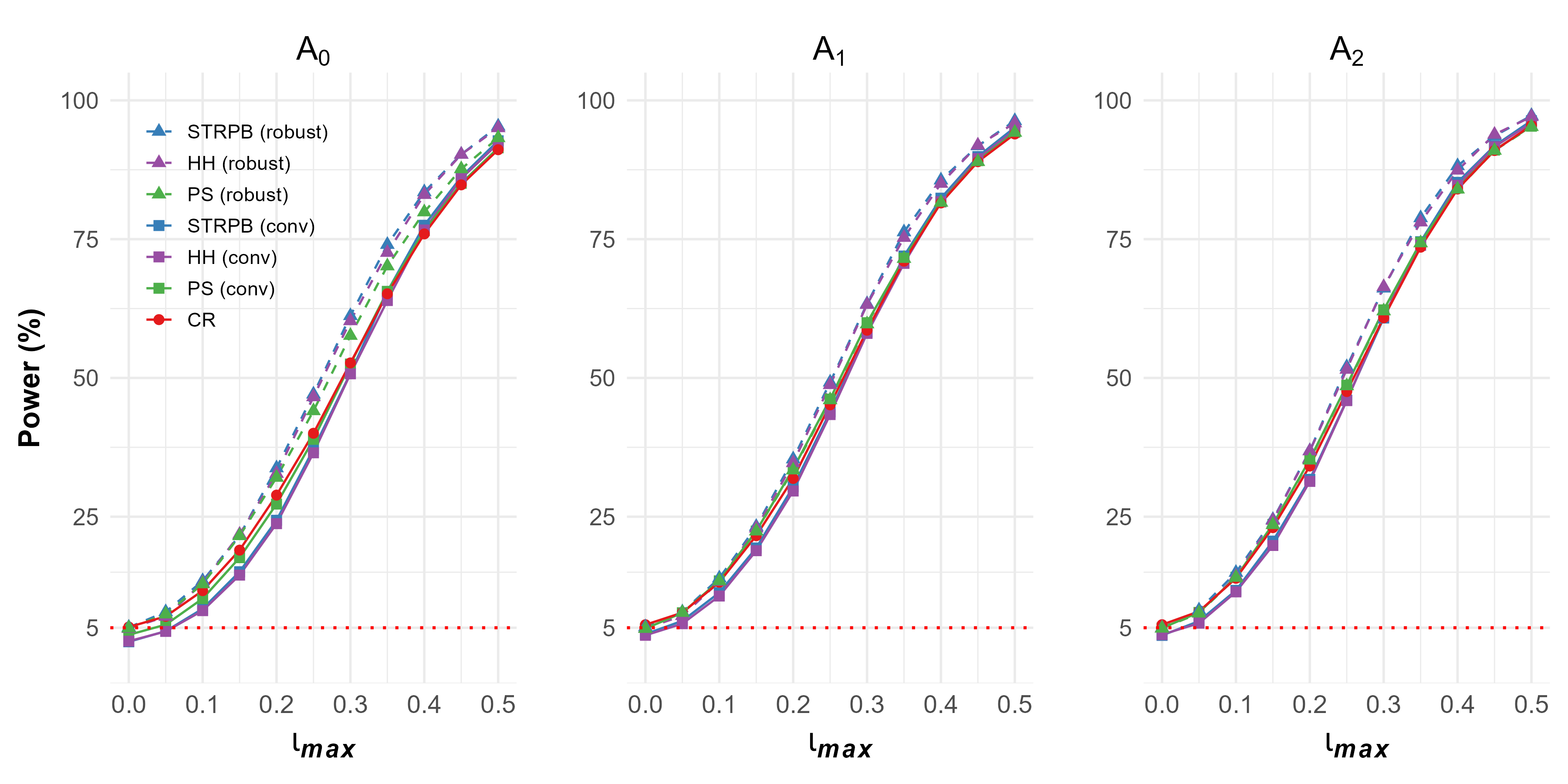}
    \caption{
   $\log$RR for Example~2, power comparison of conventional and model-robust tests 
under $(\mathcal{A}_0, \mathcal{A}_1, \mathcal{A}_2)$ across $\iota_{\max}$; 
the red dashed line denotes $\alpha=0.05$ under the null hypothesis.
}
\label{fig:sim2_logRR}
\end{figure}

\begin{figure}[H] 
  \centering
  \includegraphics[width=\textwidth]{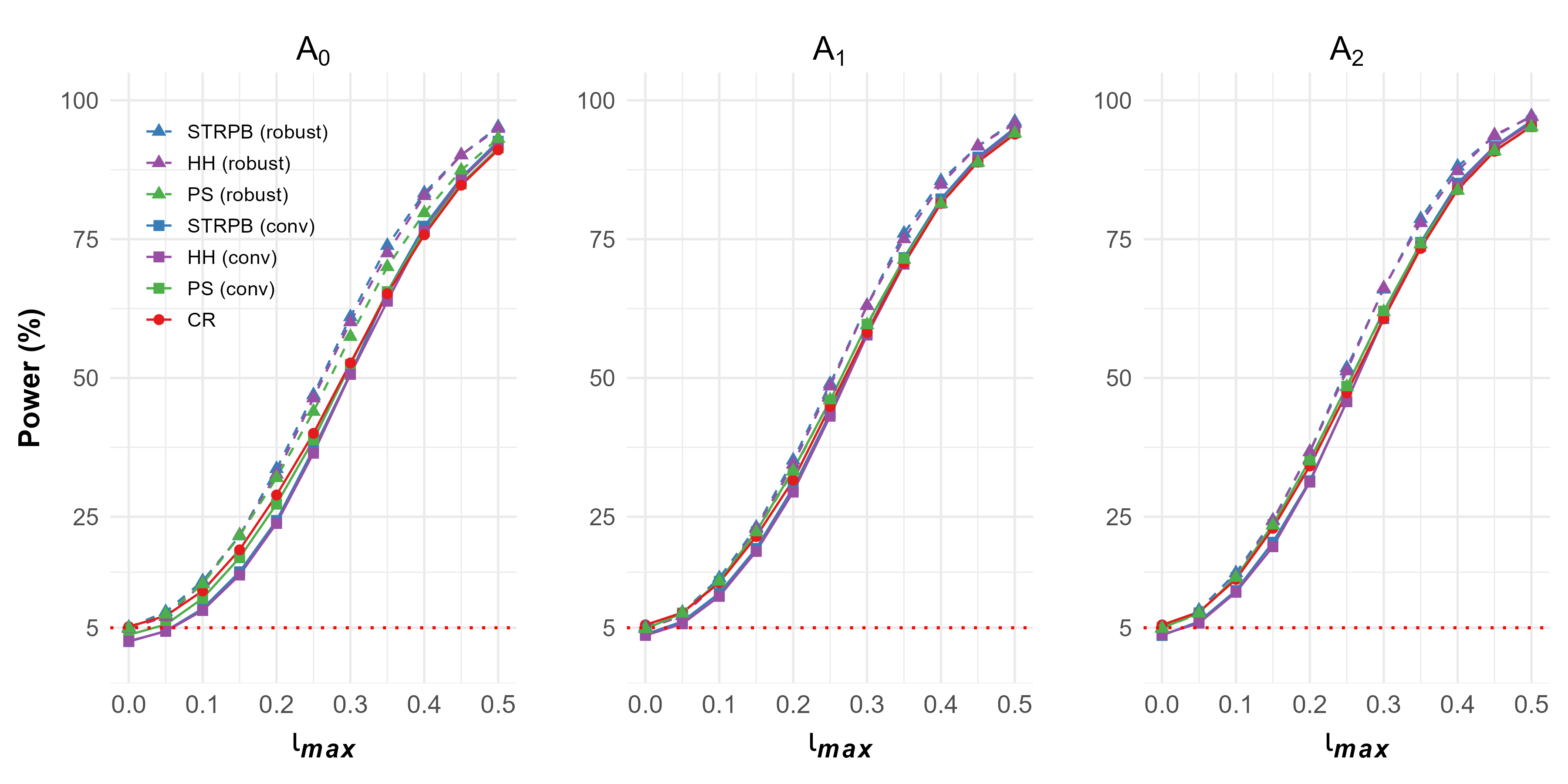}
  \caption{
    LOR for Example~2, power comparison of conventional and model-robust tests 
under $(\mathcal{A}_0, \mathcal{A}_1, \mathcal{A}_2)$ across $\iota_{\max}$; 
the red dashed line denotes $\alpha=0.05$ under the null hypothesis.
  }
  \label{fig:sim2_LOR}
\end{figure}

\begin{figure}[H] 
  \centering
  \includegraphics[width=\textwidth]{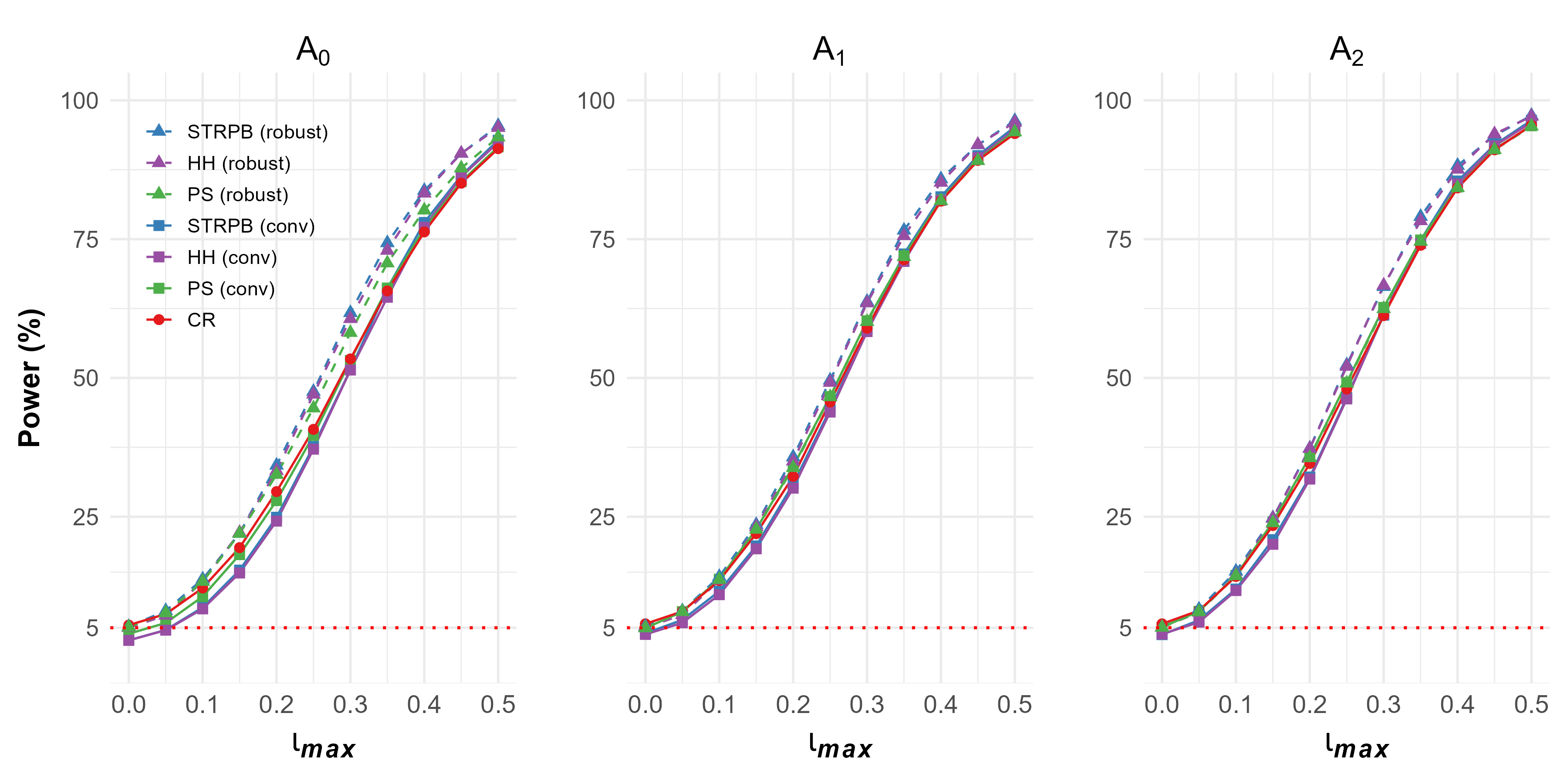}
  \caption{
    ATE for Example~2, power comparison of conventional and model-robust tests 
under $(\mathcal{A}_0, \mathcal{A}_1, \mathcal{A}_2)$ across $\iota_{\max}$; 
the red dashed line denotes $\alpha=0.05$ under the null hypothesis.
  }
  \label{fig:sim2_ATE}
\end{figure}

In this section, Tables \ref{tb:sim2_conv}, \ref{tb:sim2_LOR}, \ref{tb:sim2_ATE} and Figures \ref{fig:sim2_logRR}, \ref{fig:sim2_LOR}, \ref{fig:sim2_ATE} present additional simulation results from
Example 2. More specifically, Table~\ref{tb:sim2_conv} reports the results of conventional Wald tests across different estimands and randomization procedures; Tables \ref{tb:sim2_LOR} and \ref{tb:sim2_ATE} compare the conventional Wald test with the proposed model-robust version for the LOR and ATE metrics, respectively; Figures \ref{fig:sim2_logRR}, \ref{fig:sim2_LOR} and \ref{fig:sim2_ATE} display the statistical power of the conventional and model-robust tests under different randomization schemes as $\iota_{\max}$ increases for $\log$RR, LOR and ATE, respectively.

\begin{table}[H]
\centering
\footnotesize
\captionsetup{font=scriptsize}
\setlength{\tabcolsep}{2pt}
\renewcommand{\arraystretch}{0.4}
\caption{Results from conventional Wald tests for Example 2, the probit model: test statistics, SD ($\times \sqrt{n}$), Type I error rates (in \%) for Stage~1, Stage~2, and the overall Type I error, evaluated across different working models and randomization schemes: $(\iota_1,\iota_2)=(0,0)$.}
\begin{tabular}{llllcccccc}
\toprule
\textbf{Metric} & \textbf{Procedure} & \textbf{} & 
& \multicolumn{2}{c}{\textbf{Stage 1}} & \multicolumn{2}{c}{\textbf{Stage 2}} & \textbf{Overall} \\
\cmidrule(lr){5-6} \cmidrule(lr){7-8} \cmidrule(lr){9-9}
& & & & \textbf{Mean (SD)} & \textbf{Type I} & \textbf{Mean (SD)} & \textbf{Type I} & \textbf{Type I} \\
\midrule
\multirow{12}{*}{RR}
& \multirow{3}{*}{CR} & $\mathcal{A}_0$ & & 0.0433 (1.9857) & 4.92 & -0.0001 (1.8169) & 5.16 & 5.13 \\
& & $\mathcal{A}_1$ & & 0.0394 (1.8350) & 5.30 & -0.0006 (1.6594) & 5.16 & 5.54 \\
& & $\mathcal{A}_2$ & & 0.0384 (1.7727) & 5.29 & -0.0005 (1.6108) & 5.31 & 5.55 \\
\cmidrule{2-9}
& \multirow{3}{*}{STRPB} & $\mathcal{A}_0$ & & 0.0375 (1.7606) & 2.59 & -0.0002 (1.5993) & 3.16 & 2.47 \\
& & $\mathcal{A}_1$ & & 0.0364 (1.7021) & 3.75 & -0.0000 (1.5422) & 4.15 & 3.85 \\
& & $\mathcal{A}_2$ & & 0.0351 (1.6381) & 3.51 & -0.0001 (1.4839) & 4.10 & 3.62 \\
\cmidrule{2-9}
& \multirow{3}{*}{HH} & $\mathcal{A}_0$ & & 0.0368 (1.7593) & 2.68 & -0.0015 (1.5987) & 3.18 & 2.58 \\
& & $\mathcal{A}_1$ & & 0.0358 (1.6961) & 3.67 & -0.0013 (1.5368) & 3.81 & 3.64 \\
& & $\mathcal{A}_2$ & & 0.0351 (1.6419) & 3.79 & -0.0011 (1.4861) & 3.90 & 3.76 \\
\cmidrule{2-9}
& \multirow{3}{*}{PS} & $\mathcal{A}_0$ & & 0.0389 (1.8800) & 3.63 & 0.0014 (1.6782) & 3.78 & 3.72 \\
& & $\mathcal{A}_1$ & & 0.0380 (1.8354) & 5.21 & 0.0014 (1.6172) & 4.97 & 4.84 \\
& & $\mathcal{A}_2$ & & 0.0370 (1.7732) & 5.06 & 0.0013 (1.5671) & 4.96 & 4.89 \\
\midrule
\multirow{12}{*}{LOR}
& \multirow{3}{*}{CR} & $\mathcal{A}_0$ & & 0.0976 (4.5811) & 4.85 & -0.0003 (4.0916) & 5.24 & 5.18 \\
& & $\mathcal{A}_1$ & & 0.0889 (4.2311) & 5.18 & -0.0015 (3.7427) & 5.14 & 5.51 \\
& & $\mathcal{A}_2$ & & 0.0867 (4.0834) & 5.20 & -0.0013 (3.6324) & 5.29 & 5.51 \\
\cmidrule{2-9}
& \multirow{3}{*}{STRPB} & $\mathcal{A}_0$ & & 0.0848 (4.0654) & 2.58 & -0.0006 (3.6036) & 3.18 & 2.50 \\
& & $\mathcal{A}_1$ & & 0.0823 (3.9315) & 3.59 & -0.0001 (3.4761) & 4.19 & 3.80 \\
& & $\mathcal{A}_2$ & & 0.0794 (3.7797) & 3.47 & -0.0002 (3.3440) & 4.14 & 3.61 \\
\cmidrule{2-9}
& \multirow{3}{*}{HH} & $\mathcal{A}_0$ & & 0.0833 (4.0511) & 2.63 & -0.0033 (3.6028) & 3.18 & 2.59 \\
& & $\mathcal{A}_1$ & & 0.0811 (3.9080) & 3.61 & -0.0030 (3.4645) & 3.81 & 3.62 \\
& & $\mathcal{A}_2$ & & 0.0794 (3.7791) & 3.74 & -0.0024 (3.3487) & 3.91 & 3.76 \\
\cmidrule{2-9}
& \multirow{3}{*}{PS} & $\mathcal{A}_0$ & & 0.0876 (4.3333) & 3.60 & 0.0033 (3.7798) & 3.80 & 3.73 \\
& & $\mathcal{A}_1$ & & 0.0858 (4.2351) & 5.10 & 0.0032 (3.6474) & 4.99 & 4.78 \\
& & $\mathcal{A}_2$ & & 0.0835 (4.0882) & 4.96 & 0.0030 (3.5322) & 5.02 & 4.87 \\
\midrule
\multirow{12}{*}{ATE}
& \multirow{3}{*}{CR} & $\mathcal{A}_0$ & & 0.0239 (1.1167) & 5.12 & -0.0001 (1.0058) & 5.30 & 5.48 \\
& & $\mathcal{A}_1$ & & 0.0218 (1.0321) & 5.37 & -0.0004 (0.9200) & 5.23 & 5.70 \\
& & $\mathcal{A}_2$ & & 0.0213 (0.9967) & 5.45 & -0.0003 (0.8930) & 5.38 & 5.70 \\
\cmidrule{2-9}
& \multirow{3}{*}{STRPB} & $\mathcal{A}_0$ & & 0.0208 (0.9917) & 2.77 & -0.0001 (0.8860) & 3.19 & 2.74 \\
& & $\mathcal{A}_1$ & & 0.0202 (0.9592) & 3.81 & -0.0000 (0.8547) & 4.27 & 4.00 \\
& & $\mathcal{A}_2$ & & 0.0195 (0.9227) & 3.60 & -0.0000 (0.8224) & 4.17 & 3.76 \\
\cmidrule{2-9}
& \multirow{3}{*}{HH} & $\mathcal{A}_0$ & & 0.0204 (0.9890) & 2.89 & -0.0008 (0.8859) & 3.21 & 2.73 \\
& & $\mathcal{A}_1$ & & 0.0199 (0.9541) & 3.77 & -0.0007 (0.8519) & 3.84 & 3.80 \\
& & $\mathcal{A}_2$ & & 0.0195 (0.9231) & 3.84 & -0.0006 (0.8236) & 3.95 & 3.85 \\
\cmidrule{2-9}
& \multirow{3}{*}{PS} & $\mathcal{A}_0$ & & 0.0215 (1.0569) & 3.95 & 0.0008 (0.9292) & 3.87 & 3.91 \\
& & $\mathcal{A}_1$ & & 0.0211 (1.0329) & 5.17 & 0.0008 (0.8966) & 5.08 & 4.96 \\
& & $\mathcal{A}_2$ & & 0.0205 (0.9976) & 5.11 & 0.0007 (0.8685) & 5.05 & 5.05 \\
\bottomrule
\end{tabular}
\label{tb:sim2_conv}
\end{table}

\begin{table}[htbp]
\captionsetup{font=footnotesize}
\centering
\footnotesize
\renewcommand{\arraystretch}{0.5} 
\setlength{\tabcolsep}{4pt} 
\caption{LOR results for Example 2, the probit model: Type I error rates and power (in \%) for Stage~1, Stage~2, and the combined analysis, evaluated across different working models and randomization schemes.}
\begin{tabular}{lll *{6}{c}}
\toprule
\textbf{} & \textbf{Procedure} & \textbf{Test} &
\multicolumn{3}{c}{\textbf{Type I}} & \multicolumn{3}{c}{\textbf{Power}} \\
\cmidrule(lr){4-6}\cmidrule(lr){7-9}
& & & \textbf{Stage 1} & \textbf{Stage 2} & \textbf{All} & \textbf{Stage 1} & \textbf{Stage 2} & \textbf{All} \\
\midrule
\multicolumn{3}{r}{ } & \multicolumn{3}{c}{$(\iota_1,\iota_2)=(0,0)$} & \multicolumn{3}{c}{$(\iota_1,\iota_2)=(0.2,0.3)$} \\
\midrule

\multirow{7}{*}{$\mathcal{A}_0$}
& CR &  & 5.18 & 5.24 & 4.85 & 26.47 & 37.46 & 48.59 \\

& STRPB & conv & 2.58 & 3.18 & 2.50 & 23.72 & 36.22 & 46.83 \\
& & robust & 4.71 & 4.84 & 5.06 & 31.45 & 43.57 & 56.82 \\

& HH & conv & 2.63 & 3.18 & 2.59 & 22.75 & 35.96 & 46.77 \\
& & robust & 4.54 & 4.91 & 4.85 & 30.77 & 42.56 & 55.88 \\

& PS & conv & 3.60 & 3.80 & 3.73 & 25.00 & 37.13 & 48.16 \\
& & robust & 4.79 & 5.25 & 4.92 & 28.70 & 41.42 & 53.21 \\
\midrule

\multirow{7}{*}{$\mathcal{A}_1$}
& CR &  & 5.51 & 5.14 & 5.18 & 29.87 & 41.52 & 54.06 \\

& STRPB & conv & 3.59 & 4.19 & 3.80 & 28.62 & 41.83 & 53.51 \\
& & robust & 4.99 & 5.02 & 5.11 & 33.11 & 45.55 & 58.63 \\

& HH & conv & 3.61 & 3.81 & 3.62 & 27.42 & 41.34 & 53.57 \\
& & robust & 4.89 & 4.85 & 4.84 & 32.07 & 44.99 & 58.41 \\

& PS & conv & 5.10 & 4.99 & 4.78 & 30.31 & 43.09 & 55.41 \\
& & robust & 5.02 & 5.01 & 4.80 & 30.00 & 43.07 & 55.32 \\
\midrule

\multirow{7}{*}{$\mathcal{A}_2$}
& CR &  & 5.51 & 5.29 & 5.20 & 31.84 & 43.54 & 56.22 \\

& STRPB & conv & 3.47 & 4.14 & 3.61 & 29.86 & 44.29 & 56.11 \\
& & robust & 4.81 & 5.06 & 4.93 & 35.04 & 48.36 & 61.68 \\

& HH & conv & 3.74 & 3.91 & 3.76 & 29.27 & 43.78 & 56.44 \\
& & robust & 4.90 & 4.91 & 4.96 & 34.38 & 47.66 & 61.40 \\

& PS & conv & 5.08 & 4.97 & 4.94 & 31.78 & 45.16 & 57.69 \\
& & robust & 4.96 & 5.02 & 4.87 & 31.50 & 45.29 & 57.55 \\
\bottomrule
\end{tabular}
\label{tb:sim2_LOR}
\end{table}

\begin{table}[htbp]
\captionsetup{font=footnotesize}
\centering
\footnotesize
\renewcommand{\arraystretch}{0.5} 
\setlength{\tabcolsep}{4pt} 
\caption{ATE results for Example 2, the probit model: Type I error rates and power (in \%) for Stage~1, Stage~2, and the combined analysis, evaluated across different working models and randomization schemes.}
\begin{tabular}{lll *{6}{c}}
\toprule
\textbf{} & \textbf{Procedure} & \textbf{Test} &
\multicolumn{3}{c}{\textbf{Type I}} & \multicolumn{3}{c}{\textbf{Power}} \\
\cmidrule(lr){4-6}\cmidrule(lr){7-9}
& & & \textbf{Stage 1} & \textbf{Stage 2} & \textbf{All} & \textbf{Stage 1} & \textbf{Stage 2} & \textbf{All} \\
\midrule
\multicolumn{3}{r}{ } & \multicolumn{3}{c}{$(\iota_1,\iota_2)=(0,0)$} & \multicolumn{3}{c}{$(\iota_1,\iota_2)=(0.2,0.3)$} \\
\midrule

\multirow{7}{*}{$\mathcal{A}_0$}
& CR &  & 5.48 & 5.30 & 5.12 & 49.36 & 37.70 & 27.22 \\

& STRPB & conv & 2.77 & 3.19 & 2.74 & 24.51 & 36.59 & 47.51 \\
& & robust & 4.94 & 4.89 & 5.20 & 32.29 & 43.74 & 57.49 \\

& HH & conv & 2.89 & 3.21 & 2.73 & 23.57 & 36.17 & 47.47 \\
& & robust & 4.79 & 4.97 & 5.03 & 31.52 & 42.76 & 56.47 \\

& PS & conv & 3.95 & 3.87 & 3.91 & 25.87 & 37.44 & 48.89 \\
& & robust & 4.97 & 5.32 & 5.10 & 29.55 & 41.62 & 53.84 \\
\midrule

\multirow{7}{*}{$\mathcal{A}_1$}
& CR &  & 5.70 & 5.23 & 5.37 & 54.84 & 41.72 & 30.75 \\

& STRPB & conv & 3.81 & 4.27 & 4.00 & 29.52 & 41.99 & 54.10 \\
& & robust & 5.19 & 5.07 & 5.26 & 33.95 & 45.73 & 59.35 \\

& HH & conv & 3.77 & 3.84 & 3.80 & 28.12 & 41.56 & 54.21 \\
& & robust & 5.03 & 4.87 & 5.02 & 32.78 & 45.12 & 59.05 \\

& PS & conv & 5.28 & 5.07 & 4.93 & 31.06 & 43.24 & 56.12 \\
& & robust & 5.17 & 5.08 & 4.96 & 30.75 & 43.26 & 55.97 \\
\midrule

\multirow{7}{*}{$\mathcal{A}_2$}
& CR &  & 5.70 & 5.38 & 5.45 & 56.79 & 43.71 & 32.65 \\

& STRPB & conv & 3.60 & 4.17 & 3.76 & 30.60 & 44.51 & 56.84 \\
& & robust & 4.95 & 5.08 & 5.09 & 35.86 & 48.52 & 62.07 \\

& HH & conv & 3.84 & 3.95 & 3.85 & 30.17 & 43.98 & 56.94 \\
& & robust & 5.12 & 4.94 & 5.09 & 35.12 & 47.87 & 62.07 \\

& PS & conv & 5.23 & 5.01 & 5.11 & 32.64 & 45.42 & 58.35 \\
& & robust & 5.11 & 5.05 & 5.05 & 32.24 & 45.48 & 58.04 \\
\bottomrule
\end{tabular}
\label{tb:sim2_ATE}
\end{table}

\subsection{Hypothetical Trial Example}
In this section, Tables \ref{tb:realdata_conv}, \ref{tb:realdata_LOR} and \ref{tb:realdata_ATE} present additional results from the hypothetical trial example. More specifically, Table~\ref{tb:realdata_conv} reports the results of conventional Wald tests across different estimands and randomization procedures; Tables \ref{tb:realdata_LOR} and \ref{tb:realdata_ATE} compare the conventional Wald test with the proposed model-robust version for the LOR and ATE metrics.

\begin{table}[htbp]
\centering
\footnotesize
\captionsetup{font=scriptsize} 
\setlength{\tabcolsep}{2pt} 
\renewcommand{\arraystretch}{0.4} 
\caption{Results from conventional Wald tests for the hypothetical trial: test statistics, SD ($\times \sqrt{n}$), Type I error rates (in \%) for Stage~1, Stage~2, and the overall Type I error, evaluated across different working models and randomization schemes: $(\iota_1,\iota_2)=(0,0)$.}
\begin{tabular}{llllcccccc}
\toprule
\textbf{Metric} & \textbf{Procedure} & \textbf{} & 
& \multicolumn{2}{c}{\textbf{Stage 1}} & \multicolumn{2}{c}{\textbf{Stage 2}} & \textbf{Overall} \\
\cmidrule(lr){5-6} \cmidrule(lr){7-8} \cmidrule(lr){9-9}
& & & & \textbf{Mean (SD)} & \textbf{Type I} & \textbf{Mean (SD)} & \textbf{Type I} & \textbf{Type I} \\
\midrule
\multirow{12}{*}{$\log$RR} & \multirow{3}{*}{CR} & $\mathcal{A}_{0}$ &  & 0.1040 (4.6883) & 4.90 & -0.0013 (4.2830) & 4.86 & 4.76 \\
 &  & $\mathcal{A}_{1}$ &  & 0.1003 (4.6080) & 4.73 & 0.0019 (4.1792) & 5.12 & 4.87 \\
 &  & $\mathcal{A}_{2}$ &  & 0.0976 (4.4921) & 4.74 & 0.0019 (4.0624) & 5.18 & 5.09 \\
\cmidrule{2-9}
 & \multirow{3}{*}{STRPB} & $\mathcal{A}_{0}$ &  & 0.0960 (4.4737) & 3.99 & -0.0025 (4.0903) & 3.81 & 3.33 \\
 &  & $\mathcal{A}_{1}$ &  & 0.0960 (4.4548) & 4.36 & -0.0026 (4.0635) & 4.11 & 3.76 \\
 &  & $\mathcal{A}_{2}$ &  & 0.1017 (4.4499) & 5.05 & -0.0013 (4.0527) & 4.79 & 4.81 \\
\cmidrule{2-9}
 & \multirow{3}{*}{HH} & $\mathcal{A}_{0}$ &  & 0.0988 (4.5224) & 3.89 & -0.0001 (4.1135) & 4.30 & 4.11 \\
 &  & $\mathcal{A}_{1}$ &  & 0.0984 (4.4997) & 4.07 & -0.0000 (4.0993) & 4.74 & 4.54 \\
 &  & $\mathcal{A}_{2}$ &  & 0.1004 (4.4611) & 4.83 & 0.0004 (4.0596) & 4.64 & 4.90 \\
\cmidrule{2-9}
 & \multirow{3}{*}{PS} & $\mathcal{A}_{0}$ &  & 0.1005 (4.5501) & 4.23 & 0.0008 (4.1135) & 4.10 & 3.98 \\
 &  & $\mathcal{A}_{1}$ &  & 0.1002 (4.5288) & 4.69 & 0.0008 (4.0896) & 4.46 & 4.43 \\
 &  & $\mathcal{A}_{2}$ &  & 0.0957 (4.4654) & 4.67 & 0.0025 (4.0555) & 4.96 & 4.68 \\
\midrule
\multirow{12}{*}{LOR} & \multirow{3}{*}{CR} & $\mathcal{A}_{0}$ &  & 0.1273 (5.7768) & 4.97 & -0.0015 (5.2392) & 4.98 & 4.86 \\
 &  & $\mathcal{A}_{1}$ &  & 0.1227 (5.6778) & 4.82 & 0.0024 (5.1142) & 5.17 & 5.01 \\
 &  & $\mathcal{A}_{2}$ &  & 0.1194 (5.5356) & 4.86 & 0.0023 (4.9712) & 5.21 & 5.24 \\
\cmidrule{2-9}
 & \multirow{3}{*}{STRPB} & $\mathcal{A}_{0}$ &  & 0.1175 (5.5109) & 4.10 & -0.0030 (5.0007) & 3.89 & 3.50 \\
 &  & $\mathcal{A}_{1}$ &  & 0.1174 (5.4879) & 4.46 & -0.0032 (4.9681) & 4.16 & 3.86 \\
 &  & $\mathcal{A}_{2}$ &  & 0.1244 (5.4822) & 5.19 & -0.0016 (4.9581) & 4.83 & 4.92 \\
\cmidrule{2-9}
 & \multirow{3}{*}{HH} & $\mathcal{A}_{0}$ &  & 0.1208 (5.5749) & 4.00 & -0.0002 (5.0325) & 4.36 & 4.18 \\
 &  & $\mathcal{A}_{1}$ &  & 0.1204 (5.5461) & 4.19 & 0.0000 (5.0150) & 4.81 & 4.62 \\
 &  & $\mathcal{A}_{2}$ &  & 0.1229 (5.4960) & 4.97 & 0.0004 (4.9639) & 4.68 & 5.00 \\
\cmidrule{2-9}
 & \multirow{3}{*}{PS} & $\mathcal{A}_{0}$ &  & 0.1230 (5.6039) & 4.34 & 0.0010 (5.0305) & 4.16 & 4.08 \\
 &  & $\mathcal{A}_{1}$ &  & 0.1225 (5.5769) & 4.81 & 0.0009 (5.0016) & 4.56 & 4.56 \\
 &  & $\mathcal{A}_{2}$ &  & 0.1171 (5.5024) & 4.79 & 0.0031 (4.9601) & 5.02 & 4.90 \\
\midrule
\multirow{12}{*}{ATE} & \multirow{3}{*}{CR} & $\mathcal{A}_{0}$ &  & 0.0189 (0.8828) & 5.12 & -0.0002 (0.7813) & 5.04 & 5.08 \\
 &  & $\mathcal{A}_{1}$ &  & 0.0182 (0.8679) & 4.98 & 0.0004 (0.7637) & 5.27 & 5.24 \\
 &  & $\mathcal{A}_{2}$ &  & 0.0177 (0.8467) & 4.99 & 0.0004 (0.7424) & 5.34 & 5.41 \\
\cmidrule{2-9}
 & \multirow{3}{*}{STRPB} & $\mathcal{A}_{0}$ &  & 0.0174 (0.8418) & 4.23 & -0.0004 (0.7445) & 4.06 & 3.68 \\
 &  & $\mathcal{A}_{1}$ &  & 0.0174 (0.8385) & 4.58 & -0.0005 (0.7397) & 4.21 & 4.00 \\
 &  & $\mathcal{A}_{2}$ &  & 0.0185 (0.8377) & 5.31 & -0.0002 (0.7397) & 4.94 & 5.21 \\
\cmidrule{2-9}
 & \multirow{3}{*}{HH} & $\mathcal{A}_{0}$ &  & 0.0179 (0.8535) & 4.09 & -0.0000 (0.7509) & 4.49 & 4.36 \\
 &  & $\mathcal{A}_{1}$ &  & 0.0178 (0.8488) & 4.31 & 0.0000 (0.7482) & 5.00 & 4.82 \\
 &  & $\mathcal{A}_{2}$ &  & 0.0183 (0.8398) & 5.09 & 0.0000 (0.7393) & 4.81 & 5.18 \\
\cmidrule{2-9}
 & \multirow{3}{*}{PS} & $\mathcal{A}_{0}$ &  & 0.0182 (0.8554) & 4.45 & 0.0002 (0.7496) & 4.34 & 4.27 \\
 &  & $\mathcal{A}_{1}$ &  & 0.0182 (0.8509) & 4.90 & 0.0002 (0.7455) & 4.68 & 4.72 \\
 &  & $\mathcal{A}_{2}$ &  & 0.0174 (0.8413) & 4.97 & 0.0005 (0.7394) & 5.10 & 5.09 \\
\bottomrule
\end{tabular}
\label{tb:realdata_conv}
\end{table}

\begin{table}[htbp]
\captionsetup{font=scriptsize}
\centering
\scriptsize
\renewcommand{\arraystretch}{0.35}
\setlength{\tabcolsep}{3pt}
\setlength{\aboverulesep}{0pt}
\setlength{\belowrulesep}{0pt}
\setlength{\abovetopsep}{0pt}
\setlength{\belowbottomsep}{0pt}
\caption{LOR results for the hypothetical trial: Type I error rates and power (in $10^{-2}$) for Stage~1, Stage~2, and the combined analysis, across different working models and randomization schemes.}
\begin{tabular}{lll *{6}{c}}
\toprule
\textbf{} & \textbf{Procedure} & \textbf{Test} &
\multicolumn{3}{c}{\textbf{Type I}} & \multicolumn{3}{c}{\textbf{Power}} \\
\cmidrule(lr){4-6}\cmidrule(lr){7-9}
& & & \textbf{Stage 1} & \textbf{Stage 2} & \textbf{All} & \textbf{Stage 1} & \textbf{Stage 2} & \textbf{All} \\
\midrule
\multicolumn{3}{r}{ } & \multicolumn{3}{c}{$(\iota_1,\iota_2)=(0,0)$} & \multicolumn{3}{c}{$(\iota_1,\iota_2)=(0.2,0.4)$} \\
\midrule

\multirow{7}{*}{$\mathcal{A}_0$}
& CR &  & 4.97 & 4.98 & 4.86 & 28.42 & 42.17 & 52.52 \\

& STRPB & conv & 4.10 & 3.89 & 3.50 & 26.52 & 42.17 & 51.90 \\
& & robust & 5.11 & 5.07 & 5.06 & 30.28 & 45.16 & 56.43 \\

& HH & conv & 4.00 & 4.36 & 4.18 & 26.75 & 41.50 & 51.47 \\
& & robust & 4.83 & 4.90 & 4.82 & 30.71 & 44.70 & 55.85 \\

& PS & conv & 4.34 & 4.16 & 4.08 & 27.47 & 42.07 & 52.00 \\
& & robust & 5.37 & 4.91 & 5.07 & 30.68 & 44.95 & 55.84 \\
\midrule

\multirow{7}{*}{$\mathcal{A}_1$}
& CR &  & 4.82 & 5.17 & 5.01 & 28.99 & 44.14 & 54.23 \\

& STRPB & conv & 4.46 & 4.16 & 3.86 & 28.10 & 43.58 & 54.18 \\
& & robust & 5.08 & 4.55 & 4.55 & 30.75 & 45.66 & 57.09 \\

& HH & conv & 4.19 & 4.81 & 4.62 & 28.38 & 43.16 & 53.92 \\
& & robust & 4.82 & 5.35 & 5.30 & 31.10 & 45.44 & 56.62 \\

& PS & conv & 4.81 & 4.56 & 4.56 & 29.30 & 43.61 & 54.07 \\
& & robust & 5.38 & 5.02 & 5.18 & 31.12 & 45.31 & 56.66 \\
\midrule

\multirow{7}{*}{$\mathcal{A}_2$}
& CR &  & 4.86 & 5.21 & 5.24 & 30.53 & 45.68 & 56.85 \\

& STRPB & conv & 5.19 & 4.83 & 4.92 & 30.64 & 45.72 & 56.97 \\
& & robust & 5.17 & 4.60 & 4.58 & 30.90 & 45.77 & 57.12 \\

& HH & conv & 4.97 & 4.68 & 5.00 & 30.77 & 45.39 & 56.33 \\
& & robust & 4.90 & 5.38 & 5.33 & 31.11 & 45.45 & 56.50 \\

& PS & conv & 4.79 & 5.02 & 4.90 & 31.33 & 45.39 & 56.97 \\
& & robust & 5.35 & 5.03 & 5.21 & 31.52 & 45.41 & 57.04 \\
\bottomrule
\end{tabular}
\label{tb:realdata_LOR}
\end{table}

\begin{table}[htbp]
\captionsetup{font=footnotesize}
\centering
\footnotesize
\renewcommand{\arraystretch}{0.5} 
\setlength{\tabcolsep}{4pt} 
\caption{ATE results for the hypothetical trial: Type I error rates and power (in \%) for Stage~1, Stage~2, and the combined analysis, evaluated across different working models and randomization schemes.}
\begin{tabular}{lll *{6}{c}}
\toprule
\textbf{} & \textbf{Procedure} & \textbf{Test} &
\multicolumn{3}{c}{\textbf{Type I}} & \multicolumn{3}{c}{\textbf{Power}} \\
\cmidrule(lr){4-6}\cmidrule(lr){7-9}
& & & \textbf{Stage 1} & \textbf{Stage 2} & \textbf{All} & \textbf{Stage 1} & \textbf{Stage 2} & \textbf{All} \\
\midrule
\multicolumn{3}{r}{ } & \multicolumn{3}{c}{$(\iota_1,\iota_2)=(0,0)$} & \multicolumn{3}{c}{$(\iota_1,\iota_2)=(0.2,0.4)$} \\
\midrule

\multirow{7}{*}{$\mathcal{A}_0$}
& CR &  & 5.12 & 5.04 & 5.08 & 28.71 & 42.51 & 53.10 \\

& STRPB & conv & 4.23 & 4.06 & 3.68 & 26.83 & 42.50 & 52.36 \\
& & robust & 5.28 & 5.21 & 5.28 & 30.69 & 45.49 & 56.92 \\

& HH & conv & 4.09 & 4.49 & 4.36 & 26.99 & 41.96 & 51.99 \\
& & robust & 4.96 & 4.96 & 5.10 & 31.11 & 45.04 & 56.40 \\

& PS & conv & 4.45 & 4.34 & 4.27 & 27.65 & 42.52 & 52.36 \\
& & robust & 5.52 & 4.99 & 5.34 & 30.97 & 45.25 & 56.30 \\
\midrule

\multirow{7}{*}{$\mathcal{A}_1$}
& CR &  & 4.98 & 5.27 & 5.24 & 29.22 & 44.51 & 54.77 \\

& STRPB & conv & 4.58 & 4.21 & 4.00 & 28.45 & 43.93 & 54.64 \\
& & robust & 5.21 & 4.70 & 4.77 & 31.13 & 46.01 & 57.46 \\

& HH & conv & 4.31 & 5.00 & 4.82 & 28.82 & 43.49 & 54.36 \\
& & robust & 4.98 & 5.48 & 5.45 & 31.49 & 45.77 & 56.98 \\

& PS & conv & 4.90 & 4.68 & 4.72 & 29.55 & 43.94 & 54.48 \\
& & robust & 5.52 & 5.13 & 5.43 & 31.43 & 45.65 & 57.19 \\
\midrule

\multirow{7}{*}{$\mathcal{A}_2$}
& CR &  & 4.99 & 5.34 & 5.41 & 30.87 & 46.05 & 57.29 \\

& STRPB & conv & 5.31 & 4.94 & 5.21 & 31.06 & 46.03 & 57.42 \\
& & robust & 5.35 & 4.76 & 4.73 & 31.29 & 46.05 & 57.58 \\

& HH & conv & 5.09 & 4.81 & 5.18 & 31.28 & 45.78 & 56.94 \\
& & robust & 5.06 & 5.50 & 5.56 & 31.55 & 45.82 & 57.05 \\

& PS & conv & 4.97 & 5.10 & 5.09 & 31.73 & 45.71 & 57.38 \\
& & robust & 5.44 & 5.14 & 5.41 & 31.87 & 45.73 & 57.47 \\
\bottomrule
\end{tabular}
\label{tb:realdata_ATE}
\end{table}

\section{An additional example of estimand: the average treatment effect (ATE)}\label{subsec:ATE}

The ATE captures absolute risk differences and is widely used for policy and clinical decision making. For arm $k$,
\[
\mathrm{ATE}_k=\mu_k-\mu_0,\qquad
\bm{\delta}=\bigl(\mu_1-\mu_0,\ldots,\mu_K-\mu_0\bigr)^\top.
\]
Estimation can proceed from generalized linear working models (including logistic regression for binary outcomes), with inference carried out analogously to $\log$RR and LOR.

\begin{customlm}{14}
\label{prop:ATE}
For the ATE, $g(x)=x$ and
\[
\bm{G}_{\sf ATE}=
\bigl(- \bm{1}_K, \ \  \mathrm{diag} \{1, \ldots, 1 \bigr)
\bigr),
\qquad
\widehat{\bm{G}}_{\sf ATE}=\bm{G}_{\sf ATE}\big|_{\mu=\widehat{\mu}}.
\]
Hence Theorem~\ref{thm:contrast_normality} implies
\[
\begin{pmatrix}
\widehat{\mu}_1-\widehat{\mu}_0\\ \vdots \\ \widehat{\mu}_K-\widehat{\mu}_0
\end{pmatrix}
-
\begin{pmatrix}
\mu_1-\mu_0\\ \vdots \\ \mu_K-\mu_0
\end{pmatrix}
\xrightarrow{d} N\!\left(0,\;\bm{G}_{\sf ATE}\,\bm{\Gamma}\,\bm{G}_{\sf ATE}^\top\right).
\]
\end{customlm}

\newpage


	\end{appendices}
	
\end{document}